\documentclass{amsart}

\usepackage{amsmath}    
\usepackage{amssymb}    
\usepackage{dsfont} 	
\usepackage{bm}         
\usepackage{mathtools}  
\usepackage[hypertexnames=false]{hyperref} 
\usepackage[scientific-notation=true]{siunitx}
\usepackage{enumerate}
\usepackage{centernot}
\usepackage{pdflscape}
\usepackage{paralist}
\usepackage{stackrel}
\usepackage[english]{babel}
\usepackage[table]{xcolor}
\usepackage{setspace}
\usepackage{caption}
\usepackage{threeparttable}

\DeclareMathOperator*{\argmax}{arg\,max} 
\DeclareMathOperator*{\argmin}{arg\,min} 
\DeclarePairedDelimiter\sqfnorm{\lVert}{\rVert_F^2} 

\renewcommand{\vec}[1]{\bm{\mathbf{#1}}}

\newcommand{\calS}{\mathcal{S}}
\newcommand{\vOmega}{{\vec{\Omega}}}
\newcommand{\vS}{\vec{S}}
\newcommand{\calL}{\mathcal{L}}
\newcommand{\vT}{\vec{T}}

\usepackage[numbers]{natbib}
\usepackage{bibunits}

\def\ci{\perp\!\!\!\perp}

\usepackage{algorithm}
\usepackage{algpseudocode}

\usepackage{listings}
\usepackage{color}

\newtheorem*{heuristic*}{Heuristic Definition}
\theoremstyle{definition}

\makeatletter
\newcommand{\addresseshere}{%
  \enddoc@text\let\enddoc@text\relax
}

\definecolor{lightgrey}{rgb}{0.9,0.9,0.9}
\definecolor{darkgreen}{rgb}{0,0.6,0}
\begin{document}

\title[Blood-Based Metabolic Signatures in Alzheimer's Disease]
{Blood-Based Metabolic Signatures \\in Alzheimer's Disease}

\author[F.A.\ de Leeuw]{Francisca A. de Leeuw*$^{\dag}$}
\thanks{* Shared first authorship.}
\thanks{$^{\dag}$ Corresponding author.}
\address[Francisca A. de Leeuw]{
Alzheimer Center \and
Dept.\ of Neurology \\
Amsterdam Neuroscience \\
VU University Medical Center Amsterdam \\
Amsterdam \\
The Netherlands; \and
Dept.\ of Clinical Chemistry \\
VU University Medical Center Amsterdam \\
Amsterdam \\
The Netherlands}
\email{f.deleeuw@vumc.nl}

\author[C.F.W.\ Peeters et al.]{Carel F.W.\ Peeters*$^{\dag}$}
\address[Carel F.W.\ Peeters]{
Dept.\ of Epidemiology \& Biostatistics \\
Amsterdam Public Health research institute \\
VU University medical center Amsterdam \\
Amsterdam \\
The Netherlands}
\email{cf.peeters@vumc.nl}

\author[]{\\Maartje I.\ Kester}
\address[Maartje I.\ Kester]{
Alzheimer Center \and
Dept.\ of Neurology \\
Amsterdam Neuroscience \\
VU University Medical Center Amsterdam \\
Amsterdam \\
The Netherlands}
\email{m.kester@vumc.nl}

\author[]{Amy C.\ Harms}
\address[Amy C.\ Harms]{
Division of Analytical Biosciences \\
Leiden Academic Centre for Drug Research \\
Leiden University \\
Leiden \\
The Netherlands}
\email{a.c.harms@lacdr.leidenuniv.nl}

\author[]{Eduard A.\ Struys}
\address[Eduard A.\ Struys]{
Dept.\ of Clinical Chemistry \\
VU University medical center Amsterdam \\
Amsterdam\\
The Netherlands}
\email{E.Struys@vumc.nl}

\author[]{\\Thomas Hankemeier}
\address[Thomas Hankemeier]{
Division of Analytical Biosciences \\
Leiden Academic Centre for Drug Research \\
Leiden University \\
Leiden \\
The Netherlands}
\email{hankemeier@lacdr.leidenuniv.nl}

\author[]{Herman W.T.\ van Vlijmen}
\address[Herman W.T.\ van Vlijmen]{
Discovery Sciences \\
Janssen Research and Development \\
Beerse \\
Belgium; \and
Division of Medicinal Chemistry \\
Leiden Academic Centre for Drug Research \\
Leiden University \\
Leiden \\
The Netherlands}
\email{hvvlijme@its.jnj.com}

\author[]{\\Sven J.\ van der Lee}
\address[Sven J.\ van der Lee]{
Genetic Epidemiology Unit \\
Dept.\ of Epidemiology \\
Erasmus MC \\
Rotterdam \\
The Netherlands; \and
Alzheimer Center\\
VU University Medical Center Amsterdam \\
Amsterdam \\
The Netherlands}
\email{s.j.vanderlee@vumc.nl}

\author[]{Cornelia M.\ van Duijn}
\address[Cornelia M.\ van Duijn]{
Genetic Epidemiology Unit \\
Dept.\ of Epidemiology \\
Erasmus MC \\
Rotterdam \\
The Netherlands}
\email{c.vanduijn@erasmusmc.nl}

\author[]{Philip Scheltens}
\address[Philip Scheltens]{
Alzheimer Center \and
Dept.\ of Neurology \\
Amsterdam Neuroscience \\
VU University Medical Center Amsterdam \\
Amsterdam \\
The Netherlands}
\email{p.scheltens@vumc.nl}

\author[]{\\Ay\c{s}e Demirkan}
\address[Ay\c{s}e Demirkan]{
Genetic Epidemiology Unit \\
Dept.\ of Epidemiology \\
Erasmus MC \\
Rotterdam \\
The Netherlands; \and
Dept.\ of Human Genetics \\
Leiden University Medical Center \\
Leiden \\
The Netherlands}
\email{a.demirkan@erasmusmc.nl}

\author[]{Mark A.\ van de Wiel}
\address[Mark A.\ van de Wiel]{
Dept.\ of Epidemiology \& Biostatistics \\
Amsterdam Public Health research institute \\
VU University medical center Amsterdam \\
Amsterdam \\
The Netherlands; \and
Dept.\ of Mathematics \\
VU University Amsterdam \\
Amsterdam \\
The Netherlands}
\email{mark.vdwiel@vumc.nl}

\author[]{\\Wiesje M.\ van der Flier}
\address[Wiesje M.\ van der Flier]{
Alzheimer Center \and
Dept.\ of Neurology \\
Amsterdam Neuroscience \\
VU University Medical Center Amsterdam \\
Amsterdam \\
The Netherlands; \and
Dept.\ of Epidemiology \& Biostatistics \\
Amsterdam Public Health research institute \\
VU University medical center Amsterdam \\
Amsterdam \\
The Netherlands}
\email{WM.vdFlier@vumc.nl}

\author[]{Charlotte E.\ Teunissen}
\address[Charlotte E.\ Teunissen]{
Neurochemistry Laboratory and Biobank \\
Dept.\ of Clinical Chemistry \\
Amsterdam Neuroscience \\
VU University medical center Amsterdam \\
Amsterdam \\
The Netherlands}
\email{c.teunissen@vumc.nl}

\begin{abstract}\label{abstract}

~\\
\noindent\emph{Introduction:}
Identification of blood-based metabolic changes might provide early and easy-to-obtain biomarkers.

\noindent\emph{Methods:}
We included 127 AD patients and 121 controls with CSF-biomarker-confirmed diagnosis (cut-off tau/A$\beta_{42}$: 0.52).
Mass spectrometry platforms determined the concentrations of 53 amine, 22 organic acid, 120 lipid, and 40 oxidative stress compounds.
Multiple signatures were assessed: differential expression (nested linear models), classification (logistic regression), and regulatory (network extraction).

\noindent\emph{Results:}
Twenty-six metabolites were differentially expressed.
Metabolites improved the classification performance of clinical variables from 74\% to 79\%.
Network models identified 5 hubs of metabolic dysregulation: Tyrosine, glycylglycine, glutamine, lysophosphatic acid C18:2 and platelet activating factor C16:0.
The metabolite network for \emph{APOE} $\epsilon$4 negative AD patients was less cohesive compared to the network for \emph{APOE} $\epsilon$4 positive AD patients.

\noindent\emph{Discussion:}
Multiple signatures point to various promising peripheral markers for further validation.
The network differences in AD patients according to \emph{APOE} genotype may reflect different pathways to AD.


\bigskip
\noindent \emph{Abbreviations}:
2-AAA = 2-aminoadipic acid;
A$\beta_{42}$ = Amyloid beta peptide 42;
AD = Alzheimer's disease;
ADC = Amsterdam Dementia Cohort;
\emph{APOE} = apolipoprotein E;
AUC = Area Under the (Receiver Operating Characteristic) Curve;
BMI = body mass index;
CI = confidence interval;
CSF = cerebral spinal fluid;
DBP = diastolic blood pressure;
DNA = deoxyribonucleic acid;
EDTA = ethylenediaminetetraacetic acid;
FDR = false discovery rate;
$g$ = Earth's gravitational force;
IQR = interquartile range;
LPA = lysophosphatic acid;
PC = phosphatidylcholine;
MAP = mean arterial pressure;
MCI = mild cognitive impairment;
MS = mass spectrometry;
NIA-AA = National Institute of Aging and Alzheimer's Association;
PAF = platelet activating factor;
ROC = Receiver Operating Characteristic;
RSD$_{\mathrm{QC}}$ = relative standard deviation of quality control;
SBP = systolic blood pressure;
SCD = subjective cognitive decline;
SD = standard deviation;
SM = sphingomyelin
SMT2 = Supplementary Text 2;
TG = triglyceride
\end{abstract}
\maketitle

\begin{bibunit}
\section{Introduction}\label{SEC:Intro}
Accumulation of amyloid and tau-protein are considered the core pathological hallmarks for Alzheimer's disease (AD) \cite{PSlancet2016}, but other factors such as genetic liability, oxidative stress, inflammation and lifestyle contribute to the complex mechanism of this disease \cite{PSlancet2016, Cervellati2015,Mesa16Geronto,Alcolea15,Guerreiro2014}.
Non-invasive measurement of disease-specific biochemical changes in living patients is difficult, but may have value in terms of prognosis and identification of patients at risk for AD.

The metabolome, i.e., the collection of small-molecules that result from metabolic processes, is organized in biochemical pathways and is influenced by many internal and external factors, including genetics \cite{Holmes2008}.
Metabolomics refers to the collective quantification of these metabolites \cite{Koek11}.
Analytical methods have improved tremendously, with (targeted) mass spectrometry (MS) platforms now available for most compound classes.
In AD, metabolomics seems of utmost importance since various alterations in metabolism, e.g., higher levels of insulin and insulin resistance, are associated with an increased risk of AD \cite{Schrijvers2010}.
Moreover, the epsilon 4 ($\epsilon$4) allele of the apolipoprotein E (\emph{APOE}) gene is not only an important risk-factor for AD but is also related to alterations in lipid metabolism \cite{Rall2000,Horsburgh2000}.
Previous metabolomics studies in AD have reported alterations in lipid, antioxidant and amino acid metabolism.
However, results are not always unequivocal \cite{Trushina13,GongGinseng15,GGVG15mouse,Graham15Plasma,Ellis2015}.
This is most likely due to differences in (analytical) methods, cohort selection or context of use \cite{O'Bryant2017}.

We aim to study AD-related metabolic change from various perspectives with the use of multiple signatures in order to generate hypotheses regarding dysregulated metabolic events.
First, we evaluate shifts in the expression of individual metabolites using nested linear models.
Afterwards, we assess the classification performance of the metabolites in demarcating AD from control subjects.
Finally, we use state-of-the-art graphical modeling to explore metabolic dysregulation from a network perspective.
Additionally, we evaluate metabolic network changes according to \emph{APOE} status, to study the hypothesis that metabolic pathways are differentially dysregulated according to genotype.

\section{Methods}\label{SEC:Methods}
\subsection{Patients}\label{SEC:Patients}
\begin{sloppypar}
We selected 150 AD patients and 150 controls with available plasma from the Amsterdam Dementia Cohort (ADC) \cite{Flier14ADC}.
All subjects underwent standard cognitive screening including medical history assessment, physical-, neurological- and cognitive examination, blood sampling, lumbar puncturing, and magnetic resonance imaging.
Diagnoses were made in a multidisciplinary consensus meeting.
Until 2012, the diagnosis `probable AD' was based on the clinical criteria formulated by the NINCDS-ADRDA (National Institute of Neurological and Communicative Disorders and Stroke and the Alzheimer's Disease and Related Disorders Association) \cite{Diagnosis84}.
From 2012 onwards the criteria of the NIA-AA (National Institute on Aging-Alzheimer's Association) were used \cite{Diagnosis11}.
Subjects with subjective cognitive decline (SCD) were used as controls.
These subjects presented with memory complaints at the VUmc memory clinic, but performed normal on cognitive testing, i.e., criteria for mild cognitive impairment (MCI), dementia or psychiatric diagnosis were not fulfilled.
Clinical characteristics are provided in Table \ref{Table:ClinChars}.
All subjects gave written informed consent to use their clinical data for research purposes and to collect their blood samples for biobanking.
\end{sloppypar}

\subsection{CSF Biomarkers}\label{SEC:CSF}
\begin{sloppypar}
Amyloid beta peptide 42 (A$\beta_{42}$) and total tau (t-tau) were, for all subjects, measured in cerebrospinal fluid (CSF) using commercially available enzyme-linked immunosorbent assays (Innotest  A$\beta_{42}$ and Innotest hTAU-Ag; Innogenetics, Ghent, Belgium) \cite{Enzymes13}.
The cut-off for pathological biomarker status was defined as $\mbox{t-tau}/\mathrm{A}\beta_{42} > 0.52 $ \cite{Duits14}.
Of the 300 subjects included, 263 (136 AD patients and 127 controls) had a biomarker status in concordance with their clinical diagnosis, i.e., $\mbox{t-tau}/\mathrm{A}\beta_{42} > 0.52 $ for AD and $\mbox{t-tau}/\mathrm{A}\beta_{42} \leq 0.52 $ for controls.
These subjects were included for further analysis.
\end{sloppypar}

\subsection{\emph{APOE} Genotyping}\label{SEC:APOE}
Deoxyribonucleic acid (DNA) was isolated from 7-10ml EDTA blood.
Subsequently, samples were subjected to polymerase chain reaction.
A QIAxcel DNA Fast Analysis kit (Qiagen$^{\copyright}$, Venlo, The Netherlands) was used to check for size.
Sequencing was performed using Sanger sequencing on an ABI130XL.

\subsection{Metabolic Profiling}\label{SEC:MetProfile}
\begin{sloppypar}
Non-fasting EDTA plasma samples were, within 2 hours of collection, centrifuged at $1800 \times g$ for 10 minutes by room temperature and stored at $-80^{\circ}$C in polypropylene tubes (Sarstedt, Nurmberg, Germany).
Metabolic profiling of the samples was performed on four mass spectrometry (MS) platforms; i.e., amines, lipids and oxidative stress compounds were identified using ultra-performance liquid chromatography-tandem MS, and organic acids were analyzed with gas chromatography-MS \cite{Koek11,LipMethod,AmineMethod,QCcorrect}.
Reproducibility of individual metabolites was assessed in terms of the relative standard deviation of quality control (RSD$_{\mathrm{QC}}$) samples.
Metabolites with RSD$_{\mathrm{QC}}$ $>30$\% were deemed to fail acceptance criteria.
After QC correction, 53 amine compounds, 22 organic acid compounds, 120 lipid compounds, and 40 oxidative stress compounds were considered detected.
See \emph{Supplementary Text 1} and its accompanying tables for details on the profiling methods and detected compounds.
\end{sloppypar}

\subsection{Data Processing}\label{subsec:DatProcess}
\begin{sloppypar}
Metabolites with more than 10\% missing observations were removed, leading to the removal of 4 lipid compounds and 1 oxidative stress compound.
Three data samples (i.e., observed metabolite abundance profiles stemming from corresponding plasma samples) were removed as their (plasma) quality was deemed unsure.
These samples had many (30 or more) concentrations below the limit of detection (LOD) that could not be attributed to instrumental errors.
Twelve additional data samples were removed due to instrumental errors in one or more platforms.
Hence, we only retained data samples that were free of instrumental errors across all four different MS platforms.
The remaining missing values are attributable to concentrations failing the LOD.
These (feature-specific) missing values were imputed by half of the lowest observed value (for the corresponding metabolic feature).
The final metabolic data set thus contained $n = 263 - 3 - 12 = 248$ data samples (127 AD patients and 121 controls) and $p = 235 - 5 = 230$ metabolites.
\end{sloppypar}

The possible confounding effects of the clinical characteristics regarding anthropometrics, intoxications, comorbidities, and medication were evaluated in the expression and classification signatures demarcating the AD and control groups (see Section \ref{SEC:StatAnalysis}).
Table \ref{Table:ClinChars} contains the full list of characteristics and \emph{Table S2.1} of \emph{Supplementary Text 2 (SMT2)} contains additional information on measurement.
The missing observations on these variables ($< 6$\%) were imputed.
Continuous variables were imputed on the basis of Bayesian linear regression, polytomous variables were imputed on the basis of polytomous regression, and binary variables were imputed on the basis of logistic regression \cite{MICE}.
See \emph{Section 1} of \emph{SMT2} for additional information on data processing.

\subsection{Statistical Analysis}\label{SEC:StatAnalysis}
Differences in clinical characteristics between AD patients and controls were evaluated through Chi-square, Mann-Whitney \emph{U}, and \emph{t}-testing.
Differential metabolic expression between AD patients and controls was assessed by using nested linear models.
We tested, for each individual metabolite, whether its addition to a model containing clinical characteristics significantly contributed to model fit.
One then assesses if, conditional on the effects of the clinical characteristics, metabolic expression does indeed differ between the AD and control groups.
This entails an \emph{F}-test for nested models (see \emph{Section 2.1} of \emph{SMT2} for details).
The conditioning sets were (i) sex and age, and (ii) all clinical characteristics.
We adjusted for multiple testing by controlling the False Discovery Rate (FDR) \cite{FDR} at $.05$.

Subsequently, metabolic classification signatures for the prediction of group membership (AD or control) were constructed by way of penalized logistic regression with a Lasso-penalty \cite{LASSO}.
The Lasso-penalty enables estimation in our setting where the metabolite to sample ratio (230/248) is too high for standard logistic regression.
It also achieves automatic feature selection.
Two settings were considered: (i) the Lasso selects among the metabolites without considering the clinical characteristics; and (ii) the Lasso selects among the metabolites while the clinical characteristics go unpenalized.
The resulting models were compared to an unpenalized logistic regression model that considered only the clinical characteristics.
The optimal penalty parameter in the penalized models was determined on the basis of leave-one-out cross-validation (LOOCV) of the model likelihood.
Predictive performance of all models was assessed by way of (the comparison of) Receiver Operating Characteristic (ROC) curves and Area Under the ROC Curves (AUCs).
ROC curves and AUCs for all models were produced by 10-fold cross-validation.
See \emph{Section 2.2} of \emph{SMT2} for additional information.

A metabolic pathway can be thought of as a collection of metabolites originating from all over the metabolome, that work interdependently to regulate biochemical (disease) processes.
Hence, a pathway is a network.
We additionally employed network extraction techniques to examine regulatory signatures, i.e., dysregulation in metabolic biochemical pathways pertaining to the AD disease process.
From a network perspective, molecular pathway-dysregulation is likely characterized by the loss of normal (wanted) molecular interactions and the gain of abnormal (unwanted) molecular interactions.
From this perspective, the network topologies of the AD and control groups are expected to primarily share the same structure, while potentially differing in a number of (topological) locations of interest.
Network extraction was based on graphical modeling, more specifically, on targeted fused ridge estimation of inverse covariance (i.e., scaled partial correlation) matrices \cite{FUSED}.
This method (i) can deal with our metabolite to sample ratio (230/248, which is too high for standard graphical modeling), and (ii) explicitly takes into account that there are multiple groups of interest for which the shared network structures should be fused while the unique network structures should be distinguished.
The resulting networks are to be interpreted as conditional independence graphs, i.e, the nodes represent metabolic compounds and the edges connecting the nodes represent substantive partial correlations.
Extracted networks were subjected to subsequent analyses aimed at detecting hub compounds, group structures, and differential metabolic connections between groupings of interest.
Our efforts first juxtaposed metabolic networks for AD patients and controls.
Subsequently, we compared networks according to \emph{APOE} genotype.
See \emph{Sections 2.3 and 2.4} of \emph{SMT2} for additional detail.

\section{Results}\label{SEC:Results}
\subsection{Clinical Characteristics}\label{SEC:ClinCharacter}
Table \ref{Table:ClinChars} contains an overview of the clinical characteristics per diagnostic group.
The Mini-Mental State Examination (MMSE) score \cite{MMSE} of AD patients was lower compared to controls.
AD patients were more often carrier of at least 1 \emph{APOE} $\epsilon$4 allele.
Moreover, AD patients had a lower BMI and were less likely to have diabetes.

\begin{table}[b!]
\begin{threeparttable}
\begin{footnotesize}
\centering
\caption{Comparison clinical characteristics between AD and control groups.}
\label{Table:ClinChars}
\begin{tabular}{lllr}
\hline\hline
Characteristic                                          & AD group    & Control group & $p$-value \\ \hline
$n (\%)$                                                & 127 (51)    & 121 (49)      &           \\
                                                        &             &               &           \\
MMSE score, median (IQR)                                & 21 (5.5)    & 29 (2)        & $< .001$$^{*}$  \\
\textbf{Anthropometric}:                                &             &               &           \\
~~~Age, median (IQR)                                    & 65.1 (9.1)  & 62.7 (8)      & .548$^{*}$       \\
~~~Gender (female), $n (\%)$                            & 63 (50)     & 56 (46)       & .692$^{\dag}$       \\
~~~$\geq$ 1 \emph{APOE} $\epsilon$4 allele (yes), $n (\%)$     & 87 (69)     & 34 (28)       & $< .001$$^{\dag}$   \\
~~~MAP, mean (SD)                                       & 106.1 (11.5)& 103.9 (11.7)  & .133$^{\ddag}$       \\
~~~BMI, mean (SD)                                       & 24.2 (3.3)  & 26.27 (3.6)   & $< .001$$^{\ddag}$   \\
\textbf{Intoxications}:                                 &             &               &           \\
~~~Smoking                                              &             &               & .558$^{\dag}$       \\
~~~~~~Former, $n (\%)$                                  & 42 (33)     & 46 (38)       &           \\
~~~~~~Current, $n (\%)$                                 & 21 (17)     & 15 (12)       &           \\
~~~Alcohol (yes), $n (\%)$                              & 98 (77)     & 88 (73)       & .509$^{\dag}$       \\
\textbf{Comorbidities}:                                 &             &               &           \\
~~~Hypertension (yes), $n (\%)$                         & 37 (29)     & 33 (27)       & .854$^{\dag}$       \\
~~~Diabetes Mellitus (yes), $n (\%)$                    & 4 (3)       & 14 (12)       & .021$^{\dag}$       \\
~~~Hypercholesterolemia (yes), $n (\%)$                 & 14 (11)     & 9 (7)         & .451$^{\dag}$       \\
\textbf{Medication}:                                    &             &               &           \\
~~~Cholesterol lowering (yes), $n (\%)$                 & 31 (24)     & 22 (18)       & .298$^{\dag}$       \\
~~~Antidepressants (yes), $n (\%)$                      & 12 (9)      & 15 (12)       & .589$^{\dag}$       \\
~~~Antiplatelets (yes), $n (\%)$                        & 26 (20)     & 19 (16)       & .418$^{\dag}$       \\
\hline
\end{tabular}
\begin{tablenotes}\scriptsize
\item[] Abbreviations: AD, Alzheimer's disease; \emph{APOE}, apolipoprotein E; BMI, body mass index; IQR, interquartile range; MAP, mean arterial pressure; MMSE, Mini-Mental State Examination; SD, standard deviation.
\item[*] Mann-Whitney $U$ test.
\item[$\dag$] Pearson $\chi^{2}$ test.
\item[$\ddag$] Welch's \emph{t} test.
\end{tablenotes}
\end{footnotesize}
\end{threeparttable}
\end{table}

\subsection{Differential Expression Signature}\label{SEC:DiffExpress}
\begin{figure}[b!]
\centering
  \includegraphics[width=.94\textwidth]{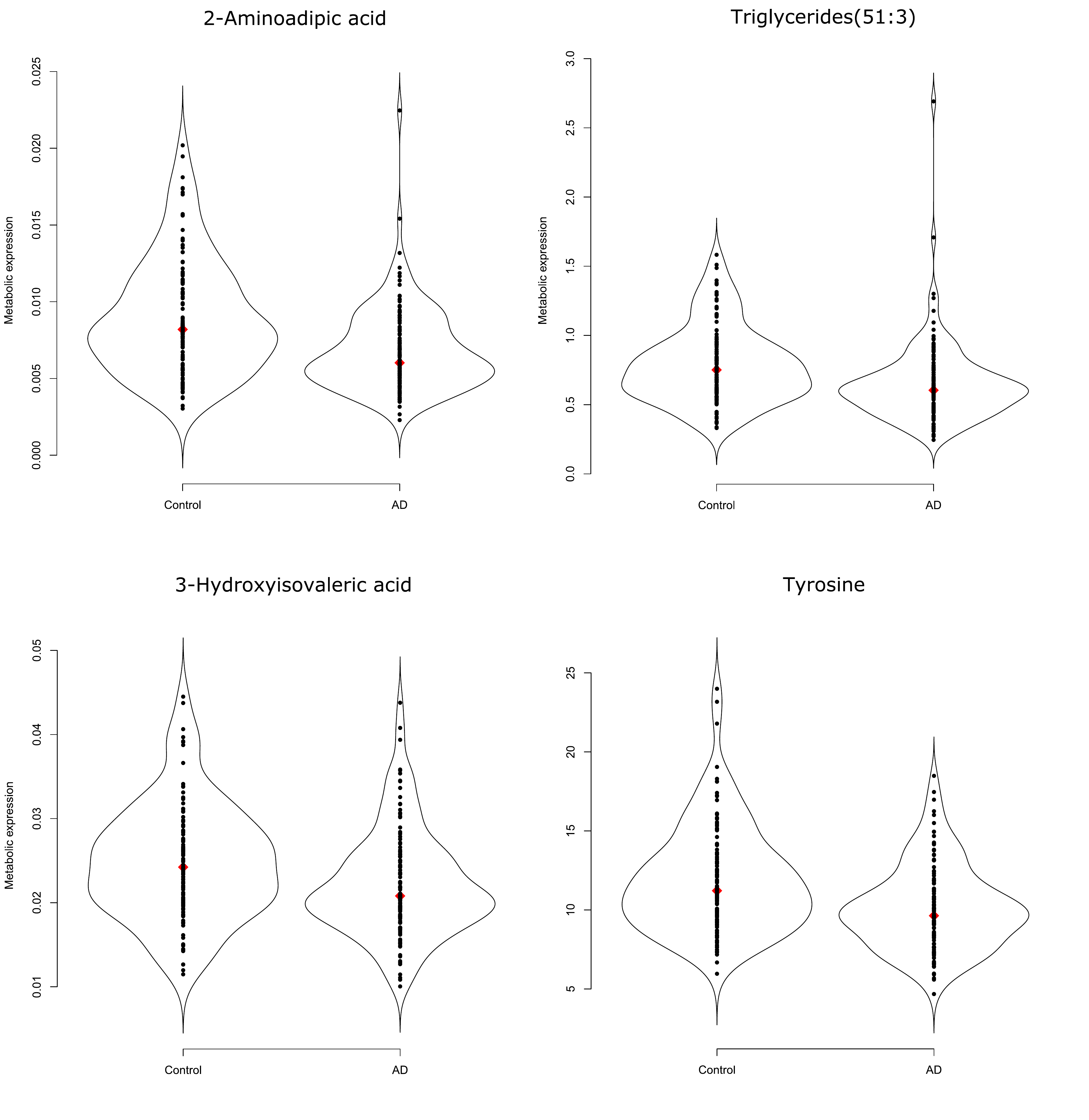}
    \caption{\footnotesize{Violin plots of the top 4 metabolites in terms of $p$-value.
    Violin plots \cite{VIOLIN} combine the familiar box plot with a kernel density to better represent the distribution of the data.
    We see relative underexpression in the AD group for all depicted metabolites.
    The associated adjusted $p$-values can be found in Table \ref{Table:DiffExpressMain}.
    The violin plots of the remaining differentially expressed metabolites can be found in \emph{Figures S2.1, S2.2, and S2.3} of \emph{SMT2}.}}
  \label{FIG:DiffExpressMain}
\end{figure}

A global test \cite{GlobalTEST} indicates that, given sex and age, the overall metabolic expression profile differs between AD patients and controls (\emph{p} = $\num{2.12e-06}$).
This difference in overall metabolic expression profile upholds when correcting for all clinical characteristics (\emph{p} = $\num{4.69e-05}$).
The metabolites listed in Table \ref{Table:DiffExpressMain} pass multiple testing correction on the \emph{F}-test for nested models (Section \ref{SEC:StatAnalysis}) with an FDR $< .05$.
The third column gives the ranking (in terms of raw $p$-value) of 52 metabolites that survive FDR correction when adjusting for sex and age only.
The fourth column analogously ranks the 26 metabolites that survive FDR correction when additionally adjusting for all clinical characteristics.
Triglycerides and amines dominate the latter compounds list.
Among its top compounds, in terms of (adjusted) $p$-value, are the amines 2-aminoadipic acid (2-AAA) and Tyrosine, the triglyceride TG(51:3), and the organic acid 3-Hydroxyisovaleric acid.
Their distributions in the AD and control groups are depicted in Figure \ref{FIG:DiffExpressMain}.
We see that these compounds are underexpressed in the AD group relative to the control group.
This relative underexpression in the AD group also holds for the remaining compounds in column 4 of Table \ref{Table:DiffExpressMain}, except for the Sphingomyelin SM(d18:1/20:1), which is overexpressed in the AD group relative to the control group (see \emph{Figures S2.1, S2.2, and S2.3} in \emph{SMT2}).

\begin{table}[t!]
\begin{scriptsize}
\caption{Differentially expressed metabolites that survive FDR adjustment.
The third column ranks (in terms of raw $p$-value) the metabolites that survive FDR correction in the model that adjusts for sex and age only.
The fourth column ranks (in terms of raw $p$-value) the metabolites that survive FDR correction in the model that adjusts for all clinical characteristics.
See \emph{Tables 2.2 and 2.3} of \emph{SMT2} for additional information.}
\centering
\label{Table:DiffExpressMain}
\begin{tabular}{llrr}
 \hline\hline
 Metabolite & Compound class & \multicolumn{2}{c}{Ranking}\\
 \hline
2-Aminoadipic acid	          	& Amines					& 1                 & 1	\\
Valine                       		& Amines					& 2                 & 16\\
Tyrosine                     		& Amines					& 3                 & 4	\\
Methyldopa                     		& Amines					& 4                 & 9	\\
Lysine                       		& Amines					& 5                 & 	\\
Methylmalonic acid		    	& Organic acids					& 6                 & 14\\
S-3-Hydroxyisobutyric acid		& Organic acids					& 7                 & 7	\\
TG(48:0)                      		& Lipids: Triglycerides				& 8                 & 21\\
TG(50:4)                      		& Lipids: Triglycerides				& 9                 & 6	\\
TG(48:2)                      		& Lipids: Triglycerides				& 10                & 13\\
TG(51:3)                      		& Lipids: Triglycerides				& 11                & 2	\\
TG(54:6)                     		& Lipids: Triglycerides				& 12                & 5	\\
TG(50:3)                      		& Lipids: Triglycerides				& 13                & 17\\
TG(50:2)                      		& Lipids: Triglycerides				& 14                & 	\\
TG(50:1)                      		& Lipids: Triglycerides				& 15                & 	\\
TG(48:1)                     		& Lipids: Triglycerides				& 16                & 25\\
TG(52:4)                      		& Lipids: Triglycerides				& 17                & 18\\
TG(48:3)                     		& Lipids: Triglycerides				& 18                & 11\\
Leucine          	            	& Amines					& 19                & 	\\
LPC(18:1)	                     	& Lipids: Lysophosphatidylcholine		& 20                & 	\\
TG(46:2)                      		& Lipids: Triglycerides				& 21                & 15\\
TG(50:0)                      		& Lipids: Triglycerides				& 22                & 	\\
TG(52:5)                      		& Lipids: Triglycerides				& 23                & 19\\
TG(52:3)                      		& Lipids: Triglycerides				& 24                & 	\\
TG(51:2)                      		& Lipids: Triglycerides				& 25                & 	\\
TG(56:8)                      		& Lipids: Triglycerides				& 26                & 8	\\
Isoleucine           	        	& Amines					& 27                & 	\\
2-hydroxybutyric acid   		& Organic acids					& 28                & 	\\
3-Hydroxyisovaleric acid		& Organic acids					& 29                & 3	\\
TG(51:1)    	                  	& Lipids: Triglycerides				& 30                & 	\\
SM(d18:1/20:1)                		& Lipids: Sphingomyelins			& 31                & 24\\
TG(52:1)    	                  	& Lipids: Triglycerides				& 32                & 	\\
8-iso-PGF2a (15-F2t-IsoP)              	& Oxidative stress: Isoprostane			& 33                & 10\\
Proline                  	     	& Amines					& 34                & 	\\
TG(54:5)              	        	& Lipids: Triglycerides				& 35                & 	\\
TG(56:7)               		       	& Lipids: Triglycerides				& 36                & 20\\
PGD2	           	            	& Lipids: Prostaglandins			& 37                & 	\\
TG(46:1)            	          	& Lipids: Triglycerides				& 38                & 	\\
PC(O-44:5)	                    	& Lipids: Plasmalogen Phosphatidylcholine	& 39                & 	\\
LPA C14:0 				& Lyso-phosphatidic acid			& 40                & 	\\
PC(O-34:1)	                    	& Lipids: Plasmalogen Phosphatidylcholine	& 41                & 	\\
LPC(20:4)	                     	& Lipids: Lysophosphatidylcholine		& 42                & 	\\
SM(d18:1/24:2)	                	& Lipids: Sphingomyelins			& 43                & 	\\
8,12-iPF2a IV		            	& Oxidative stress: Isoprostane			& 44                & 	\\
TG(46:0)             	         	& Lipids: Triglycerides				& 45                & 	\\
5-iPF2a VI		              	& Oxidative stress: Isoprostane			& 46                & 	\\
TG(52:2)        	              	& Lipids: Triglycerides				& 47                & 	\\
SM(d18:1/16:0)	                	& Lipids: Sphingomyelins			& 48                & 	\\
TG(58:10)	                     	& Lipids: Triglycerides				& 49                & 26\\
Ornithine                      		& Amines					& 50                & 22\\
Histidine                    		& Amines					& 51                & 	\\
O-Acetylserine                   	& Amines					& 		    & 12\\
SM(d18:1/23:0)				& Lipids: Sphingomyelins			& 		    & 23\\
\hline													
\end{tabular}			
\end{scriptsize}						
\end{table}

\subsection{Classification Signature}\label{SEC:Class}
\begin{figure}[h!]
\centering
  \includegraphics[width=.7\textwidth]{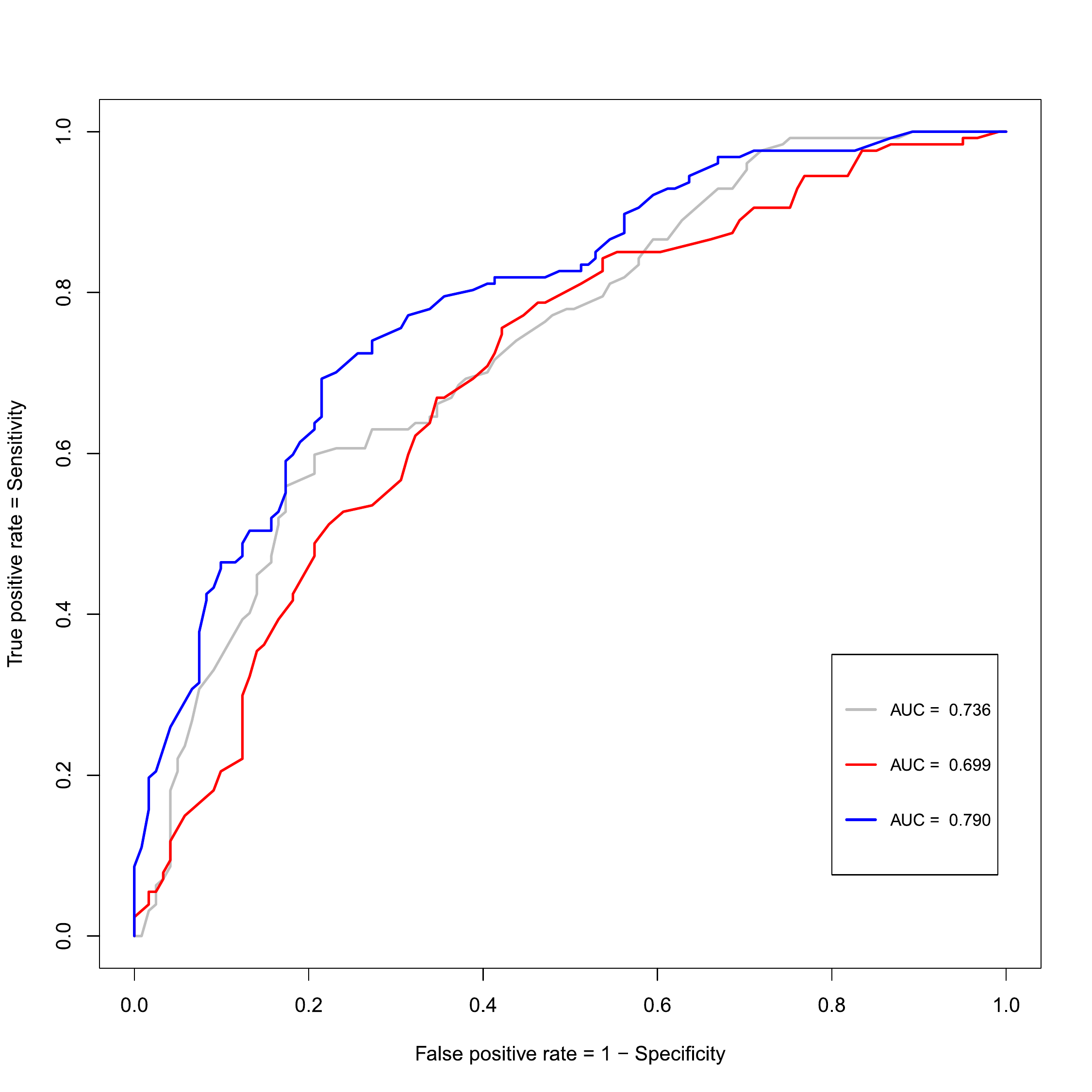}
    \caption{\footnotesize{ROC curves for the classification models.
    The grey line represents the ROC curve for the unpenalized logistic regression model that entertains the clinical characteristics only.
    The red line represents the ROC curve for the logistic model in which the Lasso performed variable selection amongst the metabolites (and that does not consider the clinical characteristics).
    The blue line represents the ROC curve of the logistic model in which the clinical characteristics are present while the Lasso may select amongst the metabolites.
    The clinical variables are listed in Table \ref{Table:ClinChars}.}}
  \label{FIG:ROCsMAIN}
\end{figure}

Subsequently, penalized logistic regression models were used to evaluate the ability of metabolites to distinguish AD patients from controls.
Classification performances can be found in Figure \ref{FIG:ROCsMAIN}.
The prediction model carrying the clinical variables only resulted in an AUC of approximately .74 (95\% bootstrap CI: $.67 - .79$).
The model that used the Lasso for selection amongst the metabolites sorts a comparable classification performance, yielding an AUC of approximately .70 (95\% bootstrap CI: $.63 - .76$).
The added value of the metabolites is reflected in the prediction model that adds a (Lasso-based) selection of metabolites to the clinical variables as it improves predictive performance, sorting a AUC of .79 (95\% bootstrap CI: $.73 - .84$).
A one-tailed bootstrap test for correlated ROC curves \cite{pROC} indicates that the AUC for this latter model is indeed higher than the AUC for the metabolites-only model ($p = .002$) and the AUC for the clinical-variables-only model ($p = .005$).
This test also indicates that the AUCs for the metabolites-only and clinical-variables-only models do not differ significantly ($p = .203$).
Metabolites consistently selected as top predictors (in terms of their absolute regression coefficient) in both penalized models that also occur in the differential expression signature are:
the amines O-Acetylserine and Methyldopa, the triglyceride TG(51:5), and the organic Methylmalonic acid.
Furthermore, oxidative stress compounds were selected by the Lasso on the basis of their predictive power, especially the prostaglandin PGD2, the isoprostane 8,12-iPF2a IV, and the nitro-fatty acid NO2-aLA (C18:3).
See \emph{Tables S2.4 and S2.5} of \emph{SMT2} for additional detail.

\subsection{Regulatory Signature}\label{SEC:Networks}
\begin{sloppypar}
Next, graphical modeling was used to explore metabolite-networks.
\emph{Section 2.3.3} of \emph{SMT2} contains visualizations of the extracted networks for AD patients and controls.
These networks convey that the strongest connections implicate metabolites from all four considered compound classes.
The metabolite-network for the control patients seems stronger locally connected (\emph{Section 2.3.4} of \emph{SMT2}), but both the AD and control networks are cohesive in the sense that they can be decomposed into clear communities (groups) of metabolites (\emph{Section 2.3.6} of \emph{SMT2}).
Hub compounds (i.e., metabolites of high regulatory importance as indicated by their centrality in a network) concur to some degree between the AD and control networks, with both having the Lyso-phosphatidic acid (LPA) C18:2 (an oxidative stress compound) as the strongest hub.
In the AD network however, as opposed to the control network, the amines Glycylglycine and Tyrosine are additionally indicated as central metabolites (\emph{Section 2.3.5} of \emph{SMT2}).
LPA C18:2, Glycylglycine and Tyrosine are amongst the metabolites whose regulatory functioning (in terms of differential connections) seems to change the most between the AD and control networks (\emph{Section 2.3.7} of \emph{SMT2}).
\end{sloppypar}

\begin{figure}[h]
\centering
  \includegraphics[width=.95\textwidth]{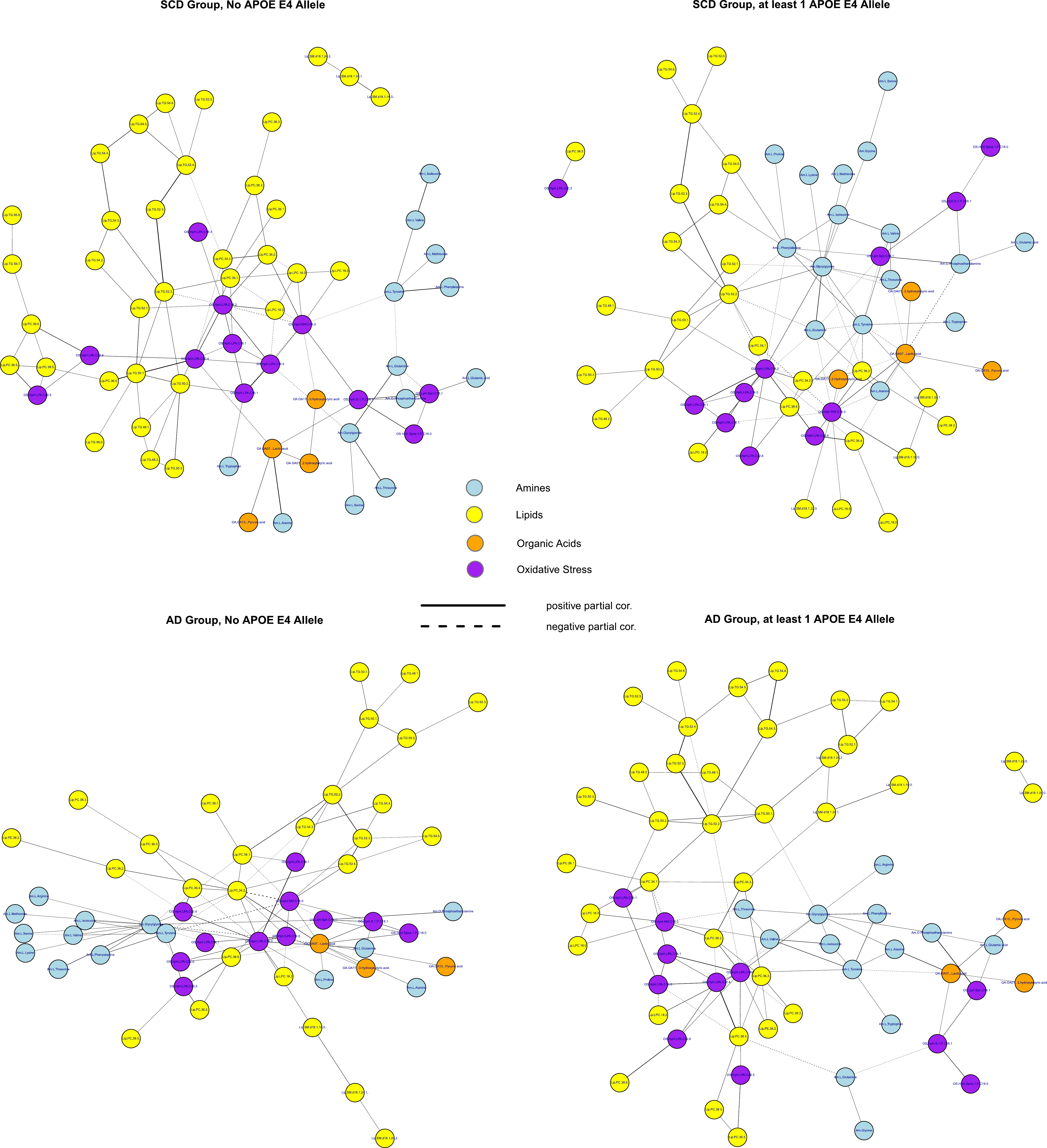}
    \caption{\footnotesize{Class-specific networks visualized with the Fruchterman-Reingold \cite{FRUCHT} algorithm.
    The upper-left panel contains the network for the control group with no \emph{APOE} $\epsilon$4 allele.
    The upper-right panel contains the network for the control group with at least 1 \emph{APOE} $\epsilon$4 allele.
    The lower-left panel represents the network for the AD group with no \emph{APOE} $\epsilon$4 allele.
    The lower-right panel represents the network for the AD group with at least 1 \emph{APOE} $\epsilon$4 allele.
    The metabolite compounds are colored according to metabolite family: Blue for amines, yellow for lipids, orange for organic acids, and purple for oxidative stress.
    Solid edges represent positive partial correlations while dashed edges represent negative partial correlations.}}
  \label{FIG:NWKSPruneMAIN}
\end{figure}

\begin{landscape}
\begin{figure}
\centering
  \includegraphics[scale = .4]{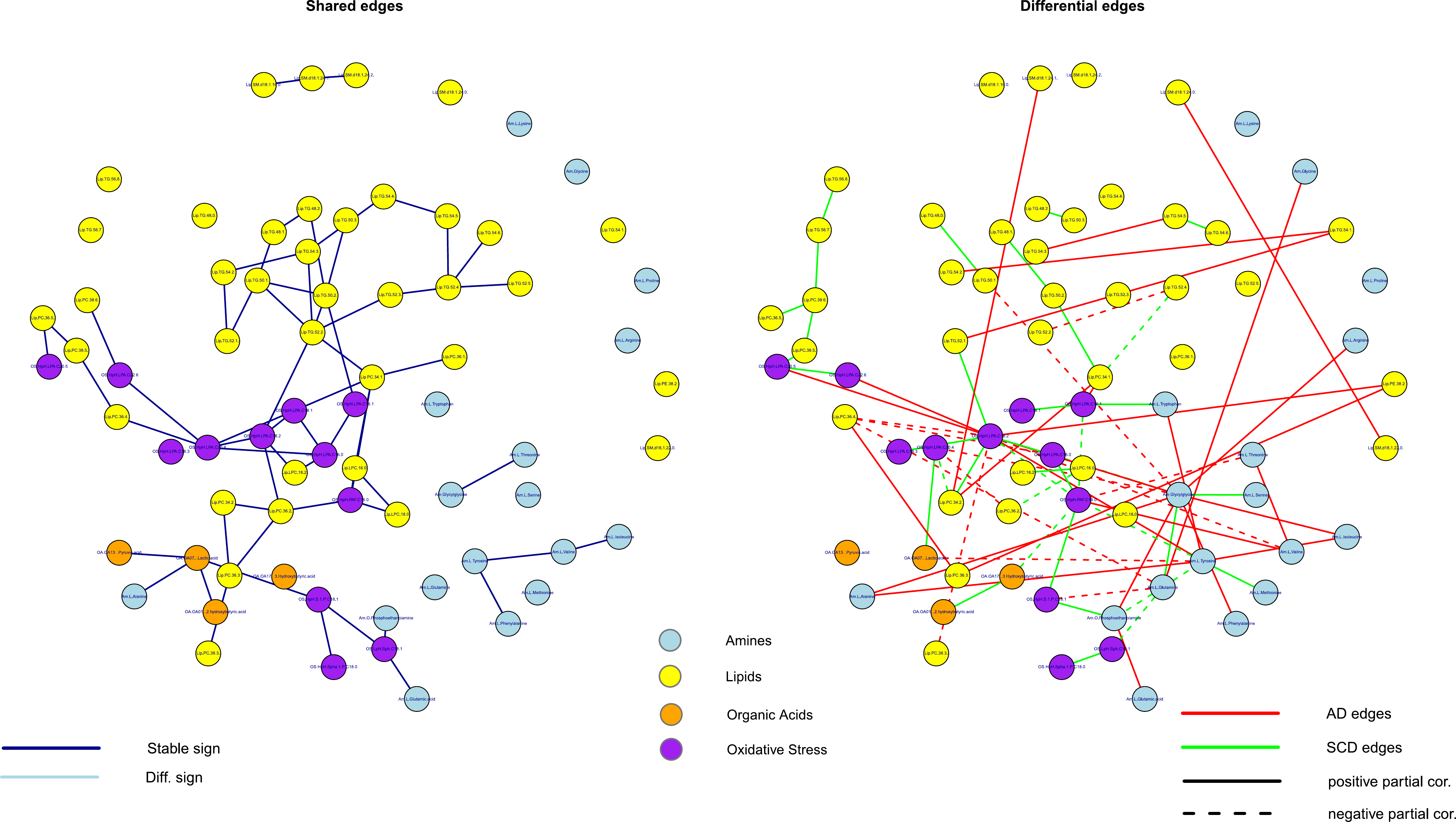}
    \caption{\footnotesize{Common and differential networks for the control group with no \emph{APOE} $\epsilon$4 allele versus the AD group with at least 1 \emph{APOE} $\epsilon$4 allele.
    The left-hand panel contains the network consisting of the edges (solid and colored blue) that are shared between these groups.
    The right-hand panel contains the network consisting of the edges that are unique for either of the groups.
    Red edges represent connections that are present in the \emph{APOE} $\epsilon$4 positive AD group only.
    Green edges represent connections that are present in the \emph{APOE} $\epsilon$4 negative control group only.
    Solid edges represent positive partial correlations while dashed edges represent negative partial correlations.
    The metabolite compounds are colored according to metabolite family: Blue for amines, yellow for lipids, orange for organic acids, and purple for oxidative stress.}}
  \label{FIG:DIFFgraphsMAIN}
\end{figure}
\end{landscape}

Overall, the AD and control networks seem to imply a shifting importance towards amine and oxidative stress compounds and their connections in the former.
This picture becomes more pronounced when the networks are stratified according to \emph{APOE} genotype (\emph{Section 2.4} of \emph{SMT2}).
Figure \ref{FIG:NWKSPruneMAIN} contains visualizations of the extracted networks for \emph{APOE} $\epsilon$4 negative controls and AD patients as well as \emph{APOE} $\epsilon$4 positive controls and AD patients.
The networks for \emph{APOE} $\epsilon$4 positive controls and \emph{APOE} $\epsilon$4 negative AD patients seem more random and less cohesive than the networks for \emph{APOE} $\epsilon$4 negative controls and \emph{APOE} $\epsilon$4 positive AD patients.
Comparing the cohesive networks for \emph{APOE} $\epsilon$4 negative controls and \emph{APOE} $\epsilon$4 positive AD patients (\emph{Section 2.4} of \emph{SMT2}) we see that all amines belong to the peripheral-structure in the former while many amines belong to the core-structure in the latter.
This might imply that biochemical functioning in the \emph{APOE} $\epsilon$4 positive AD group is more reliant on amines.
Hub compounds concur to some degree between these networks, with again (o.a.) LPA C18:2 as a strong hub.
In the network for \emph{APOE} $\epsilon$4 positive AD patients the amines Glycylglycine and Tyrosine are consistently indicated as central metabolites.
Figure \ref{FIG:DIFFgraphsMAIN} presents the networks of shared and differential connections between the \emph{APOE} $\epsilon$4 negative control and \emph{APOE} $\epsilon$4 positive AD groups.
The oxidative stress compounds LPA C18:2 and platelet activating factor (PAF) C16:0, and the amines Glycylglycine, Tyrosine, and Glutamine seem to change their regulatory function the most between the \emph{APOE} $\epsilon$4 negative control and \emph{APOE} $\epsilon$4 positive AD groups (also see \emph{Table S2.12} in \emph{SMT2}).
From the network perspective the \emph{APOE} $\epsilon$4-driven AD state can be characterized (vis-\`{a}-vis the control state without \emph{APOE} $\epsilon$4 alleles) by a loss of connections involving
PAF C16:0, a gain of connections involving Glycylglycine, and the differential wiring (both loss of normal and gain of alternative connections) of Tyrosine, Glutamine, and LPA C18:2.

\subsection{CSF Discordant Subjects}\label{SEC:CSFdiscordantsubjects}
Subjects whose clinical diagnosis was discordant from their CSF-biomarker status have an insecure disease status and where therefore excluded from the analyzes above.
A total of 37 subjects had both a complete metabolite-profile and a discordant CSF-biomarker status.
That is, these subjects were either clinically diagnosed with AD while their CSF-markers were normal ($\mbox{t-tau}/\mathrm{A}\beta_{42} \leq 0.52 $) or clinically diagnosed as normal while their CSF-markers indicated AD ($\mbox{t-tau}/\mathrm{A}\beta_{42} > 0.52 $).
For purposes of comparison we also obtained the expression and classification signatures when considering data from all $n = 285 ~(263 + 37)$ subjects with a complete metabolite-profile.
The results -- that accede to some degree with the results given in Sections \ref{SEC:DiffExpress} and \ref{SEC:Class} above -- can be found (with discussion) in \emph{Supplementary Text 3}.

\section{Discussion}\label{SEC:Discussion}
In this study, with CSF-biomarker-confirmed AD and control cases, we show that profiling metabolic alterations in AD can highlight disease-specific biochemical changes.
We assessed three metabolic signatures to highlight different aspects of metabolic change.
The expression signature shows the metabolites with relative under- or overexpression in AD versus controls.
This signature involved 26 metabolites, dominated by decreased levels of triglycerides and amines in AD.
We then evaluated classification signatures: collections of clinical and metabolite markers that can successfully demarcate AD cases from controls.
The top predictors concur (also in their sign) with metabolites found in the differential expression signature.
In addition, markers of oxidative stress were identified as strong predictors.
Lastly, graphical modeling was employed to evaluate regulatory signatures: exploratory networks of complex differential metabolite-dependencies between the AD and control groups.
Possible regulatory markers were again found in the amine and oxidative stress compound classes.
Stratifying for \emph{APOE} $\epsilon4$ status, the network for \emph{APOE} $\epsilon4$ negative AD subjects was less cohesive compared to the network for \emph{APOE} $\epsilon4$ positive AD subjects.
This suggests alternative biochemical-dysregulation involved in these patient groups.
Each signature gives a different but complementary perspective on AD-related metabolic events.
We propose the combination of these three signatures as a new approach to (i) studying the complex mechanism of metabolic change, (ii) defining characteristics involved in subtypes of AD, and (iii) selecting robust markers of interest for further research.
Below, we discuss and embed the findings related to each signature.

\subsection{Differential Expression Signature}\label{SEC:DISCUSSDiffExpress}
We show in Table \ref{Table:DiffExpressMain} that additional adjustments for clinical characteristics shortens the list and changes the ranking of metabolites that survive FDR correction.
This underlines the effects of clinical variables, such as medication, on the metabolome.
It also suggests that substantive corrections harness against overoptimistic expression signatures.
Below we will focus on the 26 metabolites listed in the expression signature adjusted for all clinical variables.
We found, in concordance with previous findings in both CSF and plasma, that AD is associated with decreased levels of amino acids and lipids \cite{Trushina13, Proitsi2016}.

Sixteen lipids, of whom fourteen triglycerides, were underexpressed, while only one lipid -- SM(d18:1/20:1) -- was overexpressed in AD.
This is in agreement with a large and recent lipidomics study that reported a decrease of most plasma lipids in AD and in particular an association of long-chain triglycerides with AD \cite{Proitsi2016}.
Moreover, supplementation of medium-chain triglycerides has been tested in AD to correct neuronal hypometabolism and might show some benefit for \emph{APOE} $\epsilon$4 negative AD patients \cite{Sharma2014}.

Multiple amino acids were also decreased in AD, amongst which 2-AAA acid (an intermediate of the lysine-pathway) and Tyrosine.
Plasma disturbances of the lysine-pathway have been suggested to differentiate controls from MCI and AD patients \cite{Trushina13}.
Decreased Tyrosine (a precursor for the neurotransmitters dopamine and norepinephrine) levels were also reported in an earlier study comparing metabolite levels in serum samples of AD patients and healthy controls \cite{Gonzalez-Dominiguez2014}.
Moreover, vanylmandellic acid -- an end-product of the Tyrosine-pathway -- was found to be elevated in the CSF of AD patients \cite{Daouk13}, suggesting disturbances of the Tyrosine-pathway.
Dopamine has been associated with cognitive control \cite{UshapeDopamine} and oral supplementation of Tyrosine has been shown to improve working memory and information-processing during demanding situations in healthy human adults \cite{HJR15Tyrosine}.
Experimental studies are needed to establish if the alterations in peripheral Tyrosine metabolism we found in our study, also affect the function of dopamine and Tyrosine in the central nervous system of AD patients.

\subsection{Classification Signature}\label{SEC:PredictPerformance}
The metabolites have added value in demarcating AD cases from controls.
This is reflected by the significant improvement of predictive performance when adding a selection of metabolites to the clinical characteristics and \emph{APOE} status.
Metabolite panels to monitor disease are of great interest for the clinic.
Especially when easy-to-obtain as with blood samples.
We here hint that a metabolite panel could be of added value to the yet available clinical variables and therefore might hold promise for use in, for example, clinical effect-monitoring.

Oxidative stress has been widely established to play a role in the pathogenesis of AD \cite{Cervellati2015}.
Defining the right markers to measure oxidative stress in vivo is however still an ongoing process, especially for peripheral markers in AD.
We found three markers of oxidative stress to have strong predictive power in demarcating AD patients from controls: the isoprostane-pathway derivatives \cite {Gao2003} 8,12-iPF-2a IV and PGD2, and the nitro fatty acid NO2-aLA (C18:3).
This result highlights again that oxidative stress is of strong influence in AD \cite{Cervellati2015,Teunissen2003}.

\subsection{Regulatory Signature}\label{SEC:DISCUSSnetworks}
\begin{sloppypar}
The network models revealed another oxidative stress marker, lysophosphatic acid (LPA C 18:2), to be one of the central players in both the AD and control networks.
It was prominently differentially related to other metabolites in AD versus control networks, perhaps representing a central player of metabolic change.
Previously, oxidized lipoproteins have been identified as a possible oxidative stressor in the brain leading to neuronal cell death in AD \cite {Darczynska1998}.
LPA is the most bioactive fraction of oxidized low density lipoprotein \cite{Shi2015}.
It has an important signal function and has been linked to the pathogenesis of AD as \emph{in vitro} results suggest they support tau phosphorylation and raise levels of $\beta$-secretase, leading to increased A$\beta$ production \cite{Shi2015, Yung2015,Sayas1999}.
Moreover, LPA's have been identified as important factors in vascular development, atherosclerosis and atherotrombogenesis \cite{Teo2009, Siess, Ken2001}.
As LPAs are a modulating factor in both AD and vascular changes it could be of special interest to further study the role of vascular factors in AD.
\end{sloppypar}

Network models for \emph{APOE} $\epsilon4$ positive AD subjects were more cohesive and less random in comparison to \emph{APOE} $\epsilon4$ negative AD subjects.
This suggests the possibility of structured, \emph{APOE} $\epsilon$4-driven changes in metabolism.
The lack of cohesiveness for the \emph{APOE} $\epsilon4$ negative AD group may be natural as this group is likely heterogeneous in disease etiology.
Hence, profiling metabolic subtypes is of interest for personalized clinical research.

\subsection{Strengths and Limitations}\label{SEC:Limits}
One strength of our study is that we used CSF biomarkers (A$\beta$ and tau) to support the clinical diagnosis of AD and controls.
This makes the metabolic alterations we describe more likely to be AD-specific.
Moreover, with the semi-targeted MS techniques here we were able to integrate data of four different compound classes and to replicate many findings from other recent metabolite studies in AD.

We note that the different signatures are not completely concordant.
This is explained by the different properties studied in each signature.
A differential expression signature explores, for individual metabolites, shifts in distribution.
A classification signature explores which conjunction of metabolites achieves an appreciable predictive performance.
A regulatory signature, then, assesses which metabolites are central in the complex network of metabolite interactions.
We pose the examination of multiple signatures as a strength as it uncovers metabolites of interest at the expression, prediction, and regulatory levels.
Assessing only the differential expression signature, for example, would imply that many metabolites of interest would go unnoticed.

Among the potential limitations of the study is the relatively small sample size ($n=248$) in comparison to the large number of metabolites studied ($p=230$).
However, we used novel statistical methods designed to account for high numbers of variables with limited case numbers.
Moreover, we used non-fasting plasma samples, while nutritional intake and medications are known to influence metabolite levels \cite{Holmes2008,Carayol2015}.
However, we corrected our results for multiple medication classes.

\subsection{Future Directions}\label{SEC:Future}
Peripheral changes in metabolism in AD are of interest because it could highlight factors that are influential in the disease-process on a systemic level.
The regulatory signature might be of added value to explore metabolic dysregulation.
However, these results are explorative and further work should focus on providing (dis)confirmation of hypotheses regarding (the effect of) network changes.
Experimental studies are needed to establish if the found alterations in peripheral metabolism are related to the function of metabolites in the central nervous system of AD patients.
In addition, effort should be directed to disentangle if these metabolic alterations are associated with AD-related risk factors and secondary changes (e.g., malnutrition, ageing, diabetes) or with AD pathology.
Moreover, integrating genomics and metabolomics could be of interest, as well as an in-depth study of the effect of patient-related and pre-analytical variation on the metabolome.
When its dynamics in terms of patient and pre-analytical influences are fully understood, it can be a powerful tool for monitoring ongoing biology.
Metabolites as identified in this study, such as for example Tyrosine and 2-AAA, could then serve as biological effect-monitoring tools in clinical trials.

\section{Conclusion}\label{SEC:Conclusion}
We show that peripheral metabolism is altered in AD patients compared to controls and between carriers and non-carriers of the \emph{APOE} $\epsilon$4 allele.
Moreover we show the added value of not only studying metabolic expression signatures, but to paint the full picture of metabolic change by also exploring classification and regulatory signatures.
These additional signatures can highlight possible prediction and regulatory markers that may be overlooked when studying expression signatures alone.
The consistent element over all signatures are the changes in the metabolism of amino acids and markers of oxidative stress.
In particular, the amino acid Tyrosine and the oxidative stress compound Lyso-phosphatidic acid C18:2 were identified as possible key players of metabolic change.
This is in concordance with previous literature describing disturbances of the tyrosine pathway in AD and oxidized lipoproteins as oxidative stressors in the AD brain \cite{Gonzalez-Dominiguez2014, Daouk13, Darczynska1998}.
Further research is needed to validate these results and to further specify their role in AD-specific metabolic alteration.

\section*{Acknowledgements}
\begin{sloppypar}
This research was supported by Janssen Pharmaceuticals Stellar Initiative:
Stellar Neurodegeneration Collaboration Project, Call 2, No.\ 3 (An Integrated MetaboloMic, Epidemiologic and genetic approach to DIscover clinically relevant biomarkers for Alzheimer's Disease: IMMEDIAD).
Research of the VUmc Alzheimer center is part of the neurodegeneration research program of Amsterdam Neuroscience.
The VUmc Alzheimer center is supported by Alzheimer Nederland and Stichting VUmc fonds.
The clinical database structure was developed with funding from Stichting Dioraphte.
F.A.d.L is appointed at the NWO-FCB project NUDAD (project number 057-14-004).

This version is a postprint of:
de Leeuw*, F.A., Peeters*, C.F.W., Kester, M.I., Harms, A.C., Struys, E.A., Hankemeier, T., van Vlijmen, H.W.T., van der Lee, S.J., van Duijn, C.M., Scheltens, P., Demirkan, A., van de Wiel, M.A., van der Flier, W.M., \& Teunissen, C.E. (2017). Blood-based metabolic signatures in Alzheimer's Disease. Alzheimer's \& Dementia: Diagnosis, Assessment \& Disease Monitoring, 8: 196--207.
This postprint is released under a Creative Commons CC BY-NC-ND 4.0 license.
\end{sloppypar}

\section*{Disclosures}
\begin{sloppypar}
F.A.d.L, C.F.W.P., M.I.K., A.C.H., E.A.S., H.W.T.v.V., S.J.v.d.L., M.A.v.d.W.\ report no relevant conflicts of interest.
T.H.\ and C.M.v.D.\ work on the CoSTREAM project (http://www.costream.eu/), a project funded by the European Union's Horizon 2020 research and innovation programme (grant agreement 667375).
A.D.\ and C.M.v.D.\ are members of PRECEDI Marie Curie exchange programme.
A.D.\ is supported by a Veni grant (2015) from ZonMw.
P.S.\ has received grant support (for the institution) from GE Healthcare, Danone Research, Piramal and MERCK.
In the past 2 years he has received consultancy/speaker fees (paid to the institution) from Lilly, GE Healthcare, Novartis, Forum, Sanofi, Nutricia.
W.M.v.d.F.\ has been an invited speaker at Boehringer Ingelheim.
Research programs of W.M.v.d.F.\ have been funded by ZonMw, NWO, EU-FP7, Alzheimer Nederland, CardioVascular Onderzoek Nederland, stichting Dioraphte, Gieskes-Strijbis fonds, Boehringer Ingelheim, Piramal Neuroimaging, Roche BV, Janssen Stellar, Combinostics.
All funding is paid to her institution.
C.E.T.\ serves on the advisory board of Fujirebio and Roche, performed contract research for IBL, Shire, Boehringer, Roche and Probiodrug; and received lecture fees from Roche and Axon Neurosciences.
\end{sloppypar}


\end{bibunit}

\newpage
\addresseshere

\cleardoublepage
\setcounter{section}{0}
\setcounter{subsection}{0}
\setcounter{equation}{0}
\setcounter{figure}{0}
\setcounter{table}{0}
\setcounter{page}{1}

\begin{center}
{\huge Research in Context}
\end{center}

\vspace{2cm}
\subsection*{1. Systematic Review:}
\begin{sloppypar}
Molecular aberrations tend to be amplified along the omics cascade.
Hence, there is increasing interest in finding biomarkers for Alzheimer's disease (AD) in peripheral fluids such as plasma.
Current study adds to a small body of literature on potential metabolite markers stemming from plasma.
\end{sloppypar}

\subsection*{2. Interpretation:}
Our data are used in a systematic effort to find differential expression, classification, and network deregulation signatures that demarcate AD from control cases.
These signatures point to certain amines and oxidative stress markers as drivers behind AD-related metabolic deregulation.

\subsection*{3. Future directions:}
\begin{sloppypar}
The results hold promise for the development of a biomarker panel.
Further studies are warranted for replication and panel development.
\end{sloppypar}

\vspace{3cm}
\begin{center}
{\huge Highlights}
\end{center}

\vspace{2cm}
\begin{enumerate}
  \item Multiple metabolic signatures point to peripheral AD markers for future validation.
  \item AD may be described by changes in the metabolism of amines and oxidative stressors.
  \item \emph{APOE} $\epsilon$4-driven AD and non-\emph{APOE} $\epsilon$4-driven AD represent different biochemical pathways.
  \item Network analyses of metabolomics data enable the study of metabolic changes in AD.
\end{enumerate}

\cleardoublepage

\renewcommand{\theequation}{S1.\arabic{equation}}
\renewcommand{\thefigure}{S1.\arabic{figure}}
\renewcommand{\thetable}{S1.\arabic{table}}
\renewcommand{\bibnumfmt}[1]{[S1.#1]}
\renewcommand{\citenumfont}[1]{S1.#1}
\renewcommand{\thesection}{\arabic{section}}

\setcounter{section}{0}
\setcounter{subsection}{0}
\setcounter{equation}{0}
\setcounter{figure}{0}
\setcounter{table}{0}
\setcounter{page}{1}

\phantomsection
\addcontentsline{toc}{section}{Supplementary Material}
\begin{center}
{\huge SUPPLEMENTARY TEXT 1\\~\\
Metabolite Analysis Methods}
\end{center}

\begin{bibunit}
\vspace{2cm}
This supplementary text contains additional information on the metabolite profiling.
Section \ref{ST1:sec:GenInfo} contains general information.
Section \ref{ST1:sec:Profile} details on the profiling methods.
Finally, Section \ref{ST1:sec:CompoundDetect} contains an overview of the metabolites considered detected.

\section{General Information}\label{ST1:sec:GenInfo}
Samples were stored at $-80^{\circ}$C until used for further analysis.
All samples were randomized and run in 5 batches which included a calibration line, quality control (QC) samples and blanks.
QC samples were analyzed every 10 samples (or every 15 samples in the oxidative stress platform).
The acquired data were evaluated using MassHunter software (Agilent) and LabSolutions software (Shimadzu).
An in-house written tool was applied that uses the QC samples to compensate for shifts in the sensitivity of the mass spectrometer throughout the batches \cite{QCcorrectS1}.
Both internal standard correction and QC correction were applied to the data set before reporting results.
All metabolites comply with the acceptance criteria of relative standard deviation QC (RSD$_{\mathrm{QC}}$) $<30$\%.

\section{Profiling}\label{ST1:sec:Profile}
\subsection{Biogenic Amine Profiling}\label{ST1:sec:Amines}
The amine platform covers amino acids and biogenic amines employing an AccQ-tag derivatization strategy adapted from the protocol supplied by Waters.
Five $\mu$L of each sample was spiked with an internal standard solution.
Then proteins were precipitated by the addition of methanol.
The supernatant was transferred to a new Eppendorf tube and taken to dryness in a vacuum centrifuge (speedvac).
The residue was reconstituted in borate buffer (pH 8.5) with 6-aminoquinolyl-\emph{N}-hydrosysuccinimidyl carbamate (AQC) reagent.
After reaction, the vials were transferred to an autosampler tray and cooled to $10^{\circ}$C until the injection.
One $\mu$L of the reaction mixture was injected into the ultra-performance liquid chromatography-tandem mass spectrometry (UPLC-MS/MS) system.

An Agilent 1290 Infinity ultra-high performance liquid chromatography (UHPLC) system with autosampler (Agilent, The Netherlands) was coupled online with a 6490 Triple quadrupole mass spectrometer (Agilent) operated using MassHunter data acquisition software (B.04.01; Agilent).
The samples were analyzed by UPLC-MS/MS using an Accq-Tag Ultra column (Waters).
The Triple quadrupole MS was used in the positive-ion electrospray mode and all analytes were monitored in dynamic Multiple Reaction Monitoring (dMRM) using nominal mass resolution \cite{AmineMethodS1}.

\subsection{Organic Acid Profiling}\label{ST1:sec:Oacids}
\begin{sloppypar}
This profiling platform, performed with gas chromatography-MS (GC-MS) technology, covered organic acids.
Sample preparation proceeded by first doing protein precipitation of 50 $\mu$L of sample with a crash solvent (MeOH/H2O) with in situ thermal desorption (ISTD) added.
After centrifugation and transferring the supernatant, the solvent was evaporated to complete dryness on the vacuum centrifuge (speedvac).
Then, two-step derivatisation procedures with oximation using methoxyamine hydrochloride (MeOX, 15 mg/mL in pyridine) as first reaction and silylation using N-Methyl-N-(trimethylsilyl)trifluoroacetamide (MSTFA) as second reaction were carried out.
After this final step the samples were transferred to the auto sampler vials and 1 $\mu$L was injected on GC-MS \cite{GCtechniqueS1}.

The metabolites were measured by gas chromatography on an Agilent Technologies 7890A equipped with an Agilent Technologies mass selective detector (MSD 5975C) and MultiPurpose Sampler (MPS, MXY016-02A, GERSTEL).
Chromatographic separations were performed on a HP-5MS UI (5\% Phenyl Methyl Silox), $30 \mathrm{m} \times 0.25 \mathrm{m}$ ID column with a film thickness of $25 \mathrm{m}$, using helium as the carrier gas at a flow rate of 1.7 mL/min.
A single-quadrupole mass spectrometer with electron impact ionization (EI, 70 eV) was used.
The mass spectrometer was operated in SCAN mode mass range 50-500.
\end{sloppypar}

\subsection{Lipid Profiling}\label{ST1:sec:Lipids}
\begin{sloppypar}
The lipid platform covers Cholesteryl ester, Ceremides, Diacylglycerols, Lysophosphatidylcholines, Lysophosphatidylethanolamine, Phosphatidylcholines, Phosphatidylethanolamines, Plasmalogen Lysophosphatidylcholines, Plasmalogen Phosphatidylcholines, Plasmalogen Phosphatidylethanolamines, Sphingomyelins, and Triglycerides.
Lipids were extracted with isopropyl alcohol (IPA).
In short, 1000 $\mu$L IPA containing calibrant and internal standards both at C4 levels were added to 10 $\mu$L serum to precipitate proteins.
After centrifugation ($12,100$ rpm, 10 min, at RT), supernatant containing the lipids was transferred to vials for Liquid chromatography-MS (LC-MS) analysis.
In total 2.5 $\mu$L was injected for analysis.
\end{sloppypar}

Chromatographic separation was achieved on an ACQUITY UPLC HSS T3 column (1.8 $\mu$m, $2.1 \times 100 \mathrm{mm}$) with a flow of 0.4 mL/min over a 16 min gradient.
The lipid analysis is performed on a UPLC-ESI-Q-TOF (Agilent 6530, Jose, CA, USA) high resolution mass spectrometer using reference mass correction.
Lipids were detected in full scan in the positive ion mode \cite{LipMethodS1}.

\subsection{Oxidative Stress Profiling}\label{ST1:sec:Ostress}
\begin{sloppypar}
The oxidative stress platform covers isoprostanes, prostaglandins, nitro-fatty acids, lyso-sphingolipids, lysophosphatidic acids, alkyl-lysophosphatidic acids and cyclic-phosphatidic acids.
One hundred and fifty $\mu$L of each sample was spiked with an internal standard solution.
The metabolite extraction is performed via liquid-liquid extraction.
To extract the compounds from the aqueous phase, butanol and ethylacetate are used.
After collection, the organic phase is concentrated by first drying and then reconstitution in a smaller volume.
After reconstitution, the extract is divided in two vials (one for each chromatography) and used for injection on UPLC-MS/MS.
The oxidative stress platform is divided in two chromatographic methods: low and high pH.
In the low pH method, isoprostanes, prostaglandins, nitro-fatty acids and lyso-sphingolipids are analyzed.
The high pH method covers lyso-sphingolipids, lysophosphatidic acids, alkyl-lysophosphatidic acids and cyclic-phosphatidic acids.
\end{sloppypar}

A Shimadzu system with three high pressure pumps (LC-30AD), a controller (CBM-20Alite), an autosampler (SIL-30AC) and an oven (CTO-30A) from Shimadzu Benelux, was coupled online with a LCMS-8050 Triple quadrupole mass spectrometer (Shimadzu) operated using LabSolutions data acquisition software (Version 5.72, Shimadzu).
The samples were analyzed by UPLC-MS/MS using a Kromasil Eternity XT C18 column (Akzo Nobel) for high pH and an Acquity BEH C18 column (Waters) for the low pH method.
The Triple quadrupole MS was used in polarity switching mode and all analytes were monitored in dynamic Multiple Reaction Monitoring (dMRM).

\section{Detected Compounds}\label{ST1:sec:CompoundDetect}
After QC correction, 53 amine compounds, 22 organic acid compounds, 120 lipid compounds, and 40 oxidative stress compounds are detected, respectively.
The detected compounds are listed in the tables below.
These tables make use of the following abbreviations: HMDB = Human Metabolome Database; ID = identifier; InChI = International Union of Pure and Applied Chemistry (IUPAC) International Chemical Identifier; Lipid Maps = LIPID Metabolites and Pathways Strategy \cite{LipidMAPSS1}.
Detected amine, organic acid, lipid, and oxidative stress compounds are listed in Tables \ref{ST1:table:Amines}, \ref{ST1:table:OrgAcids}, \ref{ST1:table:Lipids}, and \ref{ST1:table:OxSTRESS}, respectively.

\begin{table}[b!]
\begin{tiny}
\centering
\caption{Detected amine compounds.}
\label{ST1:table:Amines}
\begin{tabular}{llll}
\hline
\hline
Metabolite               & Chemical formula & HMDB ID   & InChI Key                   \\ \hline
1-Methylhistidine        & C7H11N3O2        & HMDB00001 & BRMWTNUJHUMWMS-LURJTMIESA-N \\
2-Aminoadipic acid       & C6H11NO4         & HMDB00510 & OYIFNHCXNCRBQI-UHFFFAOYSA-N \\
3-Aminoisobutyric acid   & C4H9NO2          & HMDB00452 & QWCKQJZIFLGMSD-VKHMYHEASA-N \\
3-Methoxytyramine        & C9H13NO2         & HMDB00022 & DIVQKHQLANKJQO-UHFFFAOYSA-N \\
3-Methoxytyrosine        & C10H13NO4        & HMDB01434 & PFDUUKDQEHURQC-UHFFFAOYSA-N \\
3-Methylhistidine        & C7H11N3O2        & HMDB00479 & JDHILDINMRGULE-LURJTMIESA-N \\
4-Hydroxyproline         & C5H9NO3          & HMDB00725 & PMMYEEVYMWASQN-DMTCNVIQSA-N \\
5-Hydroxylysine          & C6H14N2O3        & HMDB00450 & YSMODUONRAFBET-UHNVWZDZSA-N \\
ADMA                     & C8H18N4O2        & HMDB01539 & YDGMGEXADBMOMJ-LURJTMIESA-N \\
Alanine                  & C3H7NO2          & HMDB00161 & QNAYBMKLOCPYGJ-REOHCLBHSA-N \\
Alpha-aminobutyric acid  & C4H9NO2          & HMDB03911 & QCHPKSFMDHPSNR-UHFFFAOYSA-N \\
Arginine                 & C6H14N4O2        & HMDB00517 & ODKSFYDXXFIFQN-BYPYZUCNSA-N \\
Asparagine               & C4H8N2O3         & HMDB00168 & DCXYFEDJOCDNAF-REOHCLBHSA-N \\
Aspartic acid            & C4H7NO4          & HMDB00191 & CKLJMWTZIZZHCS-REOHCLBHSA-N \\
Carnosine                & C9H14N4O3        & HMDB00033 & CQOVPNPJLQNMDC-ZETCQYMHSA-N \\
Citrulline               & C6H13N3O3        & HMDB00904 & RHGKLRLOHDJJDR-BYPYZUCNSA-N \\
Cysteine                 & C3H7NO2S         & HMDB00574 & XUJNEKJLAYXESH-REOHCLBHSA-N \\
Dopamine                 & C8H11NO2         & HMDB00073 & VYFYYTLLBUKUHU-UHFFFAOYSA-N \\
Ethanolamine             & C2H7NO           & HMDB00149 & HZAXFHJVJLSVMW-UHFFFAOYSA-N \\
Gamma-aminobutyric acid  & C4H9NO2          & HMDB00112 & BTCSSZJGUNDROE-UHFFFAOYSA-N \\
Gamma-glutamylalanine    & C8H14N2O5        & HMDB06248 & WQXXXVRAFAKQJM-WHFBIAKZSA-N \\
Glutamic acid            & C5H9NO4          & HMDB00148 & WHUUTDBJXJRKMK-VKHMYHEASA-N \\
Glutamine                & C5H10N2O3        & HMDB00641 & ZDXPYRJPNDTMRX-VKHMYHEASA-N \\
Glutathione              & C10H17N3O6S      & HMDB00125 & RWSXRVCMGQZWBV-WDSKDSINSA-N \\
Glycine                  & C2H5NO2          & HMDB00123 & DHMQDGOQFOQNFH-UHFFFAOYSA-N \\
Glycylglycine            & C4H8N2O3         & HMDB11733 & YMAWOPBAYDPSLA-UHFFFAOYSA-N \\
Histamine                & C5H9N3           & HMDB00870 & NTYJJOPFIAHURM-UHFFFAOYSA-N \\
Histidine                & C6H9N3O2         & HMDB00177 & HNDVDQJCIGZPNO-YFKPBYRVSA-N \\
Homoserine               & C4H9NO3          & HMDB00719 & UKAUYVFTDYCKQA-VKHMYHEASA-N \\
Isoleucine               & C6H13NO2         & HMDB00172 & AGPKZVBTJJNPAG-WHFBIAKZSA-N \\
Kynurenine               & C10H12N2O3       & HMDB00684 & YGPSJZOEDVAXAB-QMMMGPOBSA-N \\
Leucine                  & C6H13NO2         & HMDB00687 & ROHFNLRQFUQHCH-YFKPBYRVSA-N \\
Lysine                   & C6H14N2O2        & HMDB00182 & KDXKERNSBIXSRK-YFKPBYRVSA-N \\
Methionine               & C5H11NO2S        & HMDB00696 & FFEARJCKVFRZRR-BYPYZUCNSA-N \\
Methionine sulfoxide     & C5H11NO3S        & HMDB02005 & QEFRNWWLZKMPFJ-UHFFFAOYSA-N \\
Methyldopa               & C10H13NO4        & HMDB11754 & CJCSPKMFHVPWAR-JTQLQIEISA-N \\
N6,N6,N6-Trimethyllysine & C9H20N2O2        & HMDB01325 & MXNRLFUSFKVQSK-QMMMGPOBSA-N \\
O-Acetylserine           & C5H9NO4          & HMDB03011 & VZXPDPZARILFQX-BYPYZUCNSA-N \\
O-Phosphoethanolamine    & C2H8NO4P         & HMDB00224 & SUHOOTKUPISOBE-UHFFFAOYSA-N \\
Ornithine                & C5H12N2O2        & HMDB00214 & AHLPHDHHMVZTML-BYPYZUCNSA-N \\
Phenylalanine            & C9H11NO2         & HMDB00159 & COLNVLDHVKWLRT-QMMMGPOBSA-N \\
Pipecolic acid           & C6H11NO2         & HMDB00716 & HXEACLLIILLPRG-YFKPBYRVSA-N \\
Proline                  & C5H9NO2          & HMDB00162 & ONIBWKKTOPOVIA-BYPYZUCNSA-N \\
Putrescine               & C4H12N2          & HMDB01414 & KIDHWZJUCRJVML-UHFFFAOYSA-N \\
Sarcosine                & C3H7NO2          & HMDB00271 & FSYKKLYZXJSNPZ-UHFFFAOYSA-N \\
SDMA                     & C8H18N4O2        & HMDB03334 & HVPFXCBJHIIJGS-LURJTMIESA-N \\
Serine                   & C3H7NO3          & HMDB00187 & MTCFGRXMJLQNBG-REOHCLBHSA-N \\
Serotonin                & C10H12N2O        & HMDB00259 & QZAYGJVTTNCVMB-UHFFFAOYSA-N \\
Taurine                  & C2H7NO3S         & HMDB00251 & XOAAWQZATWQOTB-UHFFFAOYSA-N \\
Threonine                & C4H9NO3          & HMDB00167 & AYFVYJQAPQTCCC-GBXIJSLDSA-N \\
Tryptophan               & C11H12N2O2       & HMDB00929 & QIVBCDIJIAJPQS-VIFPVBQESA-N \\
Tyrosine                 & C9H11NO3         & HMDB00158 & OUYCCCASQSFEME-QMMMGPOBSA-N \\
Valine                   & C5H11NO2         & HMDB00883 & KZSNJWFQEVHDMF-BYPYZUCNSA-N \\ \hline
\end{tabular}
\end{tiny}
\end{table}

\begin{table}[]
\begin{tiny}
\centering
\caption{Detected organic acid compounds.}
\label{ST1:table:OrgAcids}
\begin{tabular}{ll}
\hline
\hline
Metabolite                 & Identifier \\ \hline
2-hydroxybutyric acid      & HMDB00008  \\
Citric acid                & HMDB00094  \\
Glutamic acid              & HMDB00148  \\
Glycolic acid              & HMDB00115  \\
L-Lactic acid              & HMDB00190  \\
Malic acid                 & HMDB00744  \\
2-Ketoglutaric acid        & HMDB00208  \\
Succinic acid              & HMDB00254  \\
Fumaric acid               & HMDB00134  \\
Pyruvic acid               & HMDB00243  \\
Methylmalonic acid         & HMDB00202  \\
Pyroglutamic acid          & HMDB00267  \\
Isocitrate                 & HMDB00193  \\
3-hydroxybutyric acid      & HMDB00357  \\
3-Phosphoglyceric acid     & HMDB00807  \\
Aspartic acid              & HMDB00191  \\
Iminodiacetate             & HMDB11753  \\
S-3-Hydroxyisobutyric acid & HMDB00023  \\
3-Hydroxyisovaleric acid   & HMDB00754  \\
Glyceric acid              & HMDB00139  \\
Uracil                     & HMDB00300  \\
Cis-Aconitic acid          & HMDB00072  \\ \hline
\end{tabular}
\end{tiny}
\end{table}

\begin{table}[]
\begin{tiny}
\centering
\caption{Detected lipid compounds.}
\label{ST1:table:Lipids}
\begin{tabular}{lll}
\hline
Lipid class                    & Lipidmaps & Metabolite species                                 \\ \hline
Cholesteryl ester (CE)         & ST0102    & CE(18:1); CE(18:2)                                 \\
                               &           &                                                    \\
Ceremides (Cer)                & SP02      & Cer(d18:1/24:0)                                    \\
                               &           &                                                    \\
Diacylglycerol (DG)            & GL0201    & DG(36:2); DG(36:3)                                 \\
                               &           &                                                    \\
Lysophosphatidylcholine (LPC)  & GP0105    & LPC(14:0); LPC(16:0); LPC(16:1); LPC(18:0);        \\
                               &           & LPC(18:1); LPC(18:2); LPC(18:3); LPC(20:3);        \\
                               &           & LPC(20:4); LPC(20:5); LPC(22:6)                    \\
                               &           &                                                    \\
Lysophosphatidylethanolamine   & GP0205    & LPE(18:0)                                          \\
(LPE)                          &           &                                                    \\
                               &           &                                                    \\
Phosphatidylcholine (PC)       & GP0101    & PC(32:0); PC(32:1); PC(32:2); PC(34:1); PC(34:2);  \\
                               &           & PC(34:3); PC(34:4); PC(36:1); PC(36:2); PC(36:3);  \\
                               &           & PC(36:4); PC(36:5); PC(36:6); PC(38:2); PC(38:3);  \\
                               &           & PC(38:5); PC(38:6); PC(40:4); PC(40:5); PC(40:7)   \\
                               &           &                                                    \\
Phosphatidylethanolamine (PE)  & GP0201    & PE(38:2); PE(38:4)                                 \\
                               &           &                                                    \\
Plasmalogen                    & GP0106    & LPC(O-16:0); LPC(O-16:1); LPC(O-18:1)              \\
Lysophosphatidylcholine (pLPC) &           &                                                    \\
                               &           &                                                    \\
Plasmalogen                    & GP0102    & PC(O-34:1); PC(O-34:2); PC(O-34:3); PC(O-36:4);    \\
Phosphatidylcholine (pPC)      &           & PC(O-36:5); PC(O-36:6); PC(O-38:4); PC(O-38:5);    \\
                               &           & PC(O-38:6); PC(O-44:5)                             \\
                               &           &                                                    \\
Plasmalogen                    & GP0202    & PE(O-36:5); PE(O-38:5); PE(O-38:7)                 \\
Phosphatidylethanolamine (pPE) &           &                                                    \\
                               &           &                                                    \\
Sphingomyelins (SM)            & SP0301    & SM(d18:1/14:0); SM(d18:1/15:0); SM(d18:1/16:0);    \\
                               &           & SM(d18:1/16:1); SM(d18:1/18:0); SM(d18:1/18:1);    \\
                               &           & SM(d18:1/18:2); SM(d18:1/20:0); SM(d18:1/20:1);    \\
                               &           & SM(d18:1/21:0); SM(d18:1/22:0); SM(d18:1/22:1);    \\
                               &           & SM(d18:1/23:0); SM(d18:1/23:1); SM(d18:1/24:0);    \\
                               &           & SM(d18:1/24:1); SM(d18:1/24:2); SM(d18:1/25:0)     \\
                               &           &                                                    \\
Triglycerides (TG)             & GL0301    & TG(42:0); TG(44:0); TG(44:1); TG(46:0); TG(46:1);  \\
                               &           & TG(46:2); TG(48:0); TG(48:1); TG(48:2); TG(48:3);  \\
                               &           & TG(50:0); TG(50:1); TG(50:2); TG(50:3); TG(50:4);  \\
                               &           & TG(51:1); TG(51:2); TG(51:3); TG(52:0); TG(52:1);  \\
                               &           & TG(52:2); TG(52:3); TG(52:4); TG(52:5); TG(54:0);  \\
                               &           & TG(54:1); TG(54:2); TG(54:3); TG(54:4); TG(54:5);  \\
                               &           & TG(54:6); TG(56:0); TG(56:1); TG(56:2); TG(56:3);  \\
                               &           & TG(56:6); TG(56:7); TG(56:8); TG(57:2); TG(58:1);  \\
                               &           & TG(58:10); TG(58:2); TG(58:3); TG(58:8); TG(58:9); \\
                               &           & TG(60:2); TG(O-50:0)                               \\ \hline
\end{tabular}
\end{tiny}
\end{table}

\begin{table}[]
\begin{tiny}
\centering
\caption{Detected oxidative stress compounds.}
\label{ST1:table:OxSTRESS}
\begin{tabular}{lll}
\hline
Compound name             & Compound class                & Lipid Maps ID  \\ \hline
2,3-dinor-8-iso-PGF2a     & Isoprostane                   & LMFA03110010   \\
5-iPF2a VI                & Isoprostane                   & LMFA03110010   \\
8-iso-PGF2a (15-F2t-IsoP) & Isoprostane                   & LMFA03110001   \\
8,12-iPF2a IV             & Isoprostane                   & -              \\
aLPA C16:1                & Alkyl-lyso-phosphatidic acid  & -              \\
aLPA C18:1                & Alkyl-lyso-phosphatidic acid  & -              \\
cLPA C16:0                & Cyclic-lyso-phosphatidic acid & LMGP00000057   \\
cLPA C18:0                & Cyclic-lyso-phosphatidic acid & LMGP00000055   \\
cLPA C18:1                & Cyclic-lyso-phosphatidic acid & LMGP00000056   \\
cLPA C18:1                & Cyclic-lyso-phosphatidic acid & -              \\
cLPA C18:2                & Cyclic-lyso-phosphatidic acid & -              \\
cLPA C20:3                & Cyclic-lyso-phosphatidic acid & -              \\
cLPA C20:4                & Cyclic-lyso-phosphatidic acid & -              \\
iPF2a-Unknown             & -                             & -              \\
LPA C14:0                 & Lyso-phosphatidic acid        & LMGP10050007   \\
LPA C16                   & Lyso-phosphatidic acid        & LMGP10050006   \\
LPA C16:1                 & Lyso-phosphatidic acid        & LMGP10050016   \\
LPA C18                   & Lyso-phosphatidic acid        & LMGP10050005   \\
LPA C18:1                 & Lyso-phosphatidic acid        & LMGP10050008   \\
LPA C18:2                 & Lyso-phosphatidic acid        & LMGP10050017   \\
LPA C18:3                 & Lyso-phosphatidic acid        & LMGP10050023   \\
LPA C20:1                 & Lyso-phosphatidic acid        & LMGP10050026   \\
LPA C20:3                 & Lyso-phosphatidic acid        & LMGP10050028   \\
LPA C20:4                 & Lyso-phosphatidic acid        & LMGP10050013   \\
LPA C20:5                 & Lyso-phosphatidic acid        & LMGP10050033   \\
LPA C22:4                 & Lyso-phosphatidic acid        & LMGP10050020   \\
LPA C22:5                 & Lyso-phosphatidic acid        & -              \\
LPA C22:6                 & Lyso-phosphatidic acid        & LMGP10050019   \\
NO2-aLA (C18:3)           & Nitro-Fatty acid              & -              \\
NO2-LA (C18:2)            & Nitro-Fatty acid              & LMFA01120001/2 \\
NO2-OA (C18:1)            & Nitro-Fatty acid              & LMFA01120003/4 \\
PAF C16:0                 & Platelet activating factor    & LMGP01020046   \\
PGA2                      & Prostaglandins                & LMFA03010035   \\
PGD2                      & Prostaglandins                & LMFA03010004   \\
PGE2                      & Prostaglandins                & LMFA03010003   \\
PGF2a                     & Prostaglandins                & LMFA03010002   \\
S1P C18:1                 & Lyso-sphingolipid             & LMSP01050001   \\
SPH C18:1                 & Lyso-sphingolipid             & LMSP01010001   \\
SPHA C18:0                & Lyso-sphingolipid             & LMSP01020001   \\
SPHA-1-P C18:0            & Lyso-sphingolipid             & LMSP01050002   \\ \hline
\end{tabular}
\end{tiny}
\end{table}


\end{bibunit}

\cleardoublepage

\renewcommand{\theequation}{S2.\arabic{equation}}
\renewcommand{\thefigure}{S2.\arabic{figure}}
\renewcommand{\thetable}{S2.\arabic{table}}
\renewcommand{\bibnumfmt}[1]{[S2.#1]}
\renewcommand{\citenumfont}[1]{S2.#1}
\renewcommand{\thesection}{\arabic{section}}

\setcounter{section}{0}
\setcounter{subsection}{0}
\setcounter{equation}{0}
\setcounter{figure}{0}
\setcounter{table}{0}
\setcounter{page}{1}

\phantomsection
\addcontentsline{toc}{section}{Supplementary Material}
\begin{center}
{\huge SUPPLEMENTARY TEXT 2\\~\\
Statistical Analyzes: Approaches and Results}
\end{center}

\begin{bibunit}
\vspace{2cm}
This supplementary text contains additional information on the data analysis.
Section \ref{sec:DatProcess} contains information on processing the metabolite data for statistical analyzes.
Section \ref{sec:Signatures} then contains extensive information on the analyzes surrounding the differential expression, classification, and regulatory signatures.
This section also contains all obtained results.
Please note that, as this supplementary text is self-contained, there is some redundancy in presentation.

\section{Data and Data Processing}\label{sec:DatProcess}
\subsection{Data}\label{subsec:Dat}
Plasma samples of 150 subjects with Alzheimer's disease (AD) and 150 subjects with subjective cognitive decline (SCD) were available.
Subjects with SCD were used as cognitively normal controls in this study.
Of these 300 subjects 263 (136 AD and 127 SCD) had their diagnosis confirmed by cerebral spinal fluid (CSF) markers ($\mbox{t-tau}/\mathrm{A}\beta_{42} > 0.52 $ for AD diagnosis).
The 263 subjects with CSF-confirmed diagnosis were used for further study.
Metabolite concentrations in four metabolite classes were determined using four different mass spectrometry platforms: amines (53), organic acids (22), lipids (120) and oxidative stress (40) compounds (see \emph{Supplementary Text 1}).

\subsection{Data Processing}\label{subsec:DatProcess}
\begin{sloppypar}
Metabolites with more than 10\% missing observations were removed, leading to the removal of 5 metabolites (the oxidative stress compound iPF2a-Unknown, and the lipids CE(18:1), TG(57:2), TG(58:3), and PE(O-38:7)).
Three data samples (i.e., vectors of observed metabolite abundancies stemming from corresponding plasma samples) were removed as their (plasma) quality was deemed unsure.
These samples had many (30 or more) concentrations below the limit of detection (LOD) that could not be attributed to instrumental errors.
Twelve additional data samples were removed due to instrumental errors in one or more platforms.
Hence, we only retain data samples that were free of instrumental errors across all four different mass spectrometry platforms.
The remaining missing values are attributable to concentrations failing the LOD.
These (feature-specific) missing values were imputed by half of the lowest observed value (for the corresponding feature).
The final metabolic data set thus contained $n = 248$ data samples (127 AD and 121 SCD) and $p = 230$ metabolic features.
\end{sloppypar}

In addition to metabolomics data, phenotypic data (clinical and demographic characteristics such as height, weight, and APOE $\epsilon$4 allele status) were evaluated for their possible confounding effects in the expression and classification signatures demarcating the AD and SCD groups.
The missing observations on these variables (14 at most, for the height variable) were imputed.
Continuous variables were imputed on the basis of Bayesian linear regression, polytomous variables were imputed on the basis of polytomous regression, and binary variables were imputed on the basis of logistic regression \cite{MICES}.
To relief `correctional stress' on the expression and classification signatures, certain aggregational clinical measures were calculated.
The Body Mass Index (BMI) was calculated as $\mbox{weight}_{\mathrm{kg}}/\mbox{height}^{2}_{\mathrm{m}}$.
In addition, the Mean Arterial Pressure (MAP) was approximated from the systolic blood pressure (SBP) and diastolic blood pressure (DBP) by $\mbox{DBP} + \frac{1}{3}(\mathrm{SBP} - \mathrm{DBP})$.
See Table \ref{Table:clinVARS} for a full list of considered confounders.

\begin{table}[]
\begin{footnotesize}
\caption{List of clinical variables.}
\centering
\label{Table:clinVARS}
\begin{tabular}{ll}
 \hline\hline
 Variable & Measurement \\
 \hline
    \textbf{Anthropometric}:            & \\
    Age                                 & years at diagnosis \\
    Sex                                 & male or female \\
    APOE $\epsilon$4 allele status      & at least one $\epsilon$4 allele: yes, no \\
    Mean arterial pressure              & approximated by $\mbox{DBP} + \frac{1}{3}(\mathrm{SBP} - \mathrm{DBP})$ \\
    Body mass index                     & $\mbox{weight}_{\mathrm{kg}}/\mbox{height}^{2}_{\mathrm{m}}$ \\
    ~ & \\
    \textbf{Intoxications}:             &\\
    Smoking                             & status: current, former, never\\
    Alcohol                             & current consumption of: yes, no \\
    ~ & \\
    \textbf{Comorbidities}:             &\\
    Hypertension                        & present: yes, no \\
    Diabetes Mellitus                   & present: yes, no \\
    Hypercholesterolemia                & present: yes, no \\
    ~ & \\
    \textbf{Medication}:            &\\
    Cholesterol lowering medications    & usage: yes, no \\
    Antidepressant medications          & usage: yes, no \\
    Antiplatelet medications            & usage: yes, no \\
 \hline
\end{tabular}
\end{footnotesize}
\end{table}

\section{Signatures}\label{sec:Signatures}
\subsection{Differential Expression Signature}\label{subsec:DES}
\subsubsection{Approach}\label{subsubsec:DESapproach}
Differential metabolic expression between AD and SCD subjects was assessed by using nested linear models. We tested, for each individual metabolite, if its addition to a model containing the clinical characteristics (see Table \ref{Table:clinVARS}) significantly contributed to model fit.
One then assesses if, conditional on the effects of the clinical characteristics, metabolic expression does indeed differ between the AD and SCD groups.
Let $\mathrm{BG}_k$ represent the $k$th background or clinical variable and let $\mathrm{I}_\mathrm{AD}$ denote an indicator variable for AD group-membership.
We are then interested in testing the following (abusing notation somewhat) nested models
\begin{align}
\mathrm{metabolite}_j &= \beta_0 + \sum_{k = 1}^{m} \beta_k\mathrm{BG}_k + \epsilon \label{Eq:reduced}\\
\mathrm{metabolite}_j &= \beta_0 + \sum_{k = 1}^{m} \beta_k\mathrm{BG}_k + \beta_{(m + 1)}\mathrm{I}_\mathrm{AD} + \epsilon, \label{Eq:full}
\end{align}
where the reduced model in (\ref{Eq:reduced}) is clearly nested in the full model (\ref{Eq:full}).
This entails a test for nested models which, in this case, is equivalent to testing $H_0: \beta_{(m + 1)} = 0$ versus $H_a: \beta_{(m + 1)} \neq 0$.
The associated test statistic $F$ (see any standard statistics textbook) is distributed as $\mathcal{F}_{1,n-(m + 2)}$ under the null hypothesis.
The $p$-value for the observed test statistic can be obtained in reference to this distribution.

We have a multiple testing problem as we need to perform this test for each individual metabolite.
Our approach to multiple testing is by controlling the False Discovery Rate (FDR), i.e., we aim to control ``the expected proportion of falsely rejected hypotheses" \cite{FDRS}.
We control the FDR at $.05$.

\subsubsection{Results}\label{subsubsec:DESresults}
\begin{figure}[t!]
\centering
  \includegraphics[scale = .200]{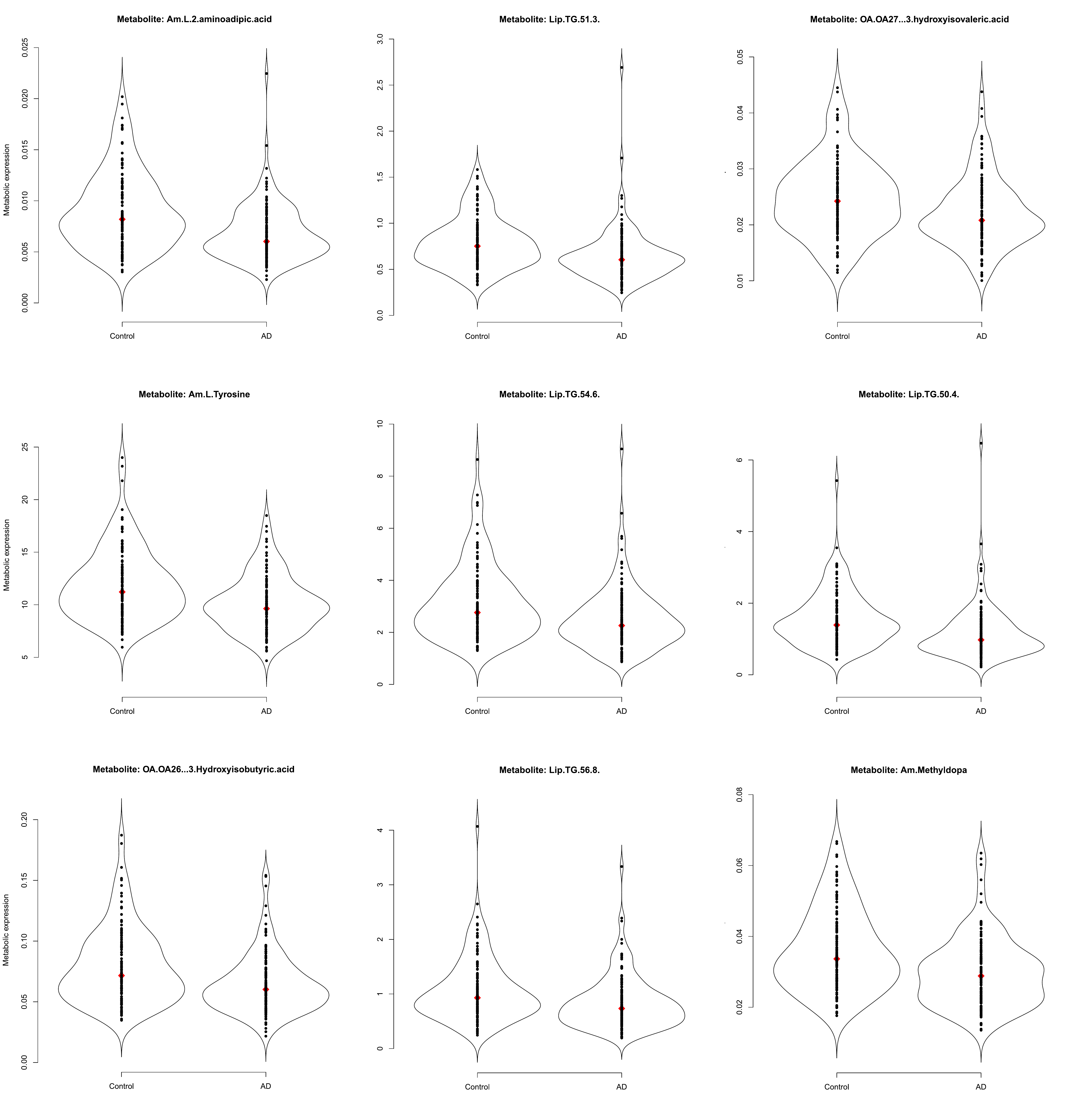}
    \caption{\footnotesize{Violin plots of (a selection of) the metabolites that survive multiple testing correction.
    Violin plots \cite{VIOLINS} combine the familiar box plot with a kernel density to better represent the distribution of the data.
    We see relative underexpression in the AD group for all depicted metabolites.
    The associated adjusted $p$-values can be found in Table \ref{Table:DiffExpress}.
    The remaining violin plots can be found in Figures \ref{Fig:DiffExpress2} and \ref{Fig:DiffExpress3}.}}
  \label{Fig:DiffExpress}
\end{figure}

\begin{figure}[t!]
\centering
  \includegraphics[scale = .200]{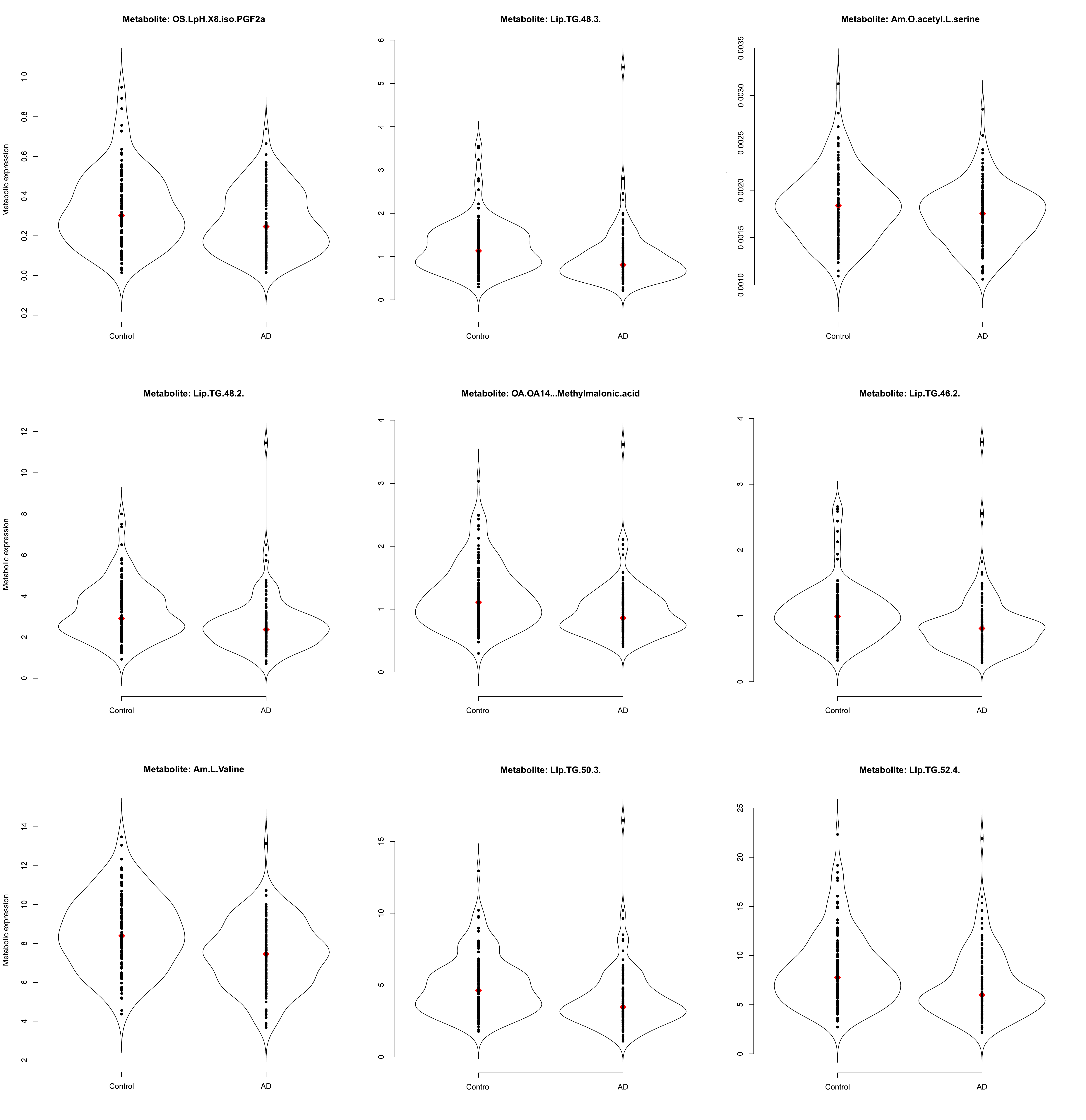}
    \caption{\footnotesize{Violin plots of (a selection of) the metabolites that survive multiple testing correction.
    Violin plots \cite{VIOLINS} combine the familiar box plot with a kernel density to better represent the distribution of the data.
    We see relative underexpression in the AD group for all depicted metabolites.
    The associated adjusted $p$-values can be found in Table \ref{Table:DiffExpress}.
    The remaining violin plots can be found in Figures \ref{Fig:DiffExpress} and \ref{Fig:DiffExpress3}.}}
  \label{Fig:DiffExpress2}
\end{figure}

\begin{figure}[t!]
\centering
  \includegraphics[scale = .200]{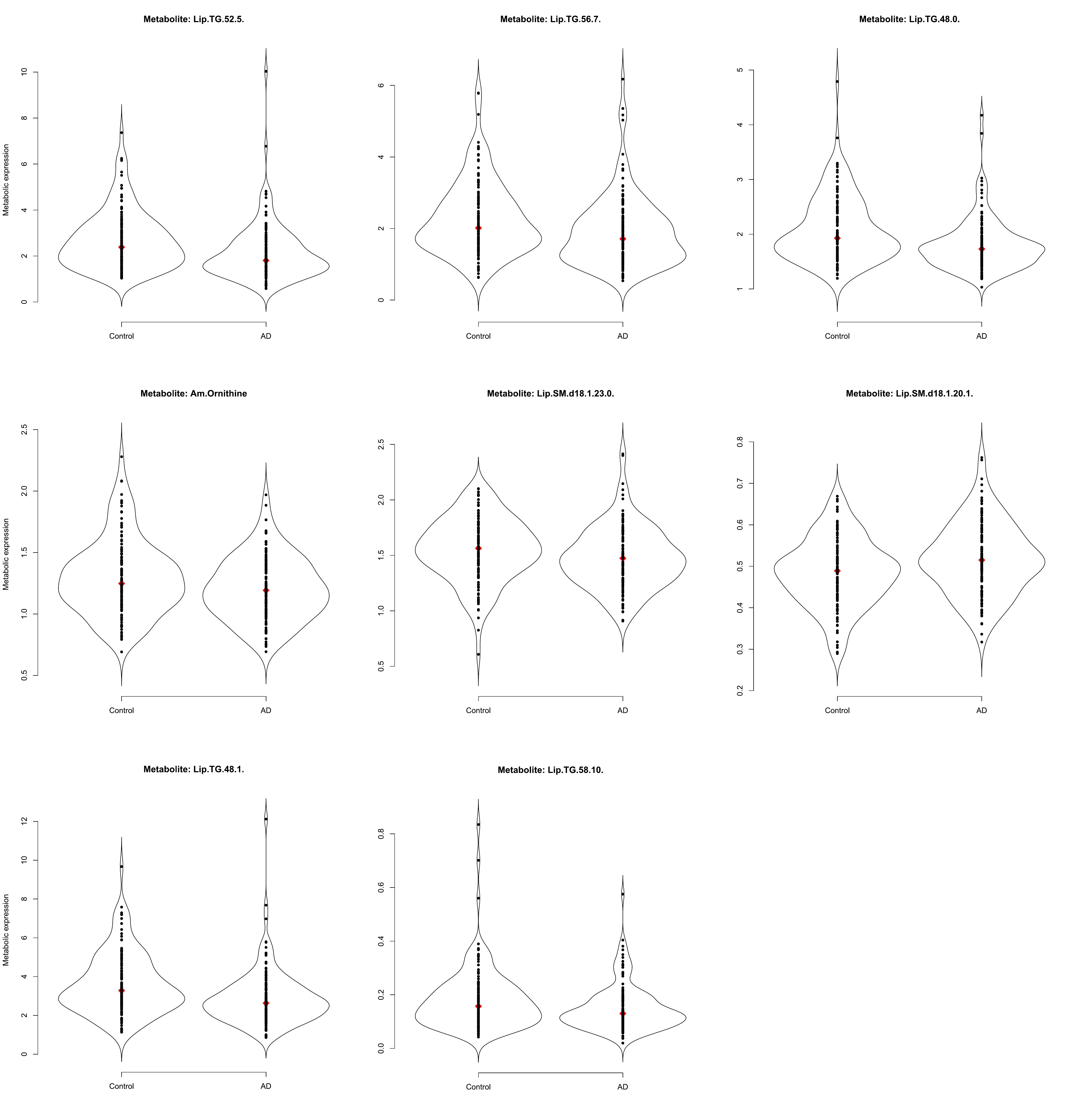}
    \caption{\footnotesize{Violin plots of (a selection of) the metabolites that survive multiple testing correction.
    Violin plots \cite{VIOLINS} combine the familiar box plot with a kernel density to better represent the distribution of the data.
    We see relative underexpression in the AD group for all depicted metabolites except SM(d18:1/20:1).
    The associated adjusted $p$-values can be found in Table \ref{Table:DiffExpress}.
    The remaining violin plots can be found in Figures \ref{Fig:DiffExpress} and \ref{Fig:DiffExpress2}.}}
  \label{Fig:DiffExpress3}
\end{figure}

\begin{sloppypar}
The metabolic features that survive multiple testing correction are listed in Table \ref{Table:DiffExpress}.
The differential distributions for these features are depicted in Figures \ref{Fig:DiffExpress}, \ref{Fig:DiffExpress2}, and \ref{Fig:DiffExpress3}.
We see that all implicated features (except for SM(d18:1/20:1)) are underexpressed in the AD group relative to the control group.
Table \ref{Table:DiffExpressSEXAGE} contains, for purposes of comparison, the list of metabolic features that survive multiple testing correction when testing nested models in which only sex and age are used as possible confounders.
We see that under less stringent corrections the list of potentially differentially expressed metabolites is longer.
Substantive corrections harness against overoptimistic expression signatures.
\end{sloppypar}

\begin{table}[t!]
\begin{footnotesize}
\caption{Differentially expressed metabolites.}
\centering
\label{Table:DiffExpress}
\begin{tabular}{llrr}
 \hline\hline
 Metabolite & Compound class & $p$-value & Adjusted $p$-value\\
 \hline
2-Aminoadipic acid	          &Amines			&.0003110134  &.03071051 \\
TG(51:3)	                  &Lipids: Triglycerides	&.0005518318  &.03071051 \\
3-Hydroxyisovaleric acid	  &Organic acids		&.0005645572  &.03071051 \\
Tyrosine        	          &Amines			&.0006614924  &.03071051 \\
TG(54:6)                          &Lipids: Triglycerides	&.0009136006  &.03071051 \\
TG(50:4)	                  &Lipids: Triglycerides	&.0010320542  &.03071051 \\
S-3-Hydroxyisobutyric acid	  &Organic acids		&.0011085893  &.03071051 \\
TG(56:8)                      	  &Lipids: Triglycerides	&.0012115039  &.03071051 \\
Methyldopa                        &Amines			&.0012272994  &.03071051 \\
8-iso-PGF2a (15-F2t-IsoP)         &Oxidative stress: Isoprostane&.0013352397  &.03071051 \\
TG(48:3)                      	  &Lipids: Triglycerides	&.0017977057  &.03703538 \\
O-Acetylserine	                  &Amines			&.0020920976  &.03703538 \\
TG(48:2)                          &Lipids: Triglycerides	&.0025274816  &.03703538 \\
Methylmalonic acid       	  &Organic acids		&.0026395625  &.03703538 \\
TG(46:2)                          &Lipids: Triglycerides	&.0027475334  &.03703538 \\
Valine                            &Amines			&.0027973120  &.03703538 \\
TG(50:3)                          &Lipids: Triglycerides	&.0031949553  &.03703538 \\
TG(52:4)                      	  &Lipids: Triglycerides	&.0034264336  &.03703538 \\
TG(52:5)                          &Lipids: Triglycerides	&.0034406945  &.03703538 \\
TG(56:7)                          &Lipids: Triglycerides	&.0034900406  &.03703538 \\
TG(48:0)                      	  &Lipids: Triglycerides	&.0035736266  &.03703538 \\
Ornithine                         &Amines			&.0036249029  &.03703538 \\
SM(d18:1/23:0)                 	  &Lipids: Sphingomyelins	&.0037035385  &.03703538 \\
SM(d18:1/20:1)	                  &Lipids: Sphingomyelins	&.0048428315  &.04491679 \\
TG(48:1)                          &Lipids: Triglycerides	&.0049569537  &.04491679 \\
TG(58:10)	                  &Lipids: Triglycerides	&.0050775504  &.04491679 \\
 \hline
\end{tabular}
\end{footnotesize}
\end{table}

\begin{table}[]
\begin{tiny}
\caption{Differentially expressed metabolites when correcting for sex and age only.}
\centering
\label{Table:DiffExpressSEXAGE}
\begin{tabular}{llrr}
 \hline\hline
 Metabolite & Compound class & $p$-value & Adjusted $p$-value\\
 \hline
2-Aminoadipic acid	          	& Amines			& 1.236871$\mathrm{e}-$07& 2.844803$\mathrm{e}-$05	\\
Valine                       		& Amines			& 3.316148$\mathrm{e}-$06& 3.494500$\mathrm{e}-$04	\\
Tyrosine                     		& Amines			& 4.558044$\mathrm{e}-$06& 3.494500$\mathrm{e}-$04	\\
Methyldopa                     		& Amines			& 7.515081$\mathrm{e}-$06& 4.321172$\mathrm{e}-$04	\\
Lysine                       		& Amines			& 2.749770$\mathrm{e}-$05& 1.264894$\mathrm{e}-$03	\\
Methylmalonic acid		    	& Organic acids			& 4.222770$\mathrm{e}-$05& 1.390688$\mathrm{e}-$03	\\
S-3-Hydroxyisobutyric acid		& Organic acids			& 4.232528$\mathrm{e}-$05& 1.390688$\mathrm{e}-$03	\\
TG(48:0)                      		& Lipids: Triglycerides		& 7.904916$\mathrm{e}-$05& 2.272663$\mathrm{e}-$03	\\
TG(50:4)                      		& Lipids: Triglycerides		& 9.542238$\mathrm{e}-$05& 2.438572$\mathrm{e}-$03	\\
TG(48:2)                      		& Lipids: Triglycerides		& 1.144591$\mathrm{e}-$04& 2.472172$\mathrm{e}-$03	\\
TG(51:3)                      		& Lipids: Triglycerides		& 1.182343$\mathrm{e}-$04& 2.472172$\mathrm{e}-$03	\\
TG(54:6)                     		& Lipids: Triglycerides		& 1.305959$\mathrm{e}-$04& 2.503088$\mathrm{e}-$03	\\
TG(50:3)                      		& Lipids: Triglycerides		& 1.629475$\mathrm{e}-$04& 2.882490$\mathrm{e}-$03	\\
TG(50:2)                      		& Lipids: Triglycerides		& 1.754559$\mathrm{e}-$04& 2.882490$\mathrm{e}-$03	\\
TG(50:1)                      		& Lipids: Triglycerides		& 2.217775$\mathrm{e}-$04& 3.400588$\mathrm{e}-$03	\\
TG(48:1)                     		& Lipids: Triglycerides		& 2.421916$\mathrm{e}-$04& 3.481504$\mathrm{e}-$03	\\
TG(52:4)                      		& Lipids: Triglycerides		& 3.157891$\mathrm{e}-$04& 4.141486$\mathrm{e}-$03	\\
TG(48:3)                     		& Lipids: Triglycerides		& 3.241163$\mathrm{e}-$04& 4.141486$\mathrm{e}-$03	\\
Leucine          	            	& Amines			& 4.124135$\mathrm{e}-$04& 4.871236$\mathrm{e}-$03	\\
LPC(18:1)	                     	& Lipids: Lysophosphatidylcholine& 4.235858$\mathrm{e}-$04& 4.871236$\mathrm{e}-$03	\\
TG(46:2)                      		& Lipids: Triglycerides		& 5.465651$\mathrm{e}-$04& 5.986189$\mathrm{e}-$03	\\
TG(50:0)                      		& Lipids: Triglycerides		& 6.518719$\mathrm{e}-$04& 6.815024$\mathrm{e}-$03	\\
TG(52:5)                      		& Lipids: Triglycerides		& 7.216611$\mathrm{e}-$04& 7.216611$\mathrm{e}-$03	\\
TG(52:3)                      		& Lipids: Triglycerides		& 1.149358$\mathrm{e}-$03& 1.101469$\mathrm{e}-$02	\\
TG(51:2)                      		& Lipids: Triglycerides		& 1.393980$\mathrm{e}-$03& 1.252523$\mathrm{e}-$02	\\
TG(56:8)                      		& Lipids: Triglycerides		& 1.451992$\mathrm{e}-$03& 1.252523$\mathrm{e}-$02	\\
Isoleucine           	        	& Amines			& 1.470353$\mathrm{e}-$03& 1.252523$\mathrm{e}-$02	\\
2-hydroxybutyric acid   		& Organic acids			& 1.704211$\mathrm{e}-$03& 1.399888$\mathrm{e}-$02	\\
3-Hydroxyisovaleric acid		& Organic acids			& 1.845997$\mathrm{e}-$03& 1.464067$\mathrm{e}-$02	\\
TG(51:1)    	                  	& Lipids: Triglycerides		& 1.996360$\mathrm{e}-$03& 1.530543$\mathrm{e}-$02	\\
SM(d18:1/20:1)                		& Lipids: Sphingomyelins	& 2.242748$\mathrm{e}-$03& 1.663974$\mathrm{e}-$02	\\
TG(52:1)    	                  	& Lipids: Triglycerides		& 2.377869$\mathrm{e}-$03& 1.709093$\mathrm{e}-$02	\\
8-iso-PGF2a (15-F2t-IsoP)              	& Oxidative stress: Isoprostane	& 2.670123$\mathrm{e}-$03& 1.860995$\mathrm{e}-$02	\\
Proline                  	     	& Amines			& 3.197010$\mathrm{e}-$03& 2.162683$\mathrm{e}-$02	\\
TG(54:5)              	        	& Lipids: Triglycerides		& 3.389655$\mathrm{e}-$03& 2.227487$\mathrm{e}-$02	\\
TG(56:7)               		       	& Lipids: Triglycerides		& 3.846194$\mathrm{e}-$03& 2.401863$\mathrm{e}-$02	\\
PGD2	           	            	& Lipids: Prostaglandins	& 3.863867$\mathrm{e}-$03& 2.401863$\mathrm{e}-$02	\\
TG(46:1)            	          	& Lipids: Triglycerides		& 4.273714$\mathrm{e}-$03& 2.586722$\mathrm{e}-$02	\\
PC(O-44:5)	                    	& Lipids: Plasmalogen Phosphatidylcholine& 4.595494$\mathrm{e}-$03& 2.683819$\mathrm{e}-$02	\\
LPA C14:0 				& Lyso-phosphatidic acid	& 4.667512$\mathrm{e}-$03& 2.683819$\mathrm{e}-$02	\\
PC(O-34:1)	                    	& Lipids: Plasmalogen Phosphatidylcholine& 5.876174$\mathrm{e}-$03& 3.296390$\mathrm{e}-$02	\\
LPC(20:4)	                     	& Lipids: Lysophosphatidylcholine& 7.163584$\mathrm{e}-$03& 3.922915$\mathrm{e}-$02	\\
SM(d18:1/24:2)	                	& Lipids: Sphingomyelins	& 7.371942$\mathrm{e}-$03& 3.943132$\mathrm{e}-$02	\\
8,12-iPF2a IV		            	& Oxidative stress: Isoprostane	& 8.208661$\mathrm{e}-$03& 4.290891$\mathrm{e}-$02	\\
TG(46:0)             	         	& Lipids: Triglycerides		& 8.513935$\mathrm{e}-$03& 4.351567$\mathrm{e}-$02	\\
5-iPF2a VI		              	& Oxidative stress: Isoprostane	& 9.211706$\mathrm{e}-$03& 4.589083$\mathrm{e}-$02	\\
TG(52:2)        	              	& Lipids: Triglycerides		& 9.377692$\mathrm{e}-$03& 4.589083$\mathrm{e}-$02	\\
SM(d18:1/16:0)	                	& Lipids: Sphingomyelins	& 9.782507$\mathrm{e}-$03& 4.687451$\mathrm{e}-$02	\\
TG(58:10)	                     	& Lipids: Triglycerides		& 1.063227$\mathrm{e}-$02& 4.861745$\mathrm{e}-$02	\\
Ornithine                      		& Amines			& 1.064486$\mathrm{e}-$02& 4.861745$\mathrm{e}-$02	\\
Histidine                    		& Amines			& 1.078039$\mathrm{e}-$02& 4.861745$\mathrm{e}-$02	\\
\hline											
\end{tabular}			
\end{tiny}							
\end{table}

\subsection{Classification Signature}\label{subsec:CS}
\subsubsection{Approach}\label{subsubsec:CSapproach}
\begin{sloppypar}
Metabolic classification signatures for the prediction of group membership (AD or SCD) were constructed by way of penalized logistic regression with a Lasso-penalty \cite{LASSOS}.
Two penalized settings were considered: (i) the Lasso selects among the metabolites while the clinical characteristics go unpenalized; and (ii) the Lasso selects among the metabolites without considering the clinical characteristics.
The resulting models were compared to an unpenalized logistic regression model that (iii) considered only the clinical characteristics.
Model estimation collides with minimizing the negative log-likelihood of the logistic model under an $\ell_1$-penalty.
The general problem can then be stated as:
\begin{equation}\label{Eq:LassoProblem}
\argmin_{\beta_0, \boldsymbol{\beta}^{u}, \boldsymbol{\beta}^{p}} \left\{ -\frac{1}{n} \mathcal{L}\left(\beta_0, \boldsymbol{\beta}^{u}, \boldsymbol{\beta}^{p};
\boldsymbol{\mathrm{y}}, \mathbf{X}^{u}, \mathbf{X}^{p}\right) + \lambda_{1}\|\boldsymbol{\beta}^{p}\|_{1}\right\},
\end{equation}
with $\mathcal{L}(\cdot)$ denoting the log-likelihood of the logistic model, $\boldsymbol{\mathrm{y}}$ the binary $n$-dimensional response vector, $\mathbf{X}^{u}$ denoting the $(n \times m)$-dimensional matrix of clinical-predictors, $\mathbf{X}^{p}$ denoting the $(n \times p)$-dimensional matrix of metabolite-predictors, $\boldsymbol{\beta}^{u}$ an $m$-dimensional vector of unpenalized regression coefficients, $\boldsymbol{\beta}^{p}$ a $p$-dimensional vector of penalized regression coefficients, and with $\beta_0$ denoting an intercept.
Lastly, $\lambda_{1}\|\cdot\|_{1}$ indicates the $\ell_1$-norm with penalty parameter $\lambda_{1}$, generally referred to as the Lasso-penalty.
The Lasso-penalty enables estimation in our setting where the feature to observation ratio (230/248) is too high for standard logistic regression.
It also achieves automatic model (i.e., feature) selection.
The problem in \ref{Eq:LassoProblem} is generally stated, in the sense that it captures all situations of interest.
In situation (iii) only the clinical predictors $\mathbf{X}^{u}$ are considered, such that the unpenalized parameters $\beta_0$ and $\boldsymbol{\beta}^{u}$ are estimated.
In situation (ii) only the metabolite-predictors $\mathbf{X}^{p}$ are considered, such that, next to $\beta_0$, the penalized parameters in $\boldsymbol{\beta}^{p}$ are estimated.
Situation (i) combines the former situations and, hence, considers the the general problem, estimating both unpenalized and penalized parameters.
The optimal penalty parameter in the penalized models was determined on the basis of leave-one-out cross-validation (LOOCV) of the model likelihood.
Predictive performance of all models was assessed by way of (the comparison of) Receiver Operating Characteristic (ROC) curves and Area Under the ROC Curves (AUCs).
ROC curves and AUCs for all models were produced by 10-fold cross-validation.

Note that the metabolic features were scaled in the classification exercises.
The (regularized) regression makes use of (in some sense) the covariance matrix of the features.
However, the variability of the features may differ substantially.
In such a situation the features with (relatively) extreme variability may drive the results.
Hence, it is appropriate to perform regularization on the standardized scale.
\end{sloppypar}

\subsubsection{Results}\label{subsubsec:CSresults}
\begin{figure}[b!]
\centering
  \includegraphics[scale=.4]{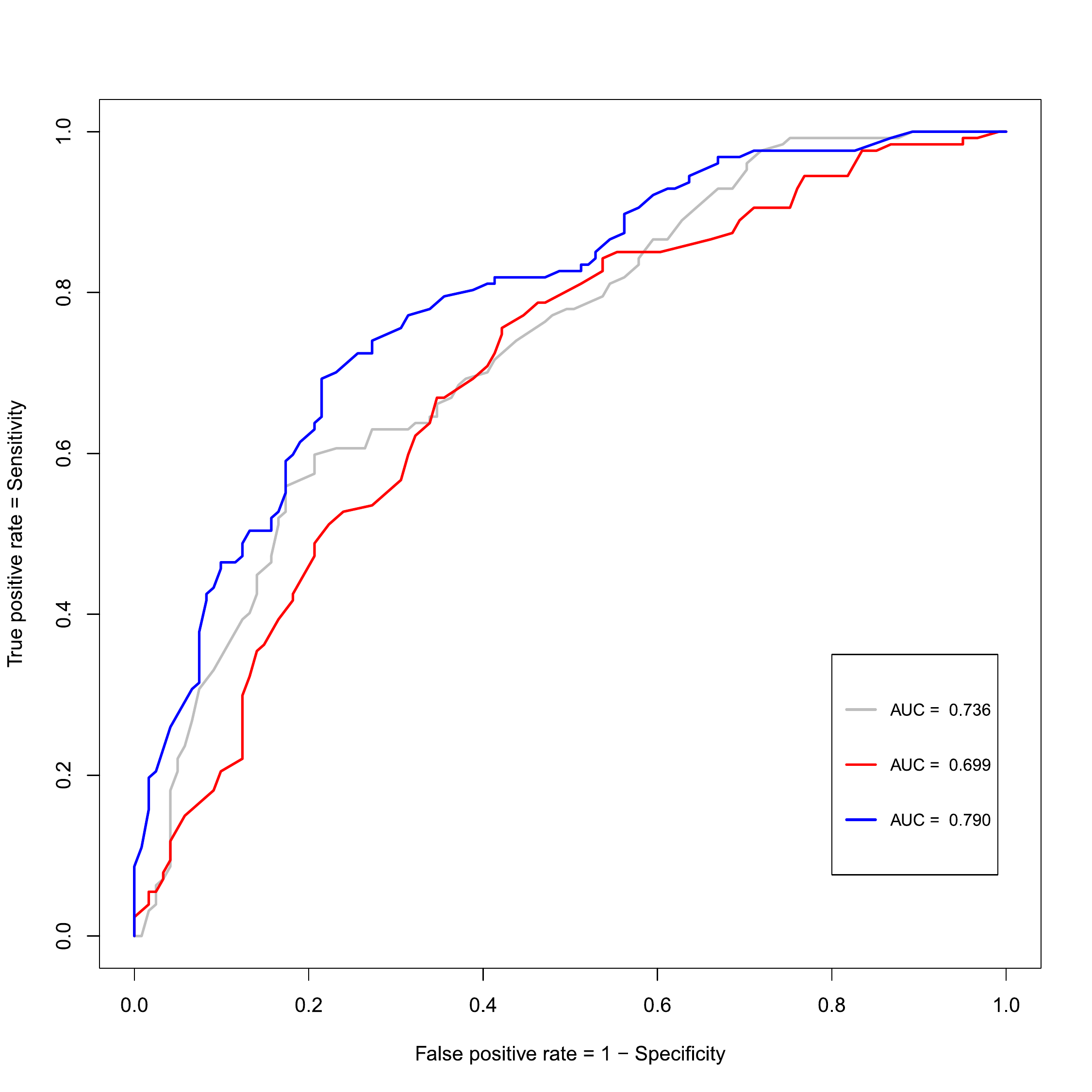}
    \caption{\footnotesize{ROC curves for the classification models.
    The grey line represents the ROC curve for the unpenalized logistic regression model that entertains the clinical characteristics only.
    The red line represents the ROC curve for the logistic model in which the Lasso performed variable selection amongst the metabolites (and that does not consider the clinical characteristics).
    The blue line represents the ROC curve of the logistic model in which the clinical characteristics are present while the Lasso may select amongst the metabolites.
    The clinical variables are listed in Table \ref{Table:clinVARS}.
    Appears as Figure 2 in the main text.}}
  \label{FIG:ROCs}
\end{figure}

Model performances can be found in Figure \ref{FIG:ROCs}.
The prediction model carrying the clinical variables only resulted in an AUC of .736.
The model that used the Lasso for selection amongst the metabolites sorts a comparable classification performance, yielding an AUC of approximately .7.
The model that adds a (Lasso-based) selection of metabolites to the clinical variables then improves predictive performance along the full false positive rate range, sorting a AUC of .79.
Table \ref{Table:ClassSITU1} contains the metabolites selected in the selection-amongst-metabolites-only situation.
Table \ref{Table:ClassSITU2} then contains the metabolites selected in the selection-amongst-metabolites-whilst-clinical-variables-present situation.
The highlighted features in these tables are also present in the differential expression signature (see Section \ref{subsec:DES}).
We see that the signs of their effects concur with the pattern of AD-associated under- and overexpression present in the differential expression signature.

\begin{table}[t!]
\begin{footnotesize}
\caption{Selected metabolites and parameter estimates when considering only metabolites as potential predictors.}
\centering
\label{Table:ClassSITU1}
\begin{tabular}{llr}
 \hline\hline
 Metabolite & Compound class & $\hat{\beta}$\\
 \hline
LPC(18:1)                      & Lipids: Lysophosphatidylcholine	&$ .431701077$  \\
PGD2                           & Oxidative stress: Prostaglandins	&$-.335722610$  \\
8,12-iPF2a IV                  & Oxidative stress: Isoprostane		&$-.321506682$  \\
\cellcolor{gray!20}O-Acetylserine& \cellcolor{gray!20}Amines		&\cellcolor{gray!20}$-.290694860$  \\
\cellcolor{gray!20}Methyldopa  & \cellcolor{gray!20}Amines		&\cellcolor{gray!20}$-.226545678$  \\
NO2-aLA (C18:3)                & Oxidative stress: Nitro-Fatty acid	&$-.214894781$  \\
\cellcolor{gray!20}Methylmalonic acid& \cellcolor{gray!20}Organic acids	&\cellcolor{gray!20}$-.211631150$  \\
\cellcolor{gray!20}TG(51:3)    & \cellcolor{gray!20}Lipids: Triglycerides&\cellcolor{gray!20}$-.202508671$  \\
\cellcolor{gray!20}Tyrosine    & \cellcolor{gray!20}Amines		&\cellcolor{gray!20}$-.192600743$  \\
Serine                         & Amines					&$ .151993825$  \\
LPC(20:4)                      & Lipids: Lysophosphatidylcholine	&$ .143911633$  \\
Arginine                       & Amines					&$ .139364338$  \\
\cellcolor{gray!20}SM(d18:1/23:0)& \cellcolor{gray!20}Lipids: Sphingomyelins&\cellcolor{gray!20}$-.138378505$  \\
Glyceric acid	               & Organic acids				&$-.107386135$  \\
Lysine                         & Amines					&$-.101945440$  \\
Glycolic acid	               & Organic acids				&$ .096537107$  \\
cLPA C18:0                     & Oxidative stress: Cyclic-lyso-phosphatidic acid&$-.072919587$  \\
SM(d18:1/18:0)                 & Lipids: Sphingomyelins			&$ .068838390$  \\
LPA C22:4 		       & Oxidative stress: Lyso-phosphatidic acid                 &$ .065264732$  \\
LPA C16                        & Oxidative stress: Lyso-phosphatidic acid&$-.064109038$  \\
3-Methoxytyramine              & Amines 				&$-.063622716$  \\
\cellcolor{gray!20}TG(54:6)    & \cellcolor{gray!20}Lipids: Triglycerides&\cellcolor{gray!20}$-.062333261$  \\
2,3-dinor-8-iso-PGF2a          & Oxidative stress: Isoprostane		&$-.059017772$  \\
PC(O-34:3)                     & Lipids: Plasmalogen Phosphatidylcholine&$-.048557268$  \\
Cis-Aconitic acid	       & Organic acids				&$-.038573969$  \\
\cellcolor{gray!20}3-Hydroxyisovaleric acid& \cellcolor{gray!20}Organic acids&\cellcolor{gray!20}$-.034646727$  \\
LPA C14:0 		       & Oxidative stress: Lyso-phosphatidic acid			&$-.033059634$  \\
\cellcolor{gray!20}2-Aminoadipic acid& \cellcolor{gray!20}Amines	&\cellcolor{gray!20}$-.028720481$  \\
PC(O-36:6)                     & Lipids: Plasmalogen Phosphatidylcholine&$-.025814804$  \\
PC(O-38:6)                     & Lipids: Plasmalogen Phosphatidylcholine&$-.021959733$  \\
Putrescine                     & Amines					&$-.017413981$  \\
TG(48:0)                       & Lipids: Triglycerides			&$-.016328393$  \\
Homoserine                     & Amines 				&$-.015906351$  \\
TG(O-50:0)                     & Lipids: Triglycerides			&$ .014668718$  \\
Carnosine                      & Amines					&$ .012064246$  \\
5-iPF2a VI                     & Oxidative stress: Isoprostane		&$-.003633138$  \\
Sarcosine                      & Amines					&$-.001072232$  \\
 \hline									
\end{tabular}							
\end{footnotesize}							
\end{table}								

\begin{table}[t!]
\begin{footnotesize}
\caption{Selected metabolites and parameter estimates when considering metabolites as potential predictors on top of the clinical variables.}
\centering
\label{Table:ClassSITU2}
\begin{tabular}{llr}
 \hline\hline
 Metabolite & Compound class & $\hat{\beta}$\\
 \hline
PGD2                           & Oxidative stress: Prostaglandins	&$-.47539723$	\\
\cellcolor{gray!20}O-Acetylserine& \cellcolor{gray!20}Amines		&\cellcolor{gray!20}$-.45299269$	\\
\cellcolor{gray!20}Methylmalonic acid& \cellcolor{gray!20}Organic acids	&\cellcolor{gray!20}$-.39360564$	\\
NO2-aLA (C18:3)                & Oxidative stress: Nitro-Fatty acid	&$-.32516790$	\\
\cellcolor{gray!20}TG(51:3)    & \cellcolor{gray!20}Lipids: Triglycerides&\cellcolor{gray!20}$-.29844463$	\\
\cellcolor{gray!20}SM(d18:1/20:1)& \cellcolor{gray!20}Lipids: Sphingomyelins&\cellcolor{gray!20}$ .29404906$	\\
\cellcolor{gray!20}3-Hydroxyisovaleric acid& \cellcolor{gray!20}Organic acids&\cellcolor{gray!20}$-.27369915$	\\
8,12-iPF2a IV                  & Oxidative stress: Isoprostane		&$-.22711647$	\\
PE(38:2)                       & Lipids: Phosphatidylethanolamine	&$-.16302211$	\\
Gamma-glutamylalanine          & Amines					&$ .15344618$	\\
LPC(18:1)                      & Lipids: Lysophosphatidylcholine	&$ .15057771$	\\
\cellcolor{gray!20}Methyldopa  & \cellcolor{gray!20}Amines		&\cellcolor{gray!20}$-.14912088$	\\
\cellcolor{gray!20}SM(d18:1/23:0)& \cellcolor{gray!20}Lipids: Sphingomyelins&\cellcolor{gray!20}$-.14714005$	\\
Putrescine                     & Amines					&$-.14137904$	\\
LPC(20:4)                      & Lipids: Lysophosphatidylcholine	&$ .12467937$	\\
\cellcolor{gray!20}8-iso-PGF2a (15-F2t-IsoP)& \cellcolor{gray!20}Oxidative stress: Isoprostane&\cellcolor{gray!20}$-.12101061$	\\
LPA C18:3 		       & Oxidative stress: Lyso-phosphatidic acid			&$-.10140491$	\\
LPA C14:0 		       & Oxidative stress: Lyso-phosphatidic acid			&$-.09396454$	\\
Uracil                         & Organic acids				&$ .09348099$	\\
Citrulline                     & Amines					&$-.08529403$	\\
Histamine                      & Amines					&$ .07351094$	\\
Glyceric acid                  & Organic acids				&$-.06673703$	\\
\cellcolor{gray!20}TG(56:8)    & \cellcolor{gray!20}Lipids: Triglycerides&\cellcolor{gray!20}$-.06152908$	\\
\cellcolor{gray!20}TG(58:10)   & \cellcolor{gray!20}Lipids: Triglycerides&\cellcolor{gray!20}$-.05699948$	\\
2,3-dinor-8-iso-PGF2a          & Oxidative stress: Isoprostane		&$-.04718506$	\\
NO2-OA (C18:1) 		       & Oxidative stress: Nitro-Fatty acid	&$-.04178329$	\\
Glycolic acid	               & Organic acids				&$ .04016606$	\\
Carnosine                      & Amines					&$ .03960024$	\\
Serine                         & Amines					&$ .03620446$	\\
SM(d18:1/18:0)                 & Lipids: Sphingomyelins			&$ .01812763$	\\
\cellcolor{gray!20}S-3-Hydroxyisobutyric acid& \cellcolor{gray!20}Organic acids&\cellcolor{gray!20}$-.01042997$	\\
 \hline
\end{tabular}
\end{footnotesize}
\end{table}

\begin{sloppypar}
Note that the Lasso selects features from all metabolite classes.
To assess if the metabolite-class has predictive power a group-regularized logistic ridge regression \cite{GRRS} was used in which the metabolite-class serves as co-data.
This analysis indicated that the metabolite-class forms weakly informative co-data.
This strengthens faith in the Lasso results.
The strongest predictor amongst the clinical variables is (naturally) APOE $\epsilon$4 allele status.
\end{sloppypar}

\subsection{Regulatory Signature}\label{subsec:RS}
\subsubsection{Graphical Modeling}\label{subsubsec:GGM}
A differential expression signature represents the features that are relatively under- or overexpressed in diagnostic groups of interest.
This signature does not have to concur with the classification signature completely, as the latter (i) chooses amongst multicollinear features and (ii) emphasizes prediction rather than shifts in location. The classification signature, in turn, is limited in its capacity to represent complex dependencies amongst the metabolites (of interest).
Hence, we seek to explore a third signature: The regulatory signature.
This signature intends to uncover deregulation in metabolic biochemical pathways as pertaining to the AD disease process.
A metabolic pathway can be thought of as a collection of metabolic features originating from all over the metabolome, that work interdependently to regulate some biochemical (disease) process.
Hence, a pathway is a network.
And a network can be represented by a graph.
We thus take interest in graphical modeling.

Graphical modeling refers to a class of probabilistic models that use graphs to express conditional (in)dependence relations between random variables.
We consider graphs $\mathcal{G} = (\mathcal{V}, \mathcal{E})$
consisting of a finite set $\mathcal{V}$ of vertices and set of
edges $\mathcal{E}$. The vertices of the graph correspond to a
collection of random variables with probability distribution
$\mathcal{P}$, i.e., $\{Y_{1},\ldots,Y_{p}\} \sim \mathcal{P}$.
Edges in $\mathcal{E}$ consist of pairs of distinct vertices such
that $Y_{j} - Y_{j'} \in \mathcal{E}$. The basic assumption is:
$\{Y_{1},\ldots,Y_{p}\} \sim \mathcal{N}_{p}(\boldsymbol{0},
\mathbf{\Sigma})$, with $\mathbf{\Sigma}$ positive definite.
Hence, we focus on Gaussian graphical modeling.

In this Gaussian case, conditional independence between a pair of
variables corresponds to zero entries in the precision matrix.
Indeed, let $\hat{\mathbf{\Sigma}}^{-1} = \hat{\mathbf{\Omega}}$ denote a generic estimate of the
precision matrix and consider its transformation to a partial
correlation matrix $\hat{\mathbf{P}}$. Then the following relations
can be shown to hold for all pairs $\{Y_{j}, Y_{j'}\} \in
\mathcal{V}$ with $j \neq j'$ \cite[see, e.g.,][]{whittakerS}:
\begin{equation}\nonumber
(\hat{\mathbf{P}})_{jj'} = 0
\Longleftrightarrow (\hat{\mathbf{\Omega}})_{jj'} = 0
\Longleftrightarrow Y_{j} \ci Y_{j'}|\mathcal{V}\setminus\{Y_{j},Y_{j'}\}
\Longleftrightarrow Y_{j} \centernot{-} Y_{j'},
\end{equation}
where $\mathcal{V}\setminus\{\cdot\}$ denotes set-minus notation and
where $\centernot{-}$ indicates the absence of an edge.
In words: A zero partial correlation implies a zero precision matrix entry which
in turn implies that the corresponding two variables are conditionally independent
(given the remaining variables) which then implies the absence of an edge
between these variables in the corresponding graph.
Such a graph can thus be interpreted as a conditional independence graph.


\subsubsection{Approach}\label{subsubsec:RSapproach}
Now, model selection efforts in Gaussian graphical models focus on determining the support of the precision matrix.
Problematic is that in situations with $p \approx n$ or $p > n$ the sample covariance matrix $\hat{\mathbf{\Sigma}} = \mathbf{S}$ is ill-behaved or singular such that it's inverse ($\mathbf{S}^{-1}$, which would constitute an estimate of the precision matrix) is unstable or does not exist.
Moreover, the metabolic features of interest are highly collinear (within the respective metabolite classes).
Hence, we need a regularized estimate of the precision matrix.
In addition, we want to take into account that our data consist of distinct classes of interest.
The first distinction is, naturally, AD versus SCD.

So, we are after jointly estimating multiple regularized precision matrices from (aggregated) high-dimensional data consisting of distinct classes. From a network perspective, molecular pathway-deregulation in the disease state is likely characterized by the loss of normal (wanted) molecular interactions and the gain of abnormal (unwanted) molecular interactions. One would thus expect the network topologies of our groups of interest to primarily share the same structure, while potentially differing in a number of (topological) locations of interest. Our regularized network extraction method takes this explicitly into account. Specifically, we employ a special case of targeted fused ridge estimation \cite{FUSEDS}, solving the following estimation problem:
\begin{equation}
  \label{eq:argmax2}
  \argmax_{\{\vOmega_g\} \in \calS_{++}^p}
  \left\{
    \calL\left(\{\vOmega_g\}; \{\vS_g\}\right)
    - \frac{\lambda}{2}\sum_g
      \sqfnorm[\big]{ \vOmega_g \!- \vT }
    - \frac{\lambda_f}{4}\sum_{\mathclap{g_1, g_2}}
      \sqfnorm[\big]{ \vOmega_{g_1} \!- \vOmega_{g_2}  }
  \right\},
\end{equation}
where the $\mathbf{S}_g$ indicate group-specific sample covariance matrices, $\lambda$ denotes a strictly positive ridge penalty, $\lambda_f$ denotes a positive fusion penalty, and $\mathbf{T}$ denotes a target matrix.
The penalty parameter $\lambda$ controls the rate of shrinkage of each precision $\vOmega_g$ towards the corresponding target $\vT$, while $\lambda_f$ determines the retainment of entry-wise similarities between $\vOmega_{g_1}$ and $\vOmega_{g_2}$ for all class pairs $g_1 \neq g_2$. For given penalties the problem can be solved with an block coordinate ascent procedure \cite{FUSEDS}, resulting in an estimated precision matrix for each class $g$. In this case $g = 1,2$.

We solve (\ref{eq:argmax2}) using the class-specific sample covariance matrices (i.e., the sample covariance matrices of the class-specific data) as the data entries.
For the target $\mathbf{T}$ we choose the (weakly informative) $p$-dimensional identity matrix $\mathbf{I}_p$.
The optimal penalty parameters were determined by the LOOCV procedure described in \cite{FUSEDS}.
The optimal penalty values were found to be $\lambda^{*} = $ 2.742348, and $\lambda_f^\mathds{*} = $ 9.867606e-22.
These penalty values emphasize individual regularization over retainment of entry-wise similarities, indicating strong differences in class-specific precision matrices.
The support of the estimated class precision matrices was determined by thresholding.
For each class-specific matrix, the 100 strongest edges (in terms of absolute partial correlations) were retained.
As metabolic networks are very dense, retaining the 100 strongest edges is assumed to give a more clear picture of the most influential regulatory players.
The retained partial correlations range, in absolute value, from .1670877 to .6412267 over the respective classes.
All analyzes were performed with the \texttt{rags2ridges} package \cite{RAGSS} in \texttt{R} \cite{RS}.

\subsubsection{Visualization}\label{subsubsec:RSvisualization}
\begin{figure}[b!]
\centering
  \includegraphics[width=.95\textwidth]{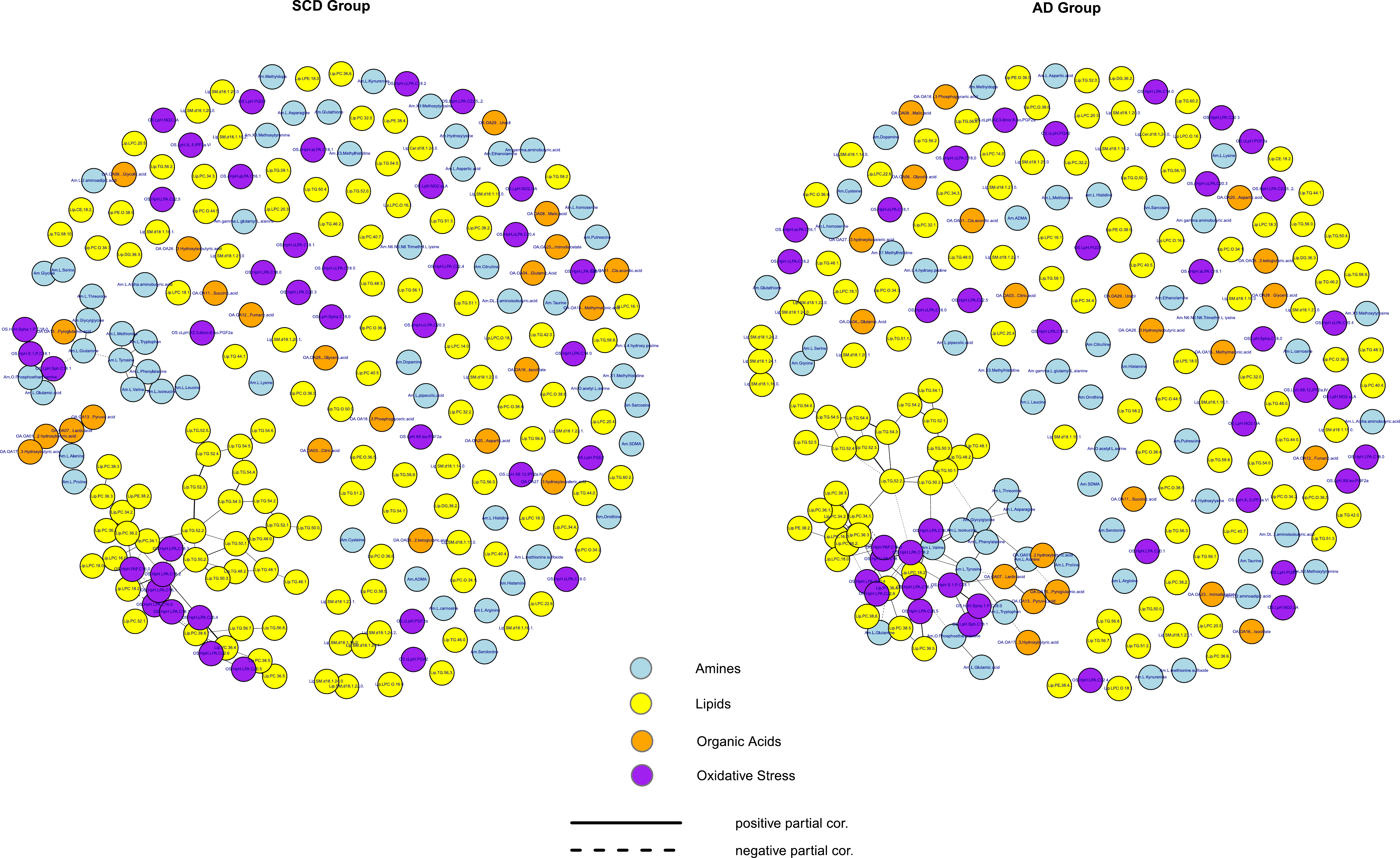}
    \caption{\footnotesize{Class-specific networks visualized with the Fruchterman-Reingold algorithm.
    The left-hand panel contains the network for the SCD group.
    The right-hand panel contains the network for the AD group.
    The metabolite compounds are colored according to metabolite family: Blue for amines, yellow for lipids, orange for organic acids, and purple for oxidative stress.
    Solid edges represent positive partial correlations while dashed edges represent negative partial correlations.}}
  \label{FIG:NWKSnoPruneSCAD}
\end{figure}

The first idea regarding the network structures represented by our class-specific precision matrices can be obtained by simple visualization.
Figure \ref{FIG:NWKSnoPruneSCAD} contains the class-specific networks visualized with the Fruchterman-Reingold (FR) algorithm \cite{FRUCHTS}.
These networks contain all metabolic features, even when they are not connected, resulting in `hairball' networks: Networks that are too tangled to be effectively visualized.
They do, however, convey that the strongest edges implicate metabolic features from all considered metabolite families.

\begin{figure}[h]
\centering
  \includegraphics[width=\textwidth]{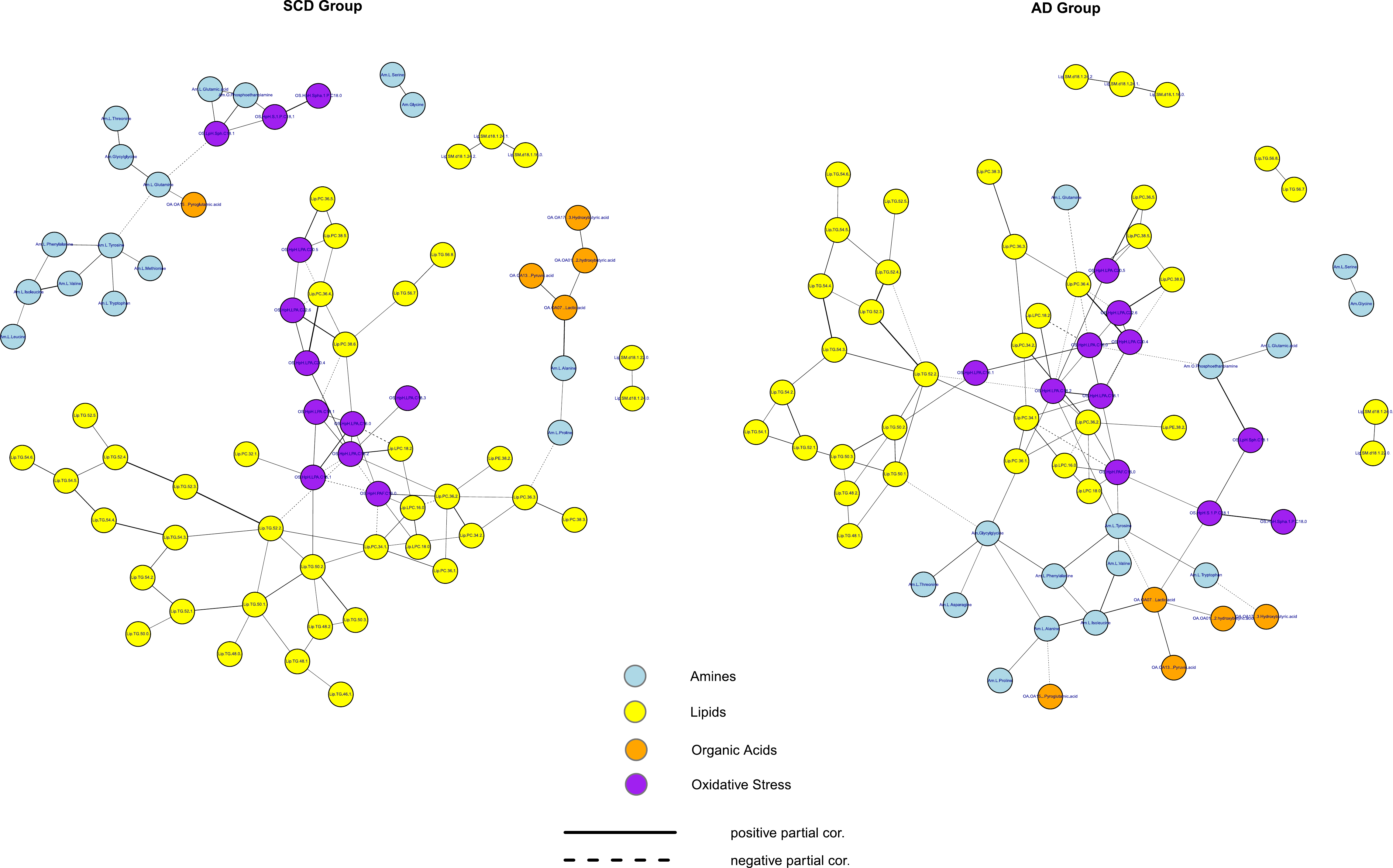}
    \caption{\footnotesize{Class-specific \emph{pruned} networks visualized with the Fruchterman-Reingold algorithm.
    The left-hand panel contains the network for the SCD group.
    The right-hand panel contains the network for the AD group.
    The metabolite compounds are colored according to metabolite family: Blue for amines, yellow for lipids, orange for organic acids, and purple for oxidative stress.
    Solid edges represent positive partial correlations while dashed edges represent negative partial correlations.}}
  \label{FIG:NWKSPruneSCAD}
\end{figure}

Figure \ref{FIG:NWKSPruneSCAD} contains the pruned class-specific networks visualized with the FR algorithm.
That is, it retains only the class-specific connected components from Figure \ref{FIG:NWKSnoPruneSCAD}.
The pruned networks more effectively represent the retained topologies.

To assess the topology more closely, it is beneficial to arrange the metabolic features in fixed coordinates over the respective groups of interest.
Figure \ref{FIG:CoordsSCAD} contains the semi-pruned class-specific networks.
They are semi-pruned as in each class-specific topology all metabolites are depicted that are present in the union of connected metabolites over all class-specific topologies.
This allows us to visualize the individual topologies with fixed metabolite-coordinates.
The FR-based coordinates for the SCD group serve as the reference coordinates for all topologies.
We see that the union of metabolic features is quite tight, suggesting that the core metabolic features for the SCD and AD groups overlap to a large extent.
At first glance the diseased state indeed seem less locally connected.
We now turn to numerical and graph theoretic assessments to support understanding of the topologies.

\subsubsection{Global Characteristics}\label{subsubsec:RSglobalCHAR}
\begin{landscape}
\begin{figure}
\centering
  \includegraphics[scale = .4]{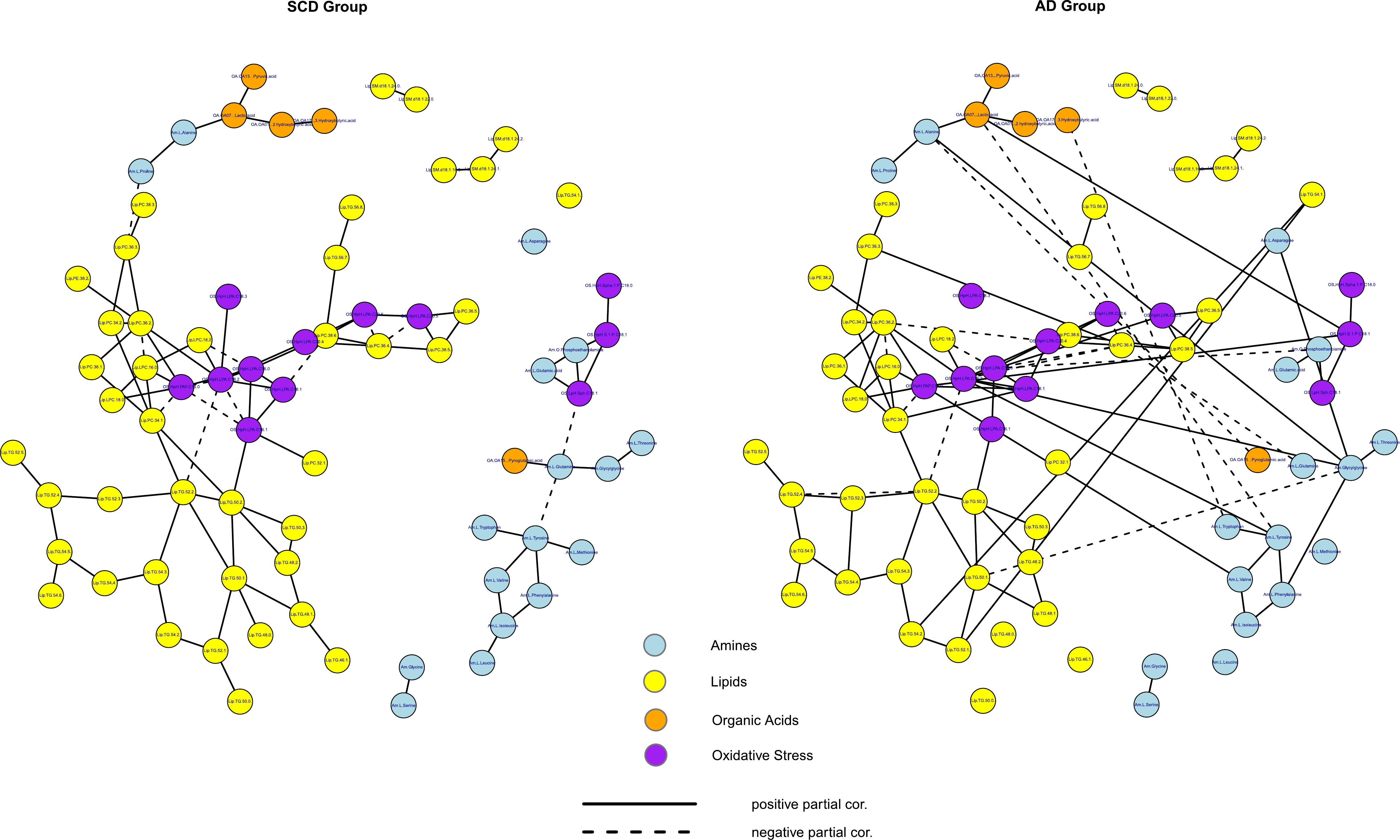}
    \caption{\footnotesize{Class-specific \emph{semi-pruned} networks visualized with the Fruchterman-Reingold algorithm.
    The left-hand panel contains the network for the SCD group.
    The right-hand panel contains the network for the AD group.
    The coordinates of the left-hand topology serve as the reference coordinates.
    The metabolite compounds are colored according to metabolite family: Blue for amines, yellow for lipids, orange for organic acids, and purple for oxidative stress.
    Solid edges represent positive partial correlations while dashed edges represent negative partial correlations.}}
  \label{FIG:CoordsSCAD}
\end{figure}
\end{landscape}

Here, we will assess some global characteristics of each class-related graph as given in Figure \ref{FIG:CoordsSCAD}.
Table \ref{Table:TRANSnoapoe} contains some global metrics for each topology.
Please note that formal definitions of all terms relating to network science as used throughout this supplement can be found in, e.g., \cite{NewmanS}.
Transitivity is a shape measure, with higher scores indicating stronger local connectivity.
Transitivity for the SCD topology is approximately .24, which is higher than the transitivity score for the AD topology ($\approx. 15$) and also higher than many other biological networks \cite[p. 200 \& Section 8.6]{NewmanS}.
Hence, the SCD topology is stronger locally connected.
Centrality \cite{FreemanS} is another shape measure, indicating the degree in which the topology resembles a maximally centralized graph (i.e., a star graph).
The more centralized a network, the more vulnerable it is, in the sense that it's connectedness hinges upon few nodes.
The centralization scores indicate that both the SCD and the AD topology are not very centralized.

\begin{figure}[b!]
\centering
  \includegraphics[width=.95\textwidth]{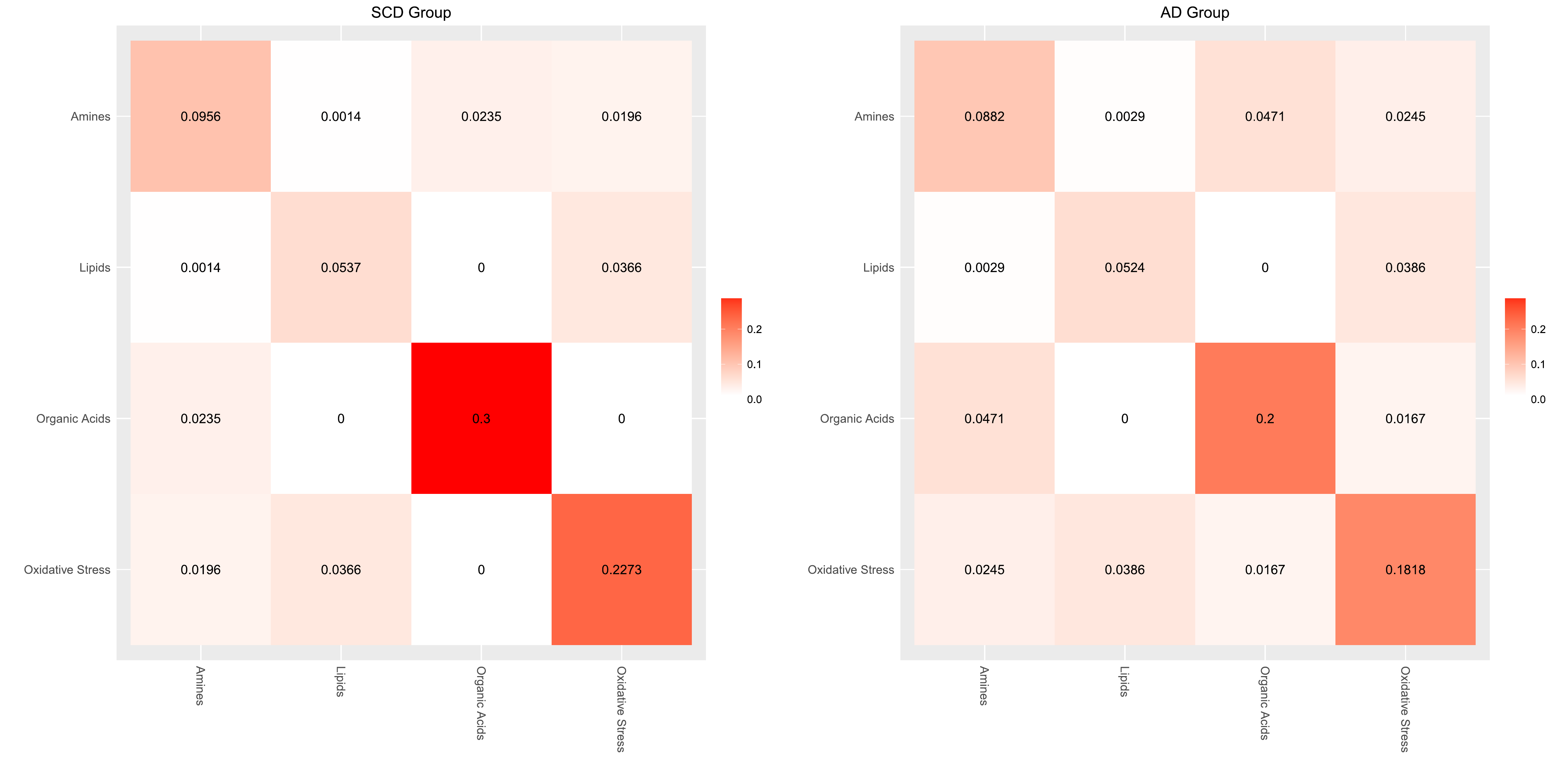}
    \caption{\footnotesize{Heatmaps of degree densities for the SCD and AD networks.
    The reported numbers represent the degree density for the (combinations of) metabolite groups.
    Degree density represents the number of connections (edges) divided by the number of possible connections.
    For example, in the network for the SCD group the proportion of actual edges relative to the number of possible edges between Amines and oxidative stress compounds is .0196.
    Note that all heatmaps received the same color key.
    Hence, the color intensities (i.e., the color-spectrum representations of the cell-numbers) are comparable over the respective heatmaps.}}
  \label{FIG:DegreeDensityNOAPOE}
\end{figure}

\begin{figure}[t!]
\centering
  \includegraphics[width=\textwidth]{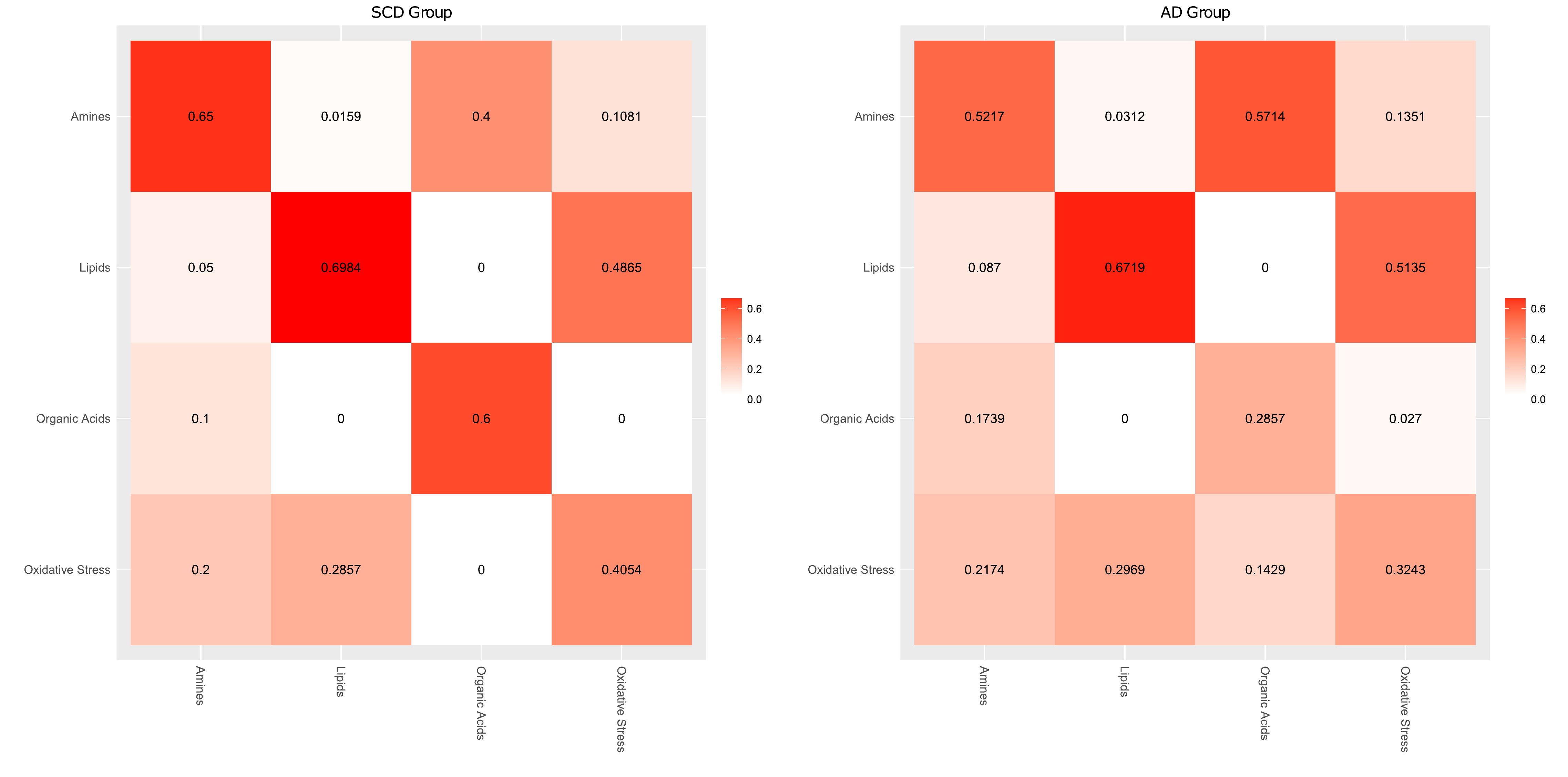}
    \caption{\footnotesize{Heatmaps of relative outdegrees for the class-specific networks.
    The reported numbers represent the relative outdegree for the (combinations of) metabolite groups.
    The relative outdegree represents the number of connections (edges) between two metabolite groups divided by the number of `outgoing' connections for one of these groups.
    For example, in the network for the SCD group the number of edges between lipid and oxidative stress compounds accounts for approximately 29\% of all edges involving lipids and approximately 49\% of all edges involving oxidative stress compounds.
    Note that all heatmaps received the same color key.
    Hence, the color intensities (i.e., the color-spectrum representations of the cell-numbers) are comparable over the respective heatmaps.
    Note that the column numbers should sum to unity.}}
  \label{FIG:ROUTDegreeDensityNOAPOE}
\end{figure}

In addition to shape metrics, there are cohesion metrics.
One such cohesion metric is `connectedness', which, for the topologies of interest, is also given in Table \ref{Table:TRANSnoapoe}.
Connectedness refers to the ``proportion of pairs of nodes that can reach each other by a path of any length" \cite{BorgattiS}.
The AD topology has a higher connectedness score than the SCD topology.
This is (in part) due to the fact that the SCD topology has a large disconnected component (consisting of amine compounds mostly).
Figure \ref{FIG:CoordsSCAD} seems to indicate that the AD topology may be characterized by an increased connection density within amine compounds and between amine and oxidative stress compounds.
This is reflected in the degree density and the relative outdegree \cite{KrumsiekS} for the respective topologies.
The between-metabolite-family and within-metabolite-family degree densities (Figure \ref{FIG:DegreeDensityNOAPOE}) suggest that the AD topology is characterized by increased amine-connections (connections in which at least 1 amine compound is present).
The relative outdegrees (Figure \ref{FIG:ROUTDegreeDensityNOAPOE}) imply that the AD topology is characterized by more connections between amines and organic acid compounds and more connections between amines and oxidative stress compounds.

\begin{table}[h]
\centering
\caption{Global metrics for the two topologies of interest (see Figure \ref{FIG:CoordsSCAD}).}
\label{Table:TRANSnoapoe}
\begin{tabular}{lrrr}
\hline
\hline
Topology    & Transitivity & Connectedness & Centralization\\ \hline
SCD group   & .2424242    &  .4864865 & .0879304 \\
AD          & .1458967    &  .6187387 & .0879304 \\
\hline
\end{tabular}
\end{table}

\subsubsection{Node Characteristics}\label{subsubsec:RSnodeCHAR}
In addition to global metrics, we assess certain characteristics of individual nodes within the topologies of interest, focusing especially on the notion of centrality.
Centrality, in general, refers to metrics regarding the most central or (functionally) important nodes in a network.
Several centrality measures are used whose formal definition can be found in, e.g., \cite{NewmanS}: degree centrality, betweenness centrality, closeness centrality, and eigenvalues centrality.
Degree centrality simply indicates the number of connections in which a node takes part.
It is indicative of the nodes that are central or influential in terms of the number of connections: more connections could imply deeper regulatory influence.
Betweenness centrality \cite{FreemanBetweenS} measures centrality in terms of information flow.
Under the assumption that information is passed over short(est) paths a node becomes central when the number of short(est) paths that pass through it is high.
Closeness centrality indicates the mean distance of a node to other nodes.
A node is central under the closeness centrality metric when it's mean distance to other nodes is low.
Note that closeness as used here reflects the sum of inverse distances \cite{CloseCentralS}, such that nodes that are close to many other nodes receive high closeness scores.
The eigenvalue centrality \cite{NewmanS} is an extension of the degree centrality.
A node's eigencentrality is based on the centrality of the nodes to which it is connected: connections to central others are weighted more heavily in the final eigencentrality score than connections to less central others.
These various centrality scores thus have different flavors and may indicate correspondingly flavored hubs (i.e., highly central nodes).

\begin{figure}[h!]
\centering
  \includegraphics[width=.97\textwidth]{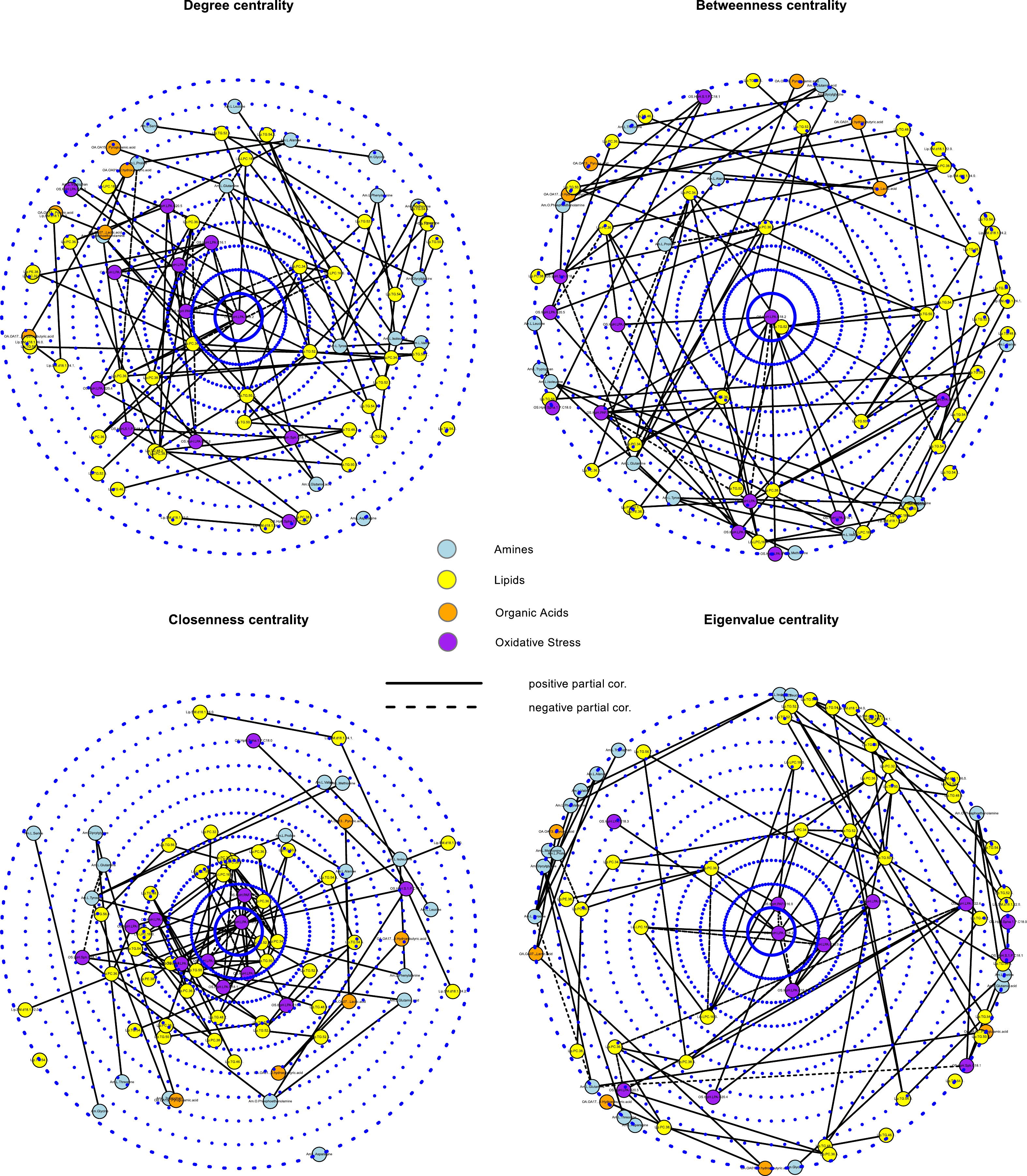}
    \caption{\footnotesize{Target plots \cite{TARGETplotsS} visualizing various centralities for the network representing the SCD group.
    The upper-left panel represents degree centralities.
    The upper-right panel represents betweenness centralities.
    The lower-left panel represents closeness centralities.
    The lower-right panel represents eigenvalue centralities.
    Note that, for each target plot, the network is the same as in the left-hand panel of Figure \ref{FIG:CoordsSCAD}.
    The topology is now however plotted to represent metabolite features according to various centrality scores.
    For example, the oxidative stress compound LPA C18:2 has the highest degree centrality and, hence, is depicted in the center of the upper-left panel.
    The metabolite compounds are again colored according to metabolite family: Blue for amines, yellow for lipids, orange for organic acids, and purple for oxidative stress.
    Solid edges represent positive partial correlations while dashed edges represent negative partial correlations.
    The metabolite features attaining the highest centrality scores are given in Table \ref{TABLE:CentralSC}.}}
  \label{FIG:CentralSC}
\end{figure}

\begin{figure}[h!]
\centering
  \includegraphics[width=.97\textwidth]{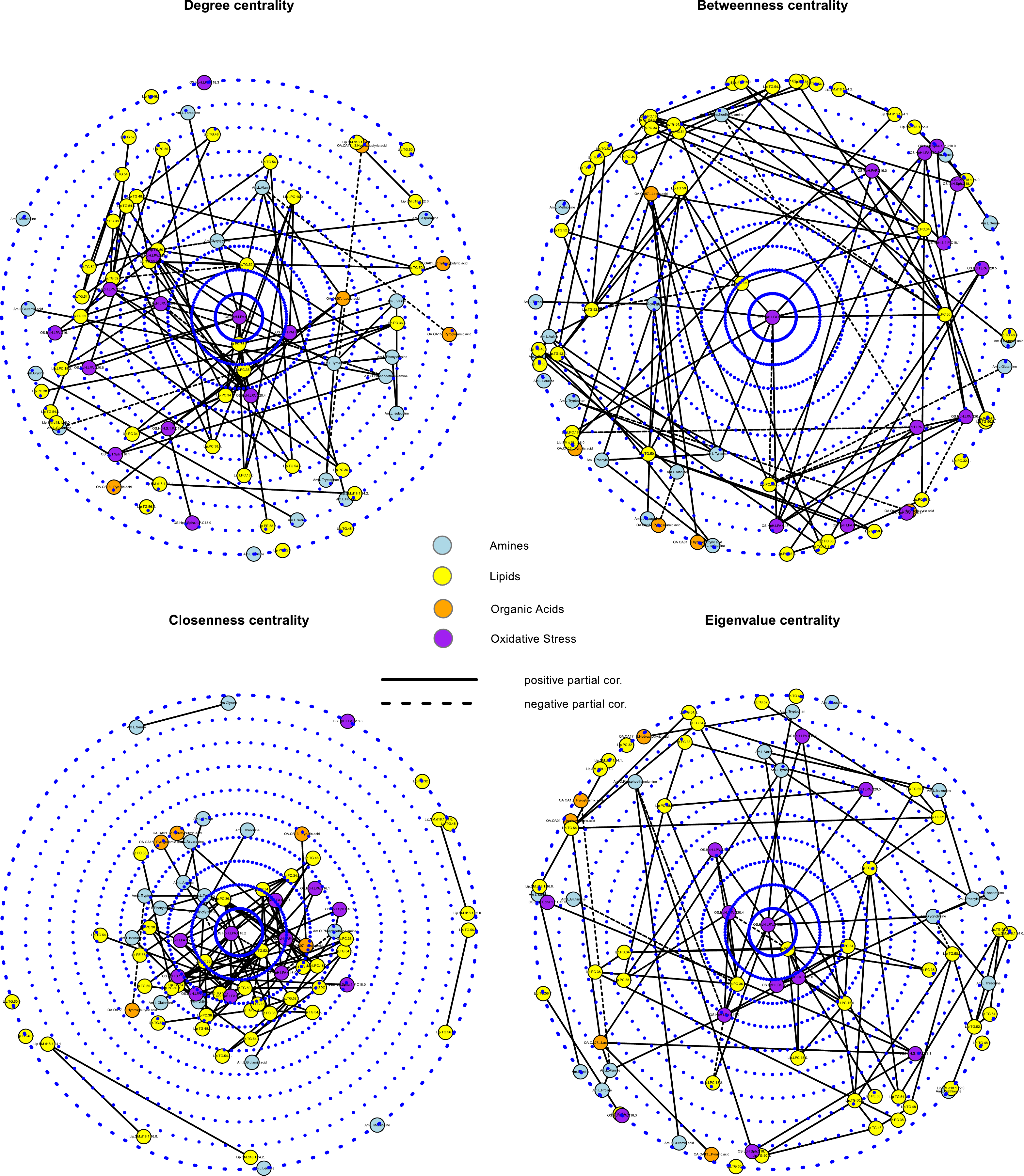}
    \caption{\footnotesize{Target plots \cite{TARGETplotsS} visualizing various centralities for the network representing the AD group.
    The upper-left panel represents degree centralities.
    The upper-right panel represents betweenness centralities.
    The lower-left panel represents closeness centralities.
    The lower-right panel represents eigenvalue centralities.
    Note that, for each target plot, the network is the same as in the right-hand panel of Figure \ref{FIG:CoordsSCAD}.
    The topology is now however plotted to represent metabolite features according to various centrality scores.
    For example, the oxidative stress compound LPA C18:2 has the highest degree centrality and, hence, is depicted in the center of the upper-left panel.
    The metabolite compounds are again colored according to metabolite family: Blue for amines, yellow for lipids, orange for organic acids, and purple for oxidative stress.
    Solid edges represent positive partial correlations while dashed edges represent negative partial correlations.
    The metabolite features attaining the highest centrality scores are given in Table \ref{TABLE:CentralAD}.}}
  \label{FIG:CentralAD}
\end{figure}

Figures \ref{FIG:CentralSC} and \ref{FIG:CentralAD} contain target plots \cite{TARGETplotsS} depicting the various centrality properties of the metabolite compounds retained in the SCD and AD topologies.
The target plots were created with the help of the \texttt{sna} and \texttt{igraph} packages in \texttt{R} \cite{SNAS, iGRAPHS, RnetworksS}.
Tables \ref{TABLE:CentralSC} and \ref{TABLE:CentralAD} then contain, to accompany these figures, the top centrality metrics for the SCD and AD topologies, respectively.
The strongest hub in the SCD topology is the oxidative stress compound LPA C18:2.
This compound sorts the highest scores on all centrality measures.
The top compounds, in terms of centrality, in the SCD topology all belong to either the oxidative stress or lipid family.
Next to LPA C18:2, the Phosphatidylcholines PC(34:1) and PC(36:2) are consistently represented as hubs by all centrality measures.
LPA C18:2 is also the strongest hub compound in the AD topology.
The top compounds, in terms of centrality, of the AD topology indeed largely overlap with the top compounds of the SCD topology.
Although the same compound may be a hub in both the AD and SCD topologies, it can still be wired very differently, i.e., it's connections may differ greatly between the two topologies (see Section \ref{subsubsec:RScommunityCHAR}).
In addition, for the AD topology the amines Glycylglycine and Tyrosine are quite consistently indicated as central compounds by the degree, betweenness, and closeness metrics.

\begin{table}[h]
\begin{scriptsize}
\centering
\caption{Centrality measures for the SCD topology.}
\label{TABLE:CentralSC}
\begin{tabular}{lrllrllrllr}
\hline
\multicolumn{2}{c}{Degree} &  & \multicolumn{2}{c}{Betweenness}     &  & \multicolumn{2}{c}{Closeness}     &  & \multicolumn{2}{c}{Eigenvalue}      \\
\cline{1-2} \cline{4-5} \cline{7-8} \cline{10-11}
LPA C18:2 		& 9 &  & LPA C18:2 		& 483.00 &  & LPA C18:2 	& .317 &  &    LPA C18:2 	& .415 \\
PC(36:2)	 	& 7 &  & TG(52:2)	 	& 454.06 &  & TG(52:2)	 	& .295 &  &    PAF C16:0	& .378 \\
PAF C16:0	 	& 7 &  & PC(36:3)	 	& 300.00 &  & PAF C16:0 	& .279 &  &    LPA C16		& .331 \\
TG(50:2)	 	& 6 &  & PC(36:2)	 	& 291.50 &  & PC(36:2)	 	& .276 &  &    LPA C16:1	& .315 \\
TG(52:2)	 	& 6 &  & Proline	 	& 220.00 &  & PC(34:1)	 	& .274 &  &    PC(36:2)		& .271 \\
PC(34:1)	 	& 6 &  & TG(50:1) 		& 202.00 &  & TG(50:2)	 	& .274 &  &    PC(34:1) 	& .238 \\
LPA C16		 	& 6 &  & PC(34:1)	 	& 180.85 &  & LPA C16:1 	& .271 &  &    LPA C18:1	& .237 \\
LPA C16:1	 	& 6 &  & Alanine	 	& 180.00 &  & LPA C16	 	& .270 &  &    LPC(16:0) 	& .235 \\
\hline
\end{tabular}
\end{scriptsize}
\end{table}

\begin{table}[h]
\begin{scriptsize}
\centering
\caption{Centrality measures for the AD topology.}
\label{TABLE:CentralAD}
\begin{tabular}{lrllrllrllr}
\hline
\multicolumn{2}{c}{Degree} &  & \multicolumn{2}{c}{Betweenness}     &  & \multicolumn{2}{c}{Closeness}     &  & \multicolumn{2}{c}{Eigenvalue}      \\
\cline{1-2} \cline{4-5} \cline{7-8} \cline{10-11}
LPA C18:2 		& 9 &  &      LPA C18:2 	& 661.87 &  & LPA C18:2 	& .395 &  &    LPA C18:2 	& .353 \\
PC(36:2)	 	& 8 &  &      TG(52:2)	 	& 533.69 &  & TG(52:2)	 	& .359 &  &    PC(36:2)		& .331 \\
TG(52:2)	 	& 7 &  &      Glycylglycine 	& 328.92 &  & PC(36:2)	 	& .345 &  &    LPA C20:4 	& .293 \\
PC(36:4)	 	& 7 &  &      Tyrosine 		& 249.87 &  & Glycylglycine 	& .336 &  &    PAF C16:0 	& .286 \\
PAF C16:0	 	& 7 &  &      TG(50:1) 		& 220.17 &  & PC(36:4)	 	& .331 &  &    LPA C18:1 	& .284 \\
Glycylglycine 		& 6 &  &      PC(36:4)	 	& 194.85 &  & PAF C16:0 	& .327 &  &    PC(36:4)		& .270 \\
PC(34:1)	 	& 6 &  &      PC(36:2)	 	& 180.92 &  & LPA C18:1 	& .325 &  &    PC(34:1) 	& .248 \\
LPA C16		 	& 6 &  &      L-Lactic acid 	& 178.38 &  & PC(34:1)	 	& .325 &  &    LPA C16 		& .218 \\
LPA C20:4	 	& 6 &  &      PC(34:1)	 	& 173.10 &  & Tyrosine 		& .322 &  &    LPA C22:6 	& .211 \\
\hline
\end{tabular}
\end{scriptsize}
\end{table}

\subsubsection{Communities}\label{subsubsec:RScommunityCHAR}
\begin{figure}[h]
\centering
  \includegraphics[width=\textwidth]{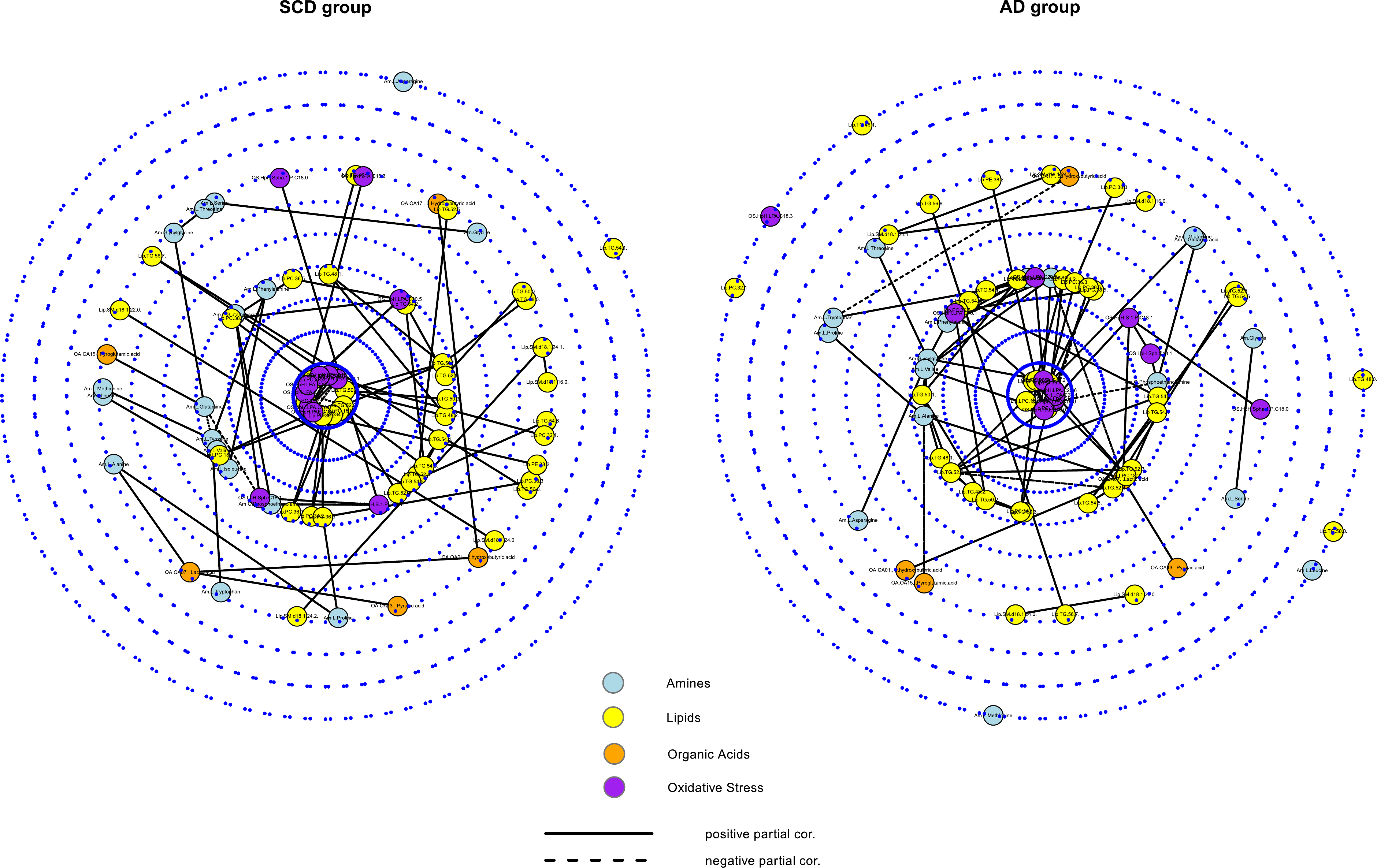}
    \caption{\footnotesize{Target plots \cite{TARGETplotsS} with the $k$-core decomposition of the SCD (left-hand panel) and AD (right-hand panel) networks.
    Note that the SCD and AD networks are the same as depicted in the panels of Figure \ref{FIG:CoordsSCAD}.
    The respective topologies are now however plotted to represent $k$-coreness.
    The features in the middle of the radial layouts then represent features in the graph-core while features that are plotted further from the center represent the peripheral features.
    The metabolite compounds are again colored according to metabolite family: Blue for amines, yellow for lipids, orange for organic acids, and purple for oxidative stress.
    Solid edges represent positive partial correlations while dashed edges represent negative partial correlations.}}
  \label{FIG:CorenessSCAD}
\end{figure}

The nodes in a network often cluster in groups: collections of nodes that are more deeply connected to each other than to nodes outside their topological environment.
There is thus interest in the detection of these groups.
We approach the question of node-grouping from two angles.
The first is the perspective of $k$-cores.
A $k$-core of a network is the maximal connected subnetwork in which all nodes have a degree of at least $k$ \cite{NewmanS}.
In this setting, the $k$-core decomposition of a topology is used as an indication of the core-periphery structure of a network.
Figure \ref{FIG:CorenessSCAD} contains the $k$-core decomposition of the SCD and AD topologies depicted in the radial layout of a target plot \cite{TARGETplotsS}.
For both panels the center represents the 3-core, the first ring of features around the center represents the 2-core, and the subsequent feature-rings represent the 1-core and 0-core, respectively.
The $k$-cores of the SCD and AD topologies are similar, although the AD topology places the amine Glycylglycine in the 2-core instead of the 1-core.

\begin{landscape}
\begin{figure}
\centering
  \includegraphics[scale = .4]{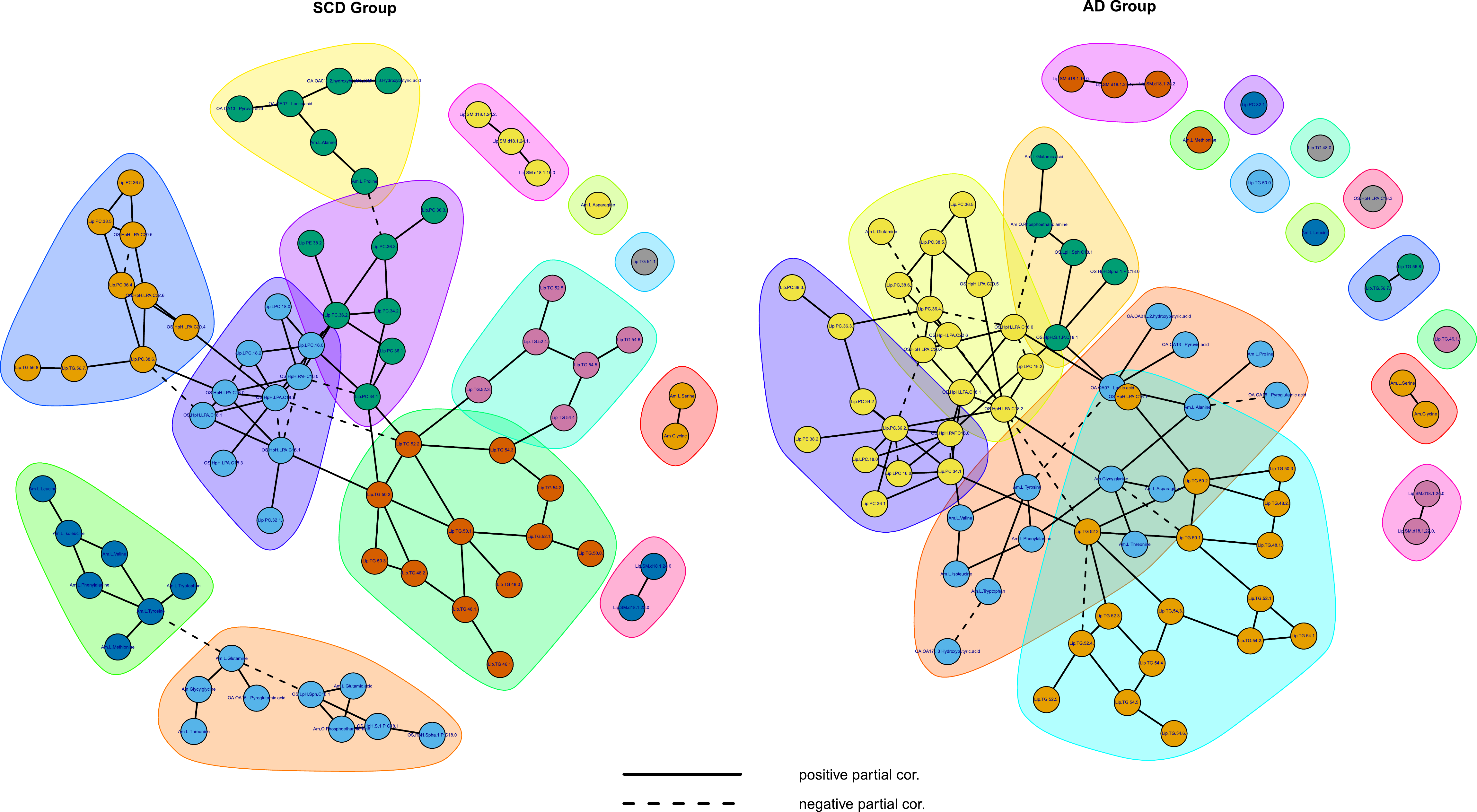}
    \caption{\footnotesize{
    Class-specific \emph{semi-pruned} networks visualized with the Fruchterman-Reingold algorithm and their community structure.
    The left-hand panel contains the network for the SCD group.
    The right-hand panel contains the network for the AD group.
    Solid edges represent positive partial correlations while dashed edges represent negative partial correlations.}}
  \label{FIG:ModuleSCAD}
\end{figure}
\end{landscape}

The second perspective on finding node-groupings is community detection.
Community detection, loosely speaking, refers to the ``search for naturally occurring groups in a network" \cite[p. 371]{NewmanS}.
A betweenness-based method of community detection is used, commonly known as the Girvan-Newman algorithm \cite{CommunityS}.
Figure \ref{FIG:ModuleSCAD} contains the same networks as Figure \ref{FIG:CoordsSCAD}, but now they are visualized to express the community structure.
The colored borders demarcate communities within the respective topologies.
Most notably, the SCD topology seems to have 2 loosely connected amine components while the AD topology seems to have a larger and more densely connected amine component that has ties to oxidative stress components.

\subsubsection{Differential Graphs}\label{subsubsec:RSdiffGraph}
\begin{landscape}
\begin{figure}
\centering
  \includegraphics[scale = .4]{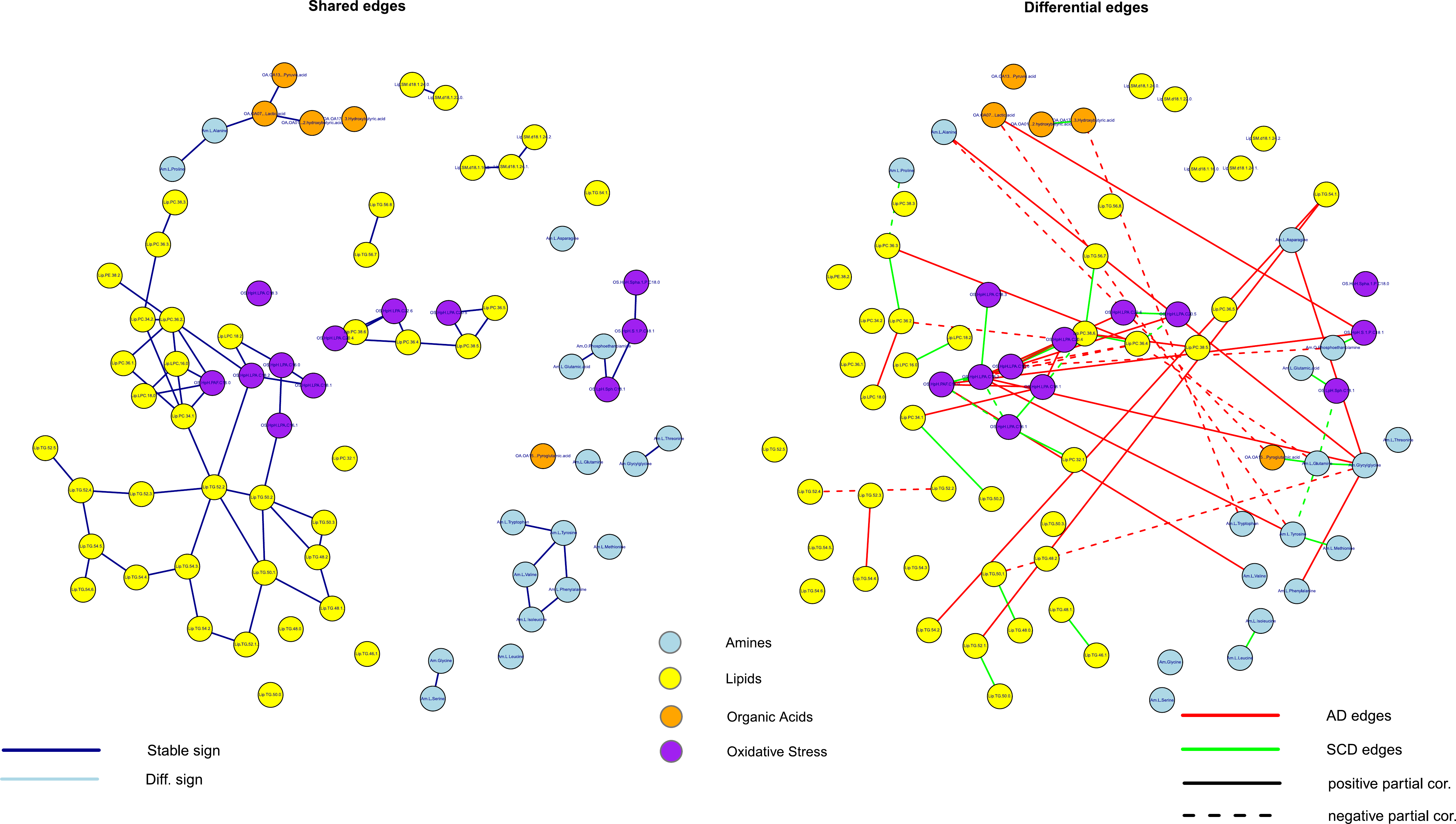}
    \caption{\footnotesize{Common and differential networks for the SCD versus AD class.
    The left-hand panel contains the network consisting of the edges (solid and colored blue) that are shared between the SCD and AD groups.
    The right-hand panel contains the network consisting of the edges that are unique for either the SCD or the AD group.
    Red edges represent connections that are present in the AD group only.
    Green edges represent connections that are present in the SCD group only.
    Solid edges represent positive partial correlations while dashed edges represent negative partial correlations.
    The metabolite compounds are colored according to metabolite family: Blue for amines, yellow for lipids, orange for organic acids, and purple for oxidative stress.
    Note that the nodes in these networks have coordinates concordant with the node-placing of Figure \ref{FIG:CoordsSCAD}.
    }}
  \label{FIG:DIFFgraphsSCAD}
\end{figure}
\end{landscape}

Section \ref{subsubsec:RSnodeCHAR} indicated that certain metabolites act as hubs in both the SCD and AD topologies.
Section \ref{subsubsec:RScommunityCHAR} then indicated that these hubs belong to different communities in the SCD and AD topologies.
Thus, while the SCD and AD topologies may contain the same hub-compounds, these compounds may be connected very differently, implying differential regulatory functioning in the respective networks.
Hence, we take interest in the networks of shared and differential connections over the SCD and AD topologies.
Figure \ref{FIG:DIFFgraphsSCAD} depicts in the left-hand panel the connections that are shared between the SCD and AD networks, and in the right-hand panel the connections that are unique to either the SCD or the AD network.
This figure is accompanied by Table \ref{TABLE:CentralDiffSCAD}, which contains the top degrees for the differential network (right-hand panel of Figure \ref{FIG:DIFFgraphsSCAD}).
These degrees indicate the compounds that are most differentially wired between the SCD and AD topologies.
We see that the regulatory functioning (in terms of connections) of the oxidative stress compounds LPA C18:2 and PAF C16:0 -- central to both the SCD and AD topologies -- is different across the SCD and AD topologies.
In addition, we see compounds that, although not central in either the SCD or AD topologies, are central in the differential network, such as the Phosphatidylcholine PC(38:6).
Moreover, we see compounds whose wiring seems to be unique to either the SCD or AD topologies.
For example, the amine Glutamine seems to be connected in the SCD topology mostly, while the amine Glycylglycine seems mostly unique to the AD topology, in which it connects with other amines and oxidative stress compounds.
Overall, the most differentially wired metabolites across the SCD and AD topologies belong predominantly to the Lyso-phosphatidic acid oxidative stress compounds, the Phosphatidylcholines, and the amine family.

\begin{table}[h]
\begin{footnotesize}
\centering
\caption{Most differentially wired metabolic features in the differential network for the SCD group versus the AD group.}
\label{TABLE:CentralDiffSCAD}
\begin{tabular}{llr}
\hline
\hline
Feature               	& Compound class &Degree     \\ \hline
LPA C18:2       	& Oxidative stress: Lyso-phosphatidic acid	&  10\\
Glycylglycine       	& Amines       					&   6\\
PC(36:4)           	& Lipids: Phosphatidylcholine 			&   6\\
PC(38:6)           	& Lipids: Phosphatidylcholine		       	&   6\\
LPA C16       		& Oxidative stress: Lyso-phosphatidic acid      &   6\\
PAF C16:0 		& Platelet activating factor 			&   6\\
Glutamine         	& Amines       					&   5\\
LPA C18:1       	& Oxidative stress: Lyso-phosphatidic acid      &   5\\
LPA C20:4       	& Oxidative stress: Lyso-phosphatidic acid      &   5\\
Tyrosine          	& Amines       					&   4\\
LPA C16:1       	& Oxidative stress: Lyso-phosphatidic acid      &   4\\
\hline
\end{tabular}
\end{footnotesize}
\end{table}

\subsection{Regulatory Signature: Including APOE $\epsilon$4 allele status}\label{subsec:RSapoe}
\subsubsection{Approach}\label{subsubsec:RSAPOEapproach}
In our graphical modeling approach we take into account that our data consist of distinct classes of interest.
In the preceding section we looked at the first natural distinction: AD versus SCD.
The second distinction is having no APOE $\epsilon$4 allele versus having at least 1 APOE $\epsilon$4 allele.
The APOE $\epsilon$4 indicator proved influential in the classification signature.
Moreover, having at least 1 APOE $\epsilon$4 allele is strongly associated with the AD disease label:
The Fisher exact test on Table \ref{Table:Classes} sorts a $p$-value of 1.553e-10, indicating that persons with AD are more likely to have at least 1 APOE $\epsilon$4 allele.
One may expect that AD with and AD without the APOE $\epsilon$4 allele represent two (somewhat) distinct disease processes.
Moreover, one could expect that for a portion of the SCD cases that have at least 1 APOE $\epsilon$4 allele, certain metabolic changes indicative of looming AD may already be present.
In this situation we thus have four classes or groups of interest (Table \ref{Table:Classes} gives the number of observations for each group).

\begin{table}[h]
\centering
\caption{Number of observations in the cross-tabulation of AD disease status and APOE $\epsilon$4 allele status.}
\label{Table:Classes}
\begin{tabular}{rrr}
\hline
\hline
                         & \multicolumn{2}{l}{At least 1 APOE $\epsilon$4 allele} \\
                         & No                & Yes               \\ \hline
\multicolumn{1}{l|}{AD}  & 40                & 87                \\
\multicolumn{1}{l|}{SCD} & 87                & 34                \\ \hline
\end{tabular}
\end{table}

The graphical modeling approach is analogous to the strategy described in Section \ref{subsubsec:RSapproach}, but now we have $g = 1,\ldots, 4$.
We solve (\ref{eq:argmax2}) using the class-specific sample covariance matrices (i.e., the sample covariance matrices of the class-specific data) as the data entries.
For the target $\mathbf{T}$ we again choose the (weakly informative) $p$-dimensional identity matrix $\mathbf{I}_p$.
The optimal penalty parameters were again determined by LOOCV \cite{FUSEDS}.
The optimal penalty values were found to be $\lambda^{*} = $ 10.02109, and $\lambda_f^\mathds{*} = $ 3.970277e-17.
These penalty values again emphasize individual regularization over retainment of entry-wise similarities, indicating strong differences in class-specific precision matrices.
For each class-specific matrix, the 100 strongest edges (in terms of absolute partial correlations) were retained.
The retained partial correlations range, in absolute value, from .1376073 to .5791377 over the respective classes.
All analyzes were again performed with the \texttt{rags2ridges} package \cite{RAGSS} in \texttt{R} \cite{RS}.

\subsubsection{Results}\label{subsubsec:RSAPOEresults}
Here, we state all results for the the group-specific networks stemming from the cross-tabulation of AD disease status APOE $\epsilon$4 allele status.
For detail on technical terms, see Section \ref{subsec:RS}.

\begin{figure}[h]
\centering
  \includegraphics[width=\textwidth]{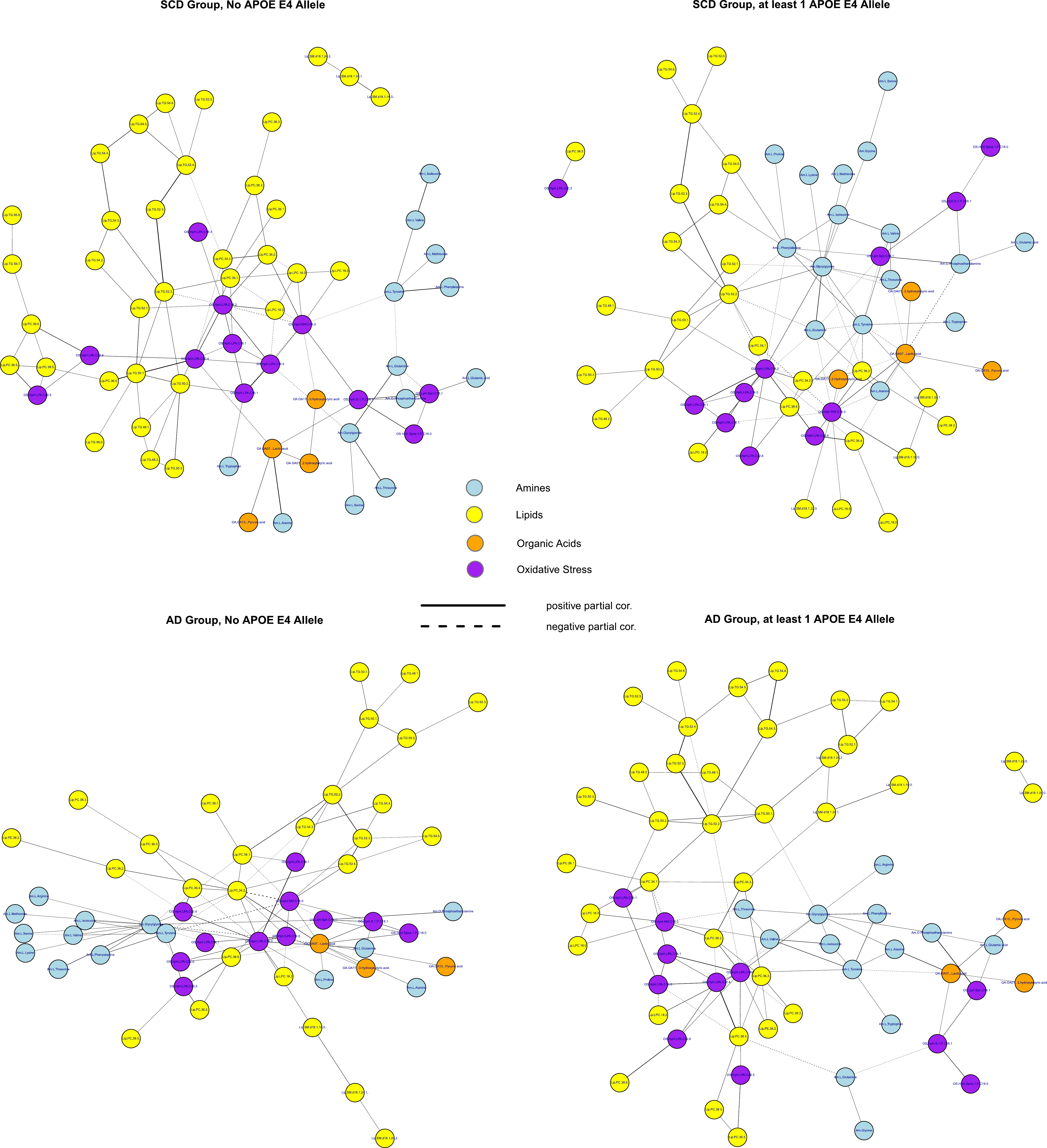}
    \caption{\footnotesize{Class-specific \emph{pruned} networks visualized with the Fruchterman-Reingold algorithm.
    The upper-left panel contains the network for the SCD group with no APOE $\epsilon$4 allele.
    The upper-right panel contains the network for the SCD group with at least 1 APOE $\epsilon$4 allele.
    The lower-left panel represents the network for the AD group with no APOE $\epsilon$4 allele.
    The lower-right panel represents the network for the AD group with at least 1 APOE $\epsilon$4 allele.
    The metabolite compounds are colored according to metabolite family: Blue for amines, yellow for lipids, orange for organic acids, and purple for oxidative stress.
    Solid edges represent positive partial correlations while dashed edges represent negative partial correlations.
    Appears as Figure 3 in the main text.}}
  \label{FIG:NWKSPrune}
\end{figure}

\begin{landscape}
\begin{figure}
\centering
  \includegraphics[scale = .4]{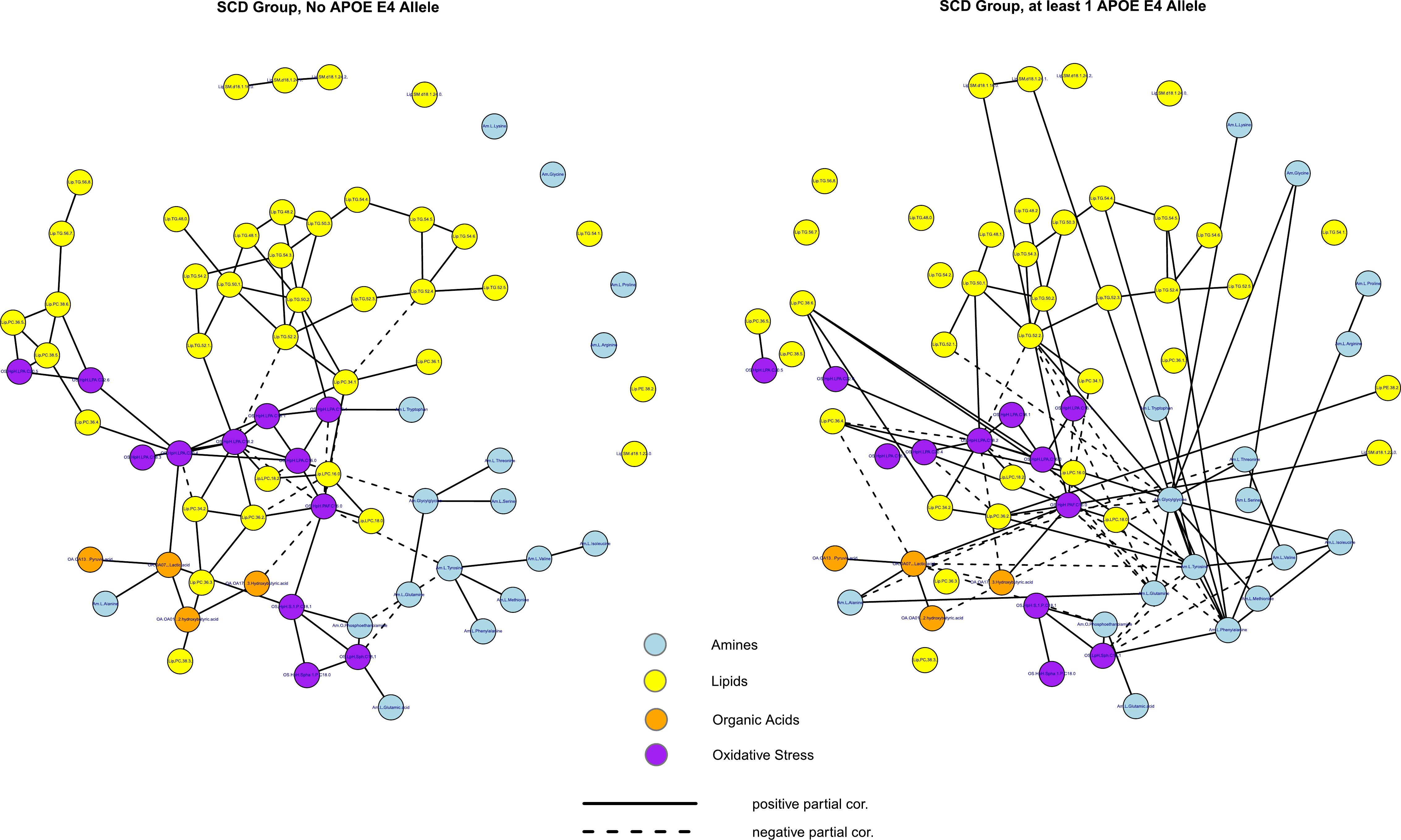}
    \caption{\footnotesize{Class-specific \emph{semi-pruned} networks visualized with the Fruchterman-Reingold algorithm.
    The left-hand panel contains the network for the SCD group with no APOE $\epsilon$4 allele.
    The right-hand panel contains the network for the SCD group with at least 1 APOE $\epsilon$4 allele.
    The coordinates of the left-hand topology serve as the reference coordinates.
    The metabolite compounds are colored according to metabolite family: Blue for amines, yellow for lipids, orange for organic acids, and purple for oxidative stress.
    Solid edges represent positive partial correlations while dashed edges represent negative partial correlations.}}
  \label{FIG:Coords1}
\end{figure}
\end{landscape}

\begin{landscape}
\begin{figure}
\centering
  \includegraphics[scale = .4]{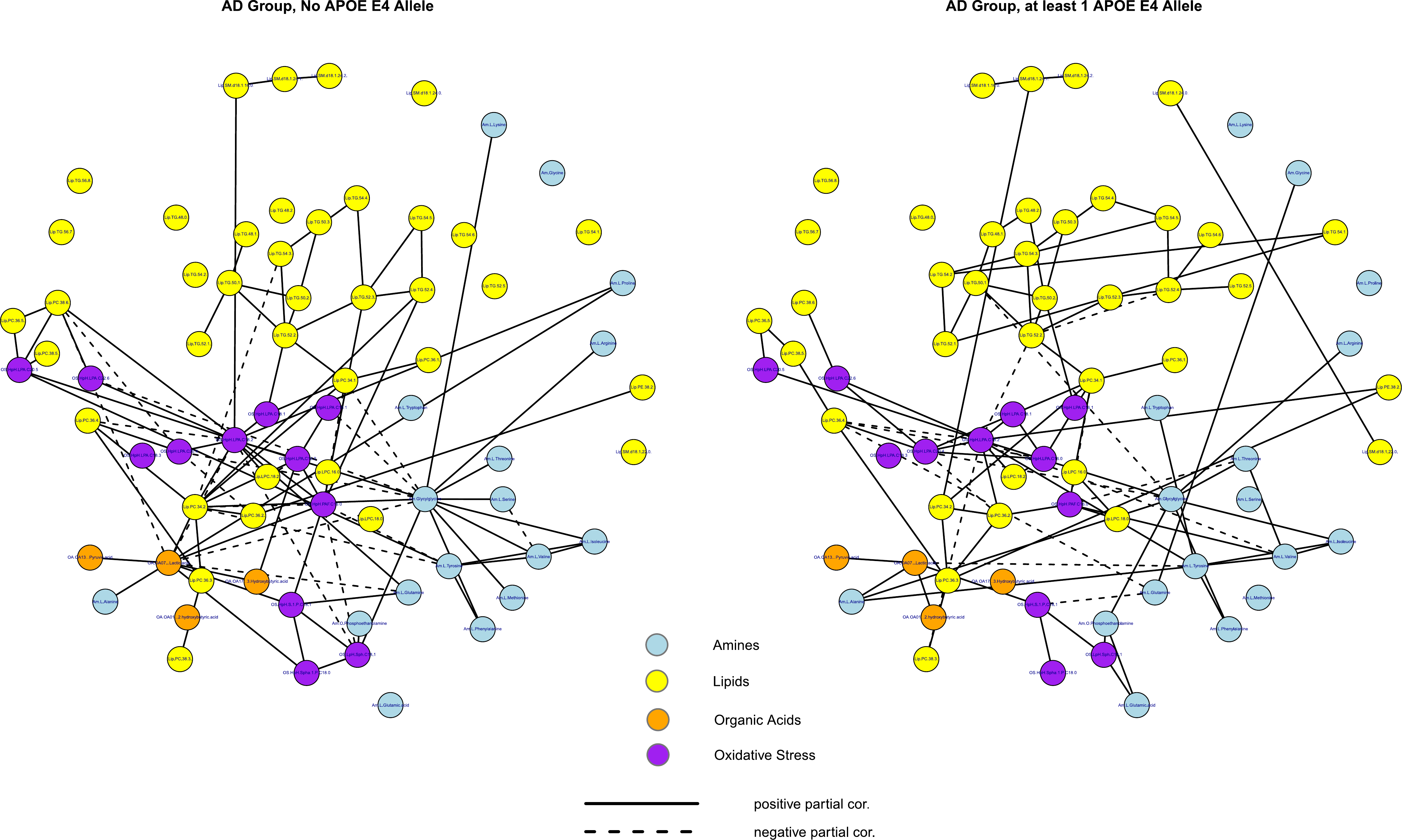}
    \caption{\footnotesize{Class-specific \emph{semi-pruned} networks visualized with the Fruchterman-Reingold algorithm.
    The left-hand panel contains the network for the AD group with no APOE $\epsilon$4 allele.
    The right-hand panel contains the network for the AD group with at least 1 APOE $\epsilon$4 allele.
    The coordinates of the left-hand topology in Figure \ref{FIG:Coords1} serve as the reference coordinates.
    The metabolite compounds are colored according to metabolite family: Blue for amines, yellow for lipids, orange for organic acids, and purple for oxidative stress.
    Solid edges represent positive partial correlations while dashed edges represent negative partial correlations.}}
  \label{FIG:Coords2}
\end{figure}
\end{landscape}

\begin{figure}[h]
\centering
  \includegraphics[width=\textwidth]{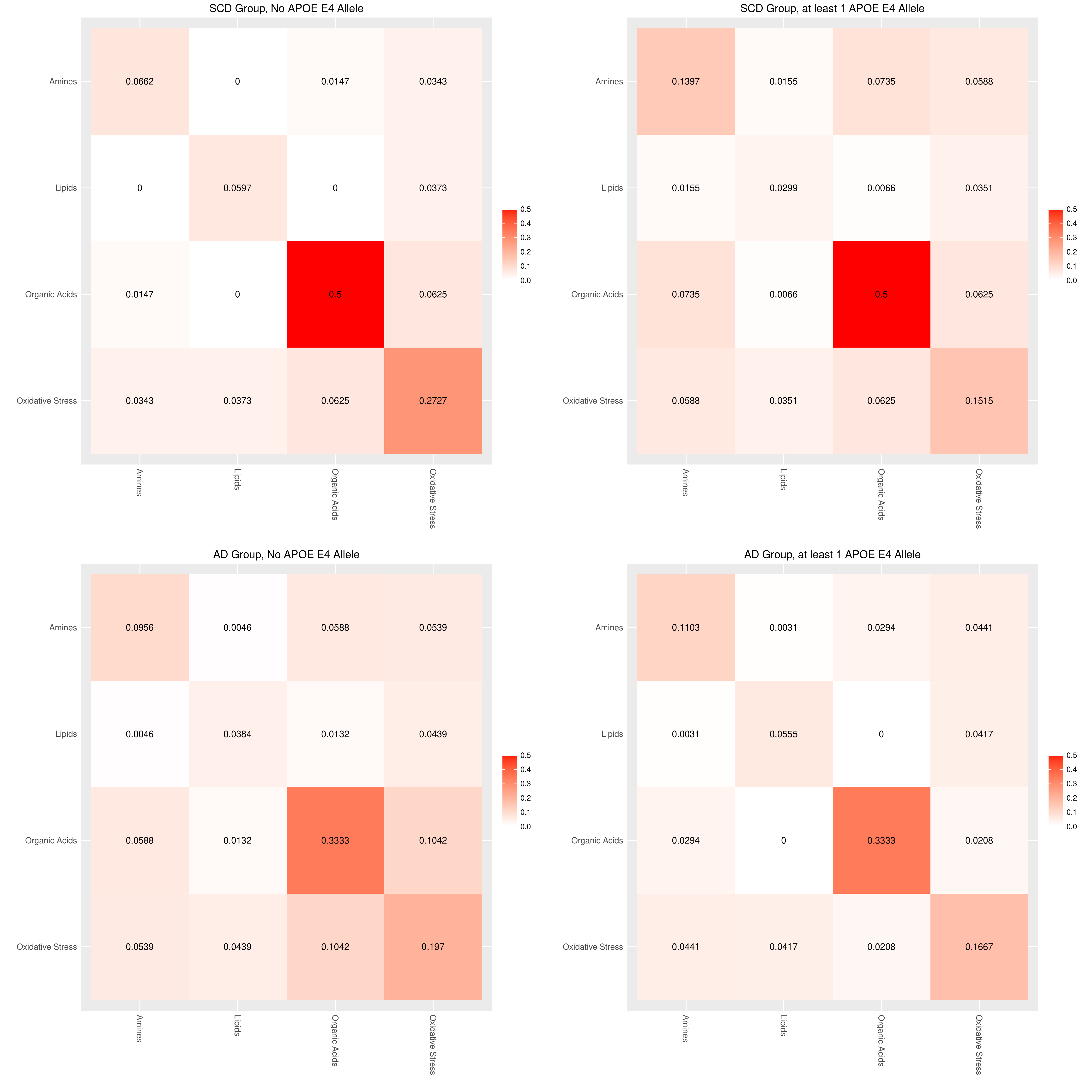}
    \caption{\footnotesize{Heatmaps of degree densities for the class-specific networks.
    The reported numbers represent the degree density for the (combinations of) metabolite groups.
    Degree density represents the number of connections (edges) divided by the number of possible connections.
    For example, in the network for the SCD group with no APOE $\epsilon$4 allele the proportion of actual edges relative to the number of possible edges between Amines and oxidative stress compounds is .0343.
    Note that all heatmaps received the same color key.
    Hence, the color intensities (i.e., the color-spectrum representations of the cell-numbers) are comparable over the respective heatmaps.}}
  \label{FIG:DegreeDensity}
\end{figure}

\begin{figure}[h]
\centering
  \includegraphics[width=\textwidth]{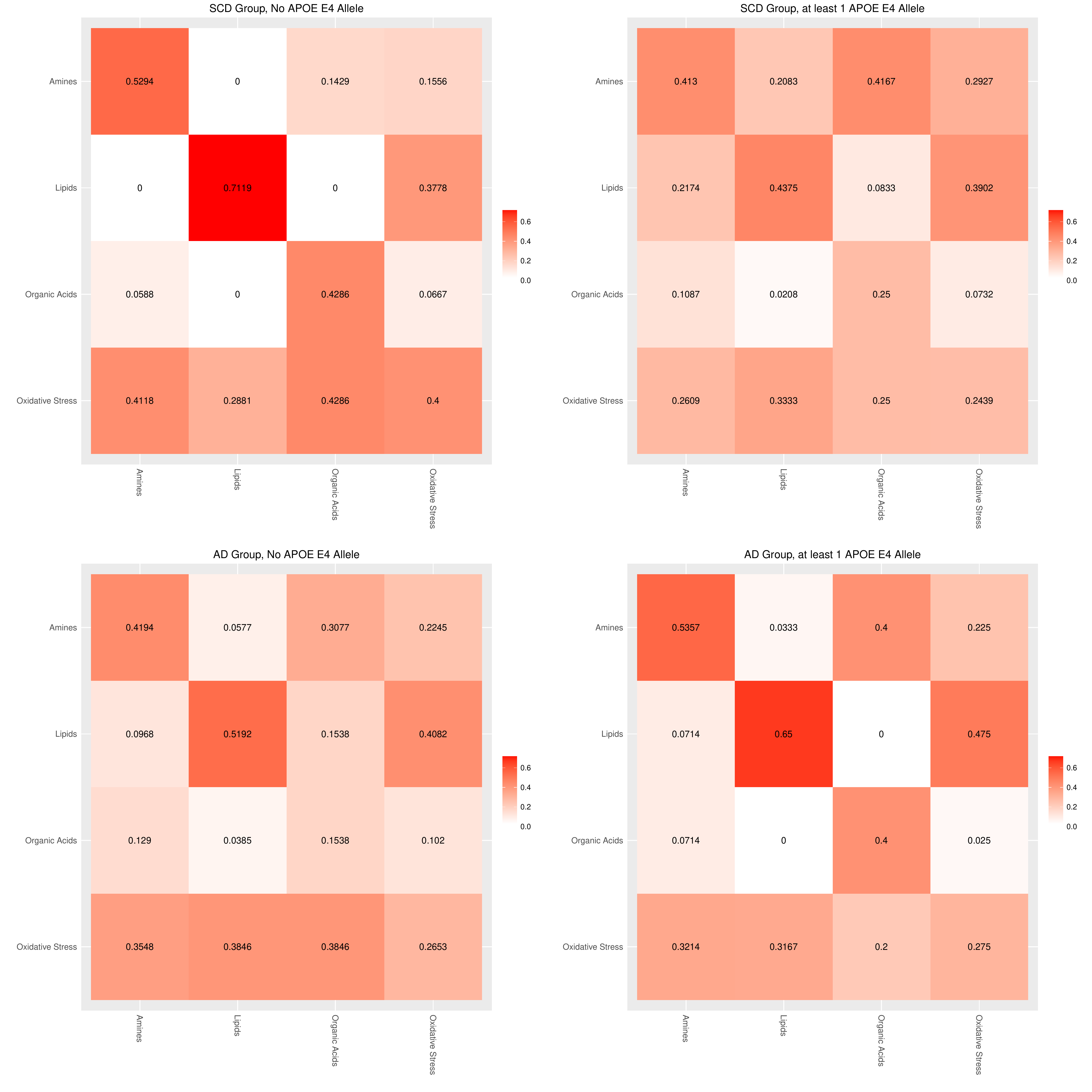}
    \caption{\footnotesize{Heatmaps of relative outdegrees for the class-specific networks.
    The reported numbers represent the relative outdegree for the (combinations of) metabolite groups.
    The relative outdegree represents the number of connections (edges) between two metabolite groups divided by the number of `outgoing' connections for one of these groups.
    For example, in the network for the SCD group with no APOE $\epsilon$4 allele the number of edges between lipid and oxidative stress compounds accounts for approximately 29\% of all edges involving lipids and approximately 38\% of all edges involving oxidative stress compounds.
    Note that all heatmaps received the same color key.
    Hence, the color intensities (i.e., the color-spectrum representations of the cell-numbers) are comparable over the respective heatmaps.
    Note that the column numbers sum to unity.}}
  \label{FIG:ROUTDegreeDensity}
\end{figure}

Figure \ref{FIG:NWKSPrune} contains the class-specific pruned networks visualized with the FR algorithm \cite{FRUCHTS}.
Figures \ref{FIG:Coords1} and \ref{FIG:Coords2} then contain the class-specific networks over the union of retained metabolites in which the FR-based coordinates of the network of the SCD group with no APOE $\epsilon$4 allele serve as reference coordinates.
As stated, the FR algorithm prefers coiled structures.
From this perspective the network for the SCD group with no APOE $\epsilon$4 allele ($\mathrm{SCD}^{\neg\epsilon4}$) and the network for the AD group with at least 1 APOE $\epsilon$4 allele ($\mathrm{AD}^{\epsilon4}$) seem the most structured ones.
Perhaps this is natural for the former group, given that these persons are, in some sense, the least at risk for developing AD, and thus should represent the normal biochemical state.
This might also be natural for the latter group, as this network may represent structured APOE $\epsilon$4-driven changes in metabolism.
The networks for the SCD group with at least 1 APOE $\epsilon$4 allele ($\mathrm{SCD}^{\epsilon4}$) and the AD group with no APOE $\epsilon$4 allele ($\mathrm{AD}^{\neg\epsilon4}$) appear more random, i.e., less modular or locally connected.
As stated, one could expect that for a portion of the $\mathrm{SCD}^{\epsilon4}$ group, certain metabolic changes indicative of looming AD may already be present.
Also, the $\mathrm{AD}^{\neg\epsilon4}$ group may represent various alternative AD disease processes.
In short, both these latter groups are likely heterogeneous.
Table \ref{Table:TRANS} then contains some global characteristics of each class-related graph as given in Figures \ref{FIG:Coords1} and \ref{FIG:Coords2}.
These metrics corroborate to some degree the assessment made above: The topology for the $\mathrm{SCD}^{\neg\epsilon4}$ group is most strongly locally connected, and the topologies for the $\mathrm{SCD}^{\neg\epsilon4}$ group and the $\mathrm{AD}^{\epsilon4}$ group are most cohesive (overall).
Figures \ref{FIG:DegreeDensity} and \ref{FIG:ROUTDegreeDensity} also indicate that the topologies for the $\mathrm{SCD}^{\epsilon4}$ group and the $\mathrm{AD}^{\neg\epsilon4}$ group are more diffuse.
Moreover, they indicate that the topology for the $\mathrm{AD}^{\epsilon4}$ group can be characterized (vis-\`{a}-vis the $\mathrm{SCD}^{\neg\epsilon4}$ topology) by increased connection density within the amine compounds and between the amine and oxidative stress compounds.

\begin{table}[h]
\begin{footnotesize}
\centering
\caption{Global metrics for the four topologies of interest.}
\label{Table:TRANS}
\begin{tabular}{lrrr}
\hline
\hline
Topology    & Transitivity & Connectedness & Centralization\\ \hline
SCD group with no APOE $\epsilon$4 allele               & .2136986    &  .7134809 & .1055901 \\
SCD group with at least 1 APOE $\epsilon$4 allele       & .142315     &  .6201207 & .1496894 \\
AD group with no APOE $\epsilon$4 allele                & .1758621    &  .5758551 & .1937888 \\
AD group with at least 1 APOE $\epsilon$4 allele        & .1480519    &  .7126761 & .1496894 \\ \hline
\end{tabular}
\end{footnotesize}
\end{table}

\begin{figure}[h]
\centering
  \includegraphics[width=.97\textwidth]{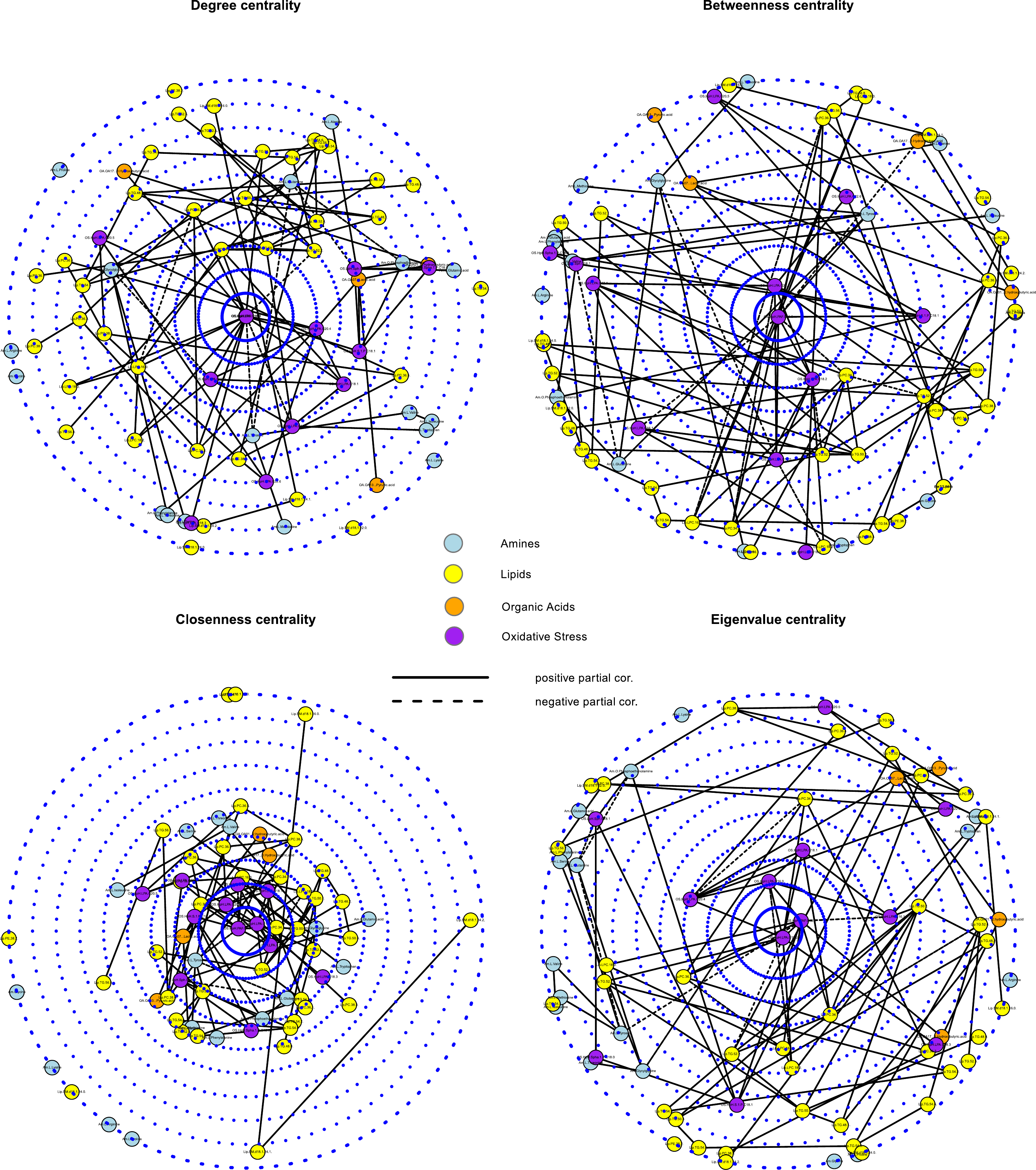}
    \caption{\footnotesize{Target plots visualizing various centralities for the network representing the SCD group with no APOE $\epsilon$4 allele.
    The upper-left panel represents degree centralities.
    The upper-right panel represents betweenness centralities.
    The lower-left panel represents closeness centralities.
    The lower-right panel represents eigenvalue centralities.
    Note that, for each target plot, the network is the same as in the left-hand panel of \ref{FIG:Coords1}.
    The topology is now however plotted to represent metabolite features according to various centrality scores.
    For example, the oxidative stress compounds LPA.C18.2 and PAF.C16.0 have the highest degree centrality and, hence, are depicted in the center of the upper-left panel.
    The metabolite compounds are again colored according to metabolite family: Blue for amines, yellow for lipids, orange for organic acids, and purple for oxidative stress.
    Solid edges represent positive partial correlations while dashed edges represent negative partial correlations.
    The metabolite features attaining the highest centrality scores are given in Table \ref{TABLE:CentralSCnoAPOE}.}}
  \label{FIG:CentralSCnoAPOE}
\end{figure}

\begin{figure}[h]
\centering
  \includegraphics[width=.97\textwidth]{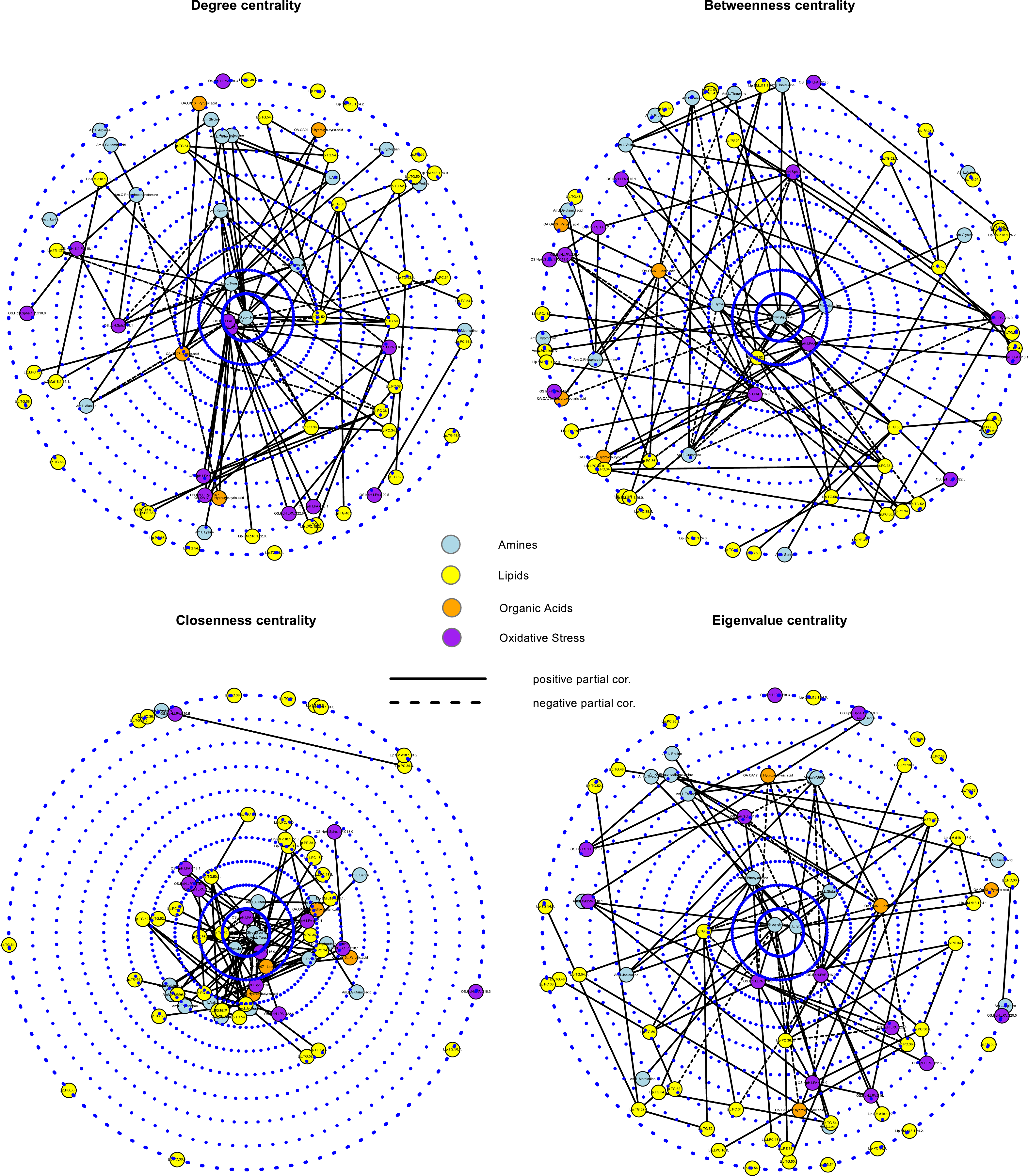}
    \caption{\footnotesize{Target plots visualizing various centralities for the network representing the SCD group with at least 1 APOE $\epsilon$4 allele.
    The upper-left panel represents degree centralities.
    The upper-right panel represents betweenness centralities.
    The lower-left panel represents closeness centralities.
    The lower-right panel represents eigenvalue centralities.
    Note that, for each target plot, the network is the same as in the right-hand panel of \ref{FIG:Coords1}.
    The topology is now however plotted to represent metabolite features according to various centrality scores.
    For example, the Amine Glycylglycine has the highest degree centrality and, hence, it is depicted in the center of the upper-left panel.
    The metabolite compounds are again colored according to metabolite family: Blue for amines, yellow for lipids, orange for organic acids, and purple for oxidative stress.
    Solid edges represent positive partial correlations while dashed edges represent negative partial correlations.
    The metabolite features attaining the highest centrality scores are given in Table \ref{TABLE:CentralSCAPOE}.}}
  \label{FIG:CentralSCAPOE}
\end{figure}

\begin{figure}[h]
\centering
  \includegraphics[width=.97\textwidth]{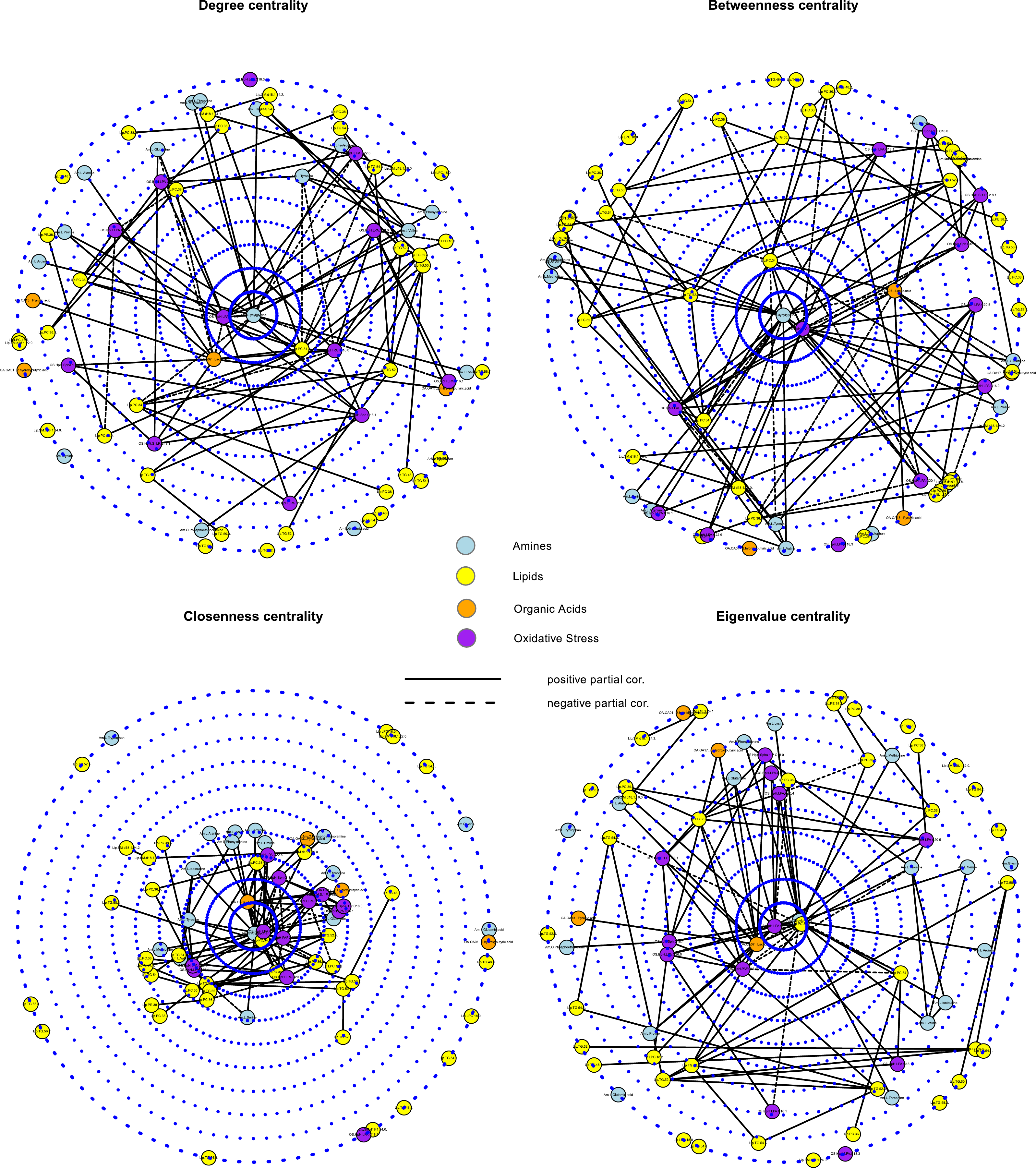}
    \caption{\footnotesize{Target plots visualizing various centralities for the network representing the AD group with no APOE $\epsilon$4 allele.
    The upper-left panel represents degree centralities.
    The upper-right panel represents betweenness centralities.
    The lower-left panel represents closeness centralities.
    The lower-right panel represents eigenvalue centralities.
    Note that, for each target plot, the network is the same as in the left-hand panel of \ref{FIG:Coords2}.
    The topology is now however plotted to represent metabolite features according to various centrality scores.
    For example, the Amine Glycylglycine has the highest degree centrality and, hence, it is depicted in the center of the upper-left panel.
    The metabolite compounds are again colored according to metabolite family: Blue for amines, yellow for lipids, orange for organic acids, and purple for oxidative stress.
    Solid edges represent positive partial correlations while dashed edges represent negative partial correlations.
    The metabolite features attaining the highest centrality scores are given in Table \ref{TABLE:CentralADnoAPOE}.}}
  \label{FIG:CentralADnoAPOE}
\end{figure}

\begin{figure}[h]
\centering
  \includegraphics[width=.97\textwidth]{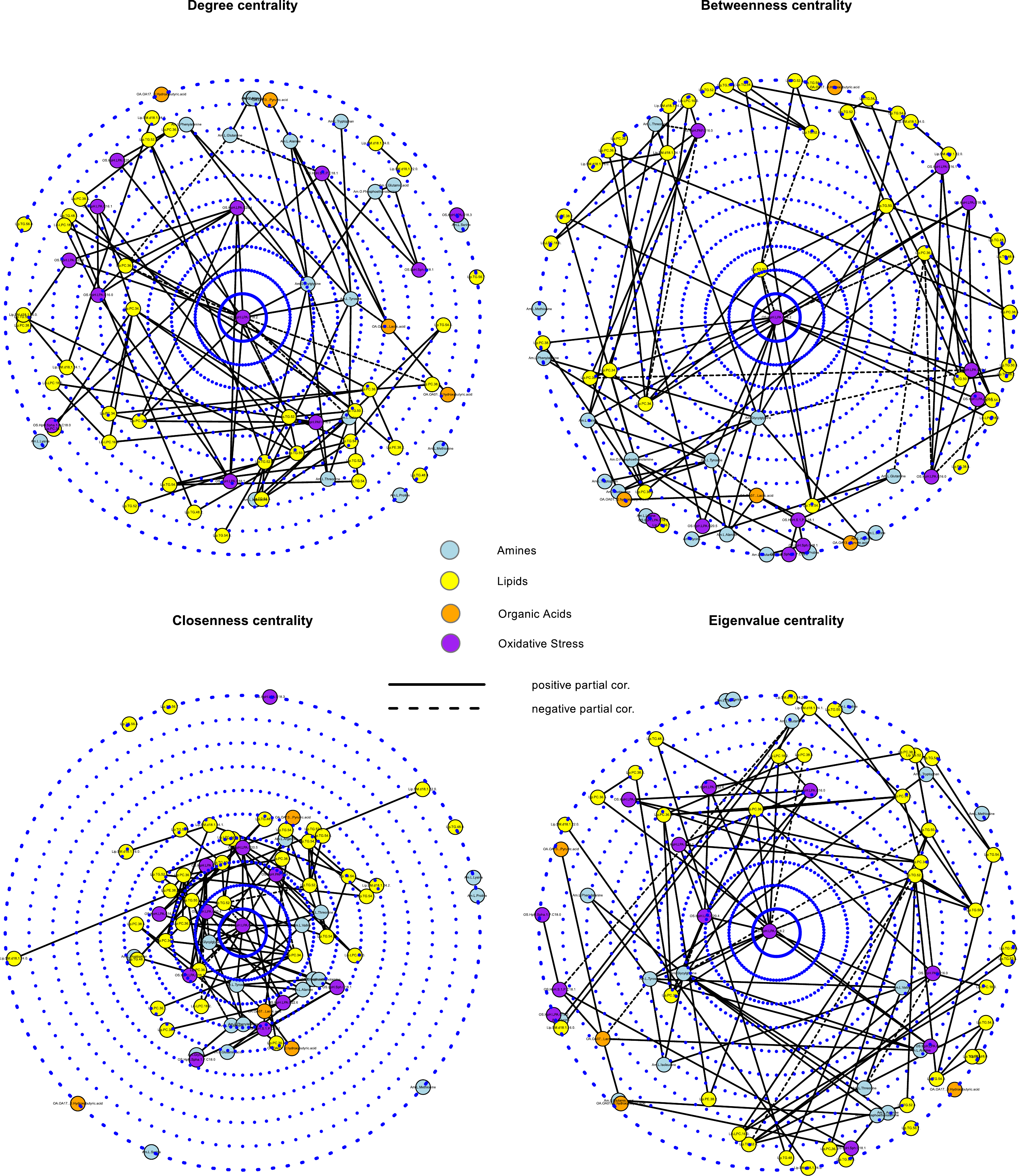}
    \caption{\footnotesize{Target plots visualizing various centralities for the network representing the AD group at least 1 APOE $\epsilon$4 allele.
    The upper-left panel represents degree centralities.
    The upper-right panel represents betweenness centralities.
    The lower-left panel represents closeness centralities.
    The lower-right panel represents eigenvalue centralities.
    Note that, for each target plot, the network is the same as in the right-hand panel of \ref{FIG:Coords2}.
    The topology is now however plotted to represent metabolite features according to various centrality scores.
    For example, the oxidative stress compound LPA.C18.2 has the highest degree centrality and, hence, it is depicted in the center of the upper-left panel.
    The metabolite compounds are again colored according to metabolite family: Blue for amines, yellow for lipids, orange for organic acids, and purple for oxidative stress.
    Solid edges represent positive partial correlations while dashed edges represent negative partial correlations.
    The metabolite features attaining the highest centrality scores are given in Table \ref{TABLE:CentralADAPOE}.}}
  \label{FIG:CentralADAPOE}
\end{figure}

Figures \ref{FIG:CentralSCnoAPOE}, \ref{FIG:CentralSCAPOE}, \ref{FIG:CentralADnoAPOE}, and \ref{FIG:CentralADAPOE} contain target plots depicting the various centrality properties of the metabolite compounds retained in the $\mathrm{SCD}^{\neg\epsilon4}$, $\mathrm{SCD}^{\epsilon4}$, $\mathrm{AD}^{\neg\epsilon4}$, and $\mathrm{AD}^{\epsilon4}$ topologies, respectively.
Tables \ref{TABLE:CentralSCnoAPOE}, \ref{TABLE:CentralSCAPOE}, \ref{TABLE:CentralADnoAPOE}, and \ref{TABLE:CentralADAPOE} then contain the top centrality metrics for these topologies.
The top compounds, in terms of centrality, in the $\mathrm{SCD}^{\neg\epsilon4}$ topology, all belong to either the oxidative stress or lipid family.
The $\mathrm{SCD}^{\epsilon4}$ and $\mathrm{AD}^{\neg\epsilon4}$ topologies are indeed, as also indicated in Table \ref{Table:TRANS}, more centralized, having more compounds with a high degree.
The centrality picture for these topologies is more diffuse in terms of compound-families (i.e., all compound families are represented).
The top compounds in the $\mathrm{AD}^{\epsilon4}$ topology largely overlap with the to compounds of the $\mathrm{SCD}^{\neg\epsilon4}$ topology.
However, $\mathrm{AD}^{\epsilon4}$ topology the amines Glycylglycine and Tyrosine are consistently indicated as central compounds by all centrality measures.
Interestingly, these amines are also consistently marked as central in the $\mathrm{SCD}^{\epsilon4}$ topology.

\begin{figure}[h]
\centering
  \includegraphics[width=\textwidth]{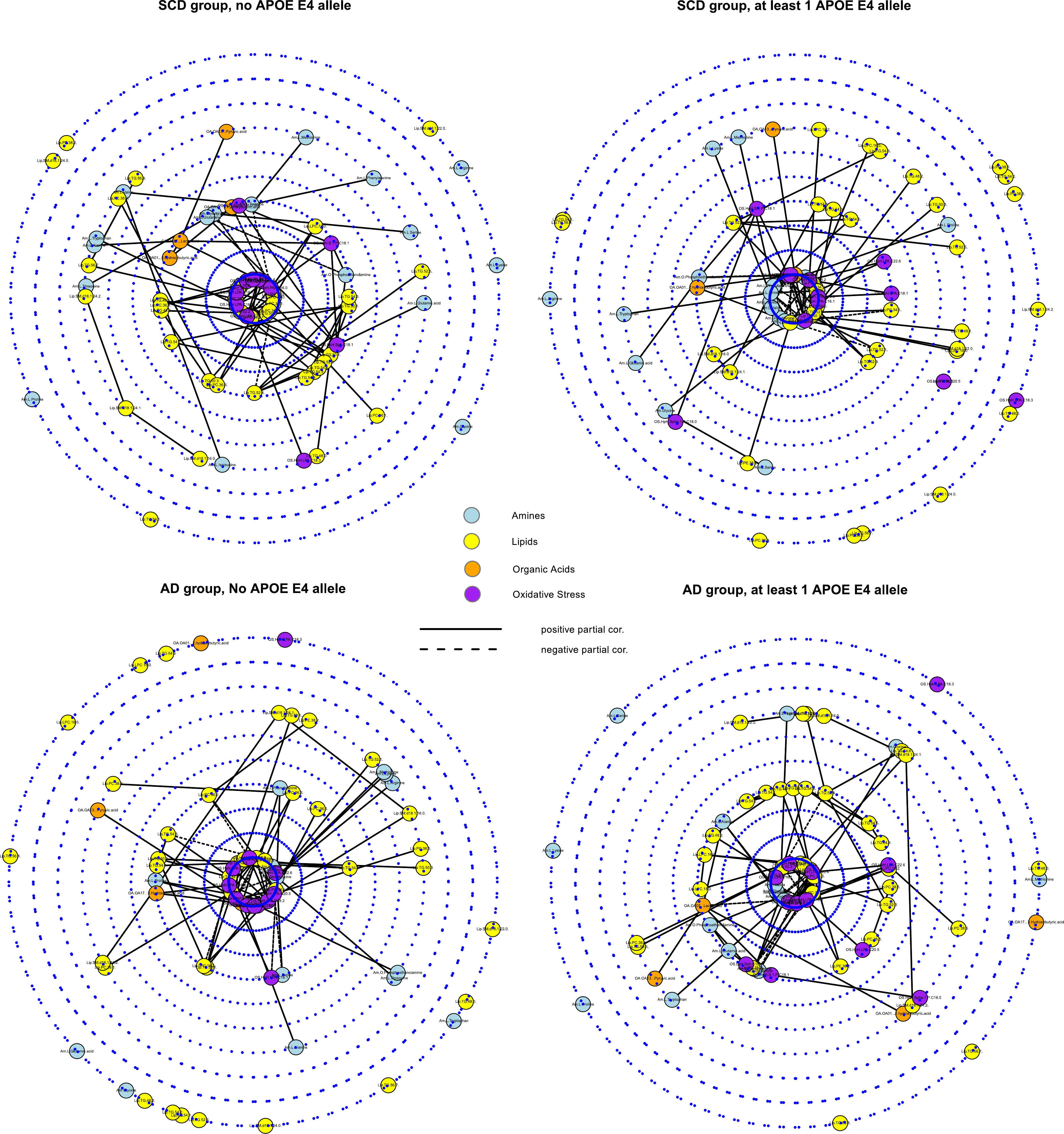}
    \caption{\footnotesize{Target plots \cite{TARGETplotsS} depicting $k$-core decompositions.
    The upper-left panel contains the network for the SCD group with no APOE $\epsilon$4 allele.
    The upper-right panel contains the network for the SCD group with at least 1 APOE $\epsilon$4 allele.
    The lower-left panel represents the network for the AD group with no APOE $\epsilon$4 allele.
    The lower-right panel represents the network for the AD group with at least 1 APOE $\epsilon$4 allele.
    Note that the respective topologies are now plotted to represent coreness.
    The features in the middle of the radial layouts then represent features in the graph-core while features that are plotted further from the center then represent the peripheral features.
    The metabolite compounds are again colored according to metabolite family: Blue for amines, yellow for lipids, orange for organic acids, and purple for oxidative stress.
    Solid edges represent positive partial correlations while dashed edges represent negative partial correlations.
    }}
  \label{FIG:Coreness}
\end{figure}

\begin{landscape}
\begin{figure}
\centering
  \includegraphics[scale = .4]{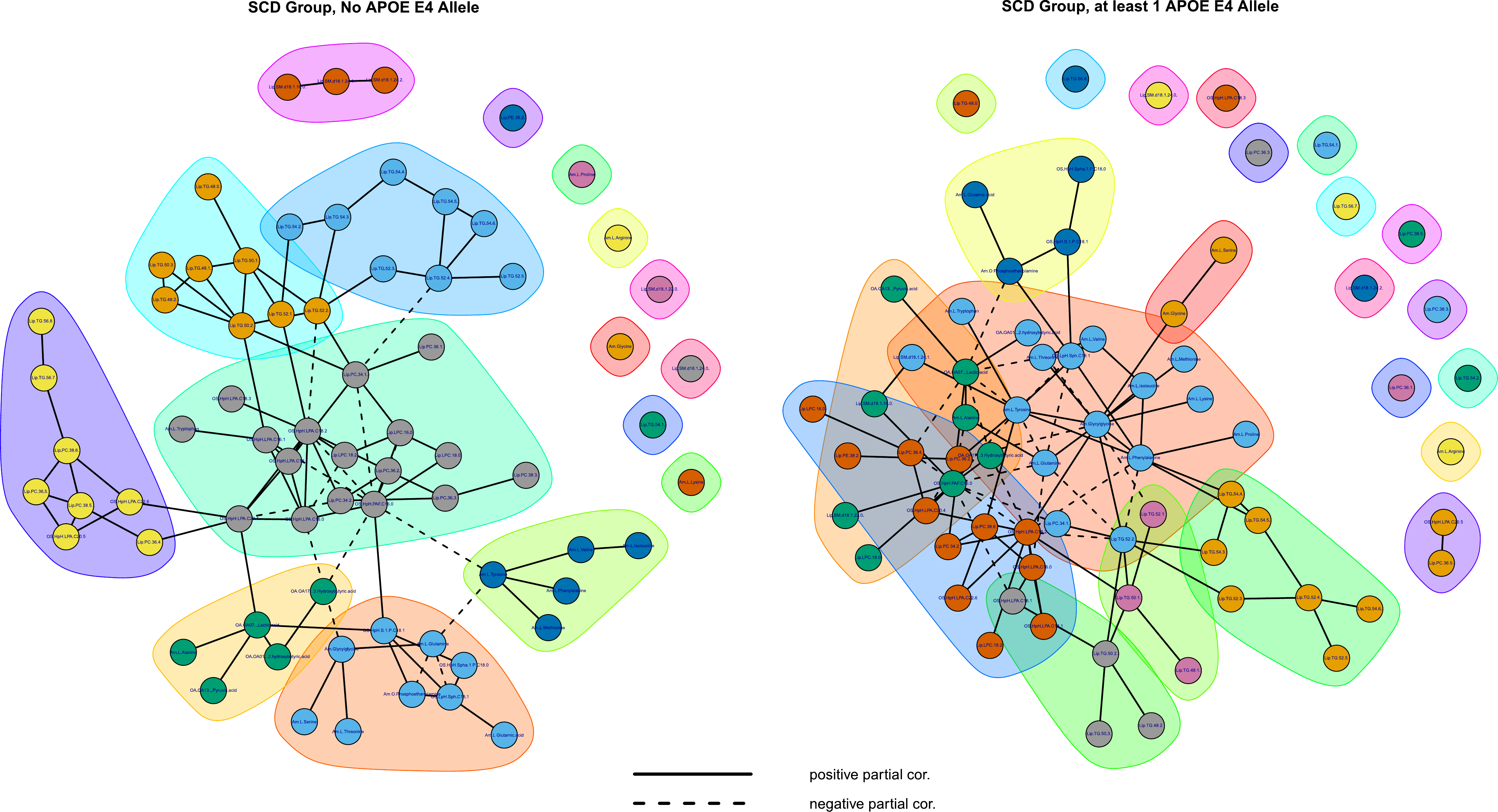}
    \caption{\footnotesize{Class-specific \emph{semi-pruned} networks visualized with their community structure.
    The left-hand panel contains the network for the SCD group with no APOE $\epsilon$4 allele.
    The right-hand panel contains the network for the SCD group with at least 1 APOE $\epsilon$4 allele.
    Solid edges represent positive partial correlations while dashed edges represent negative partial correlations.}}
  \label{FIG:Module1}
\end{figure}
\end{landscape}

\begin{landscape}
\begin{figure}
\centering
  \includegraphics[scale = .4]{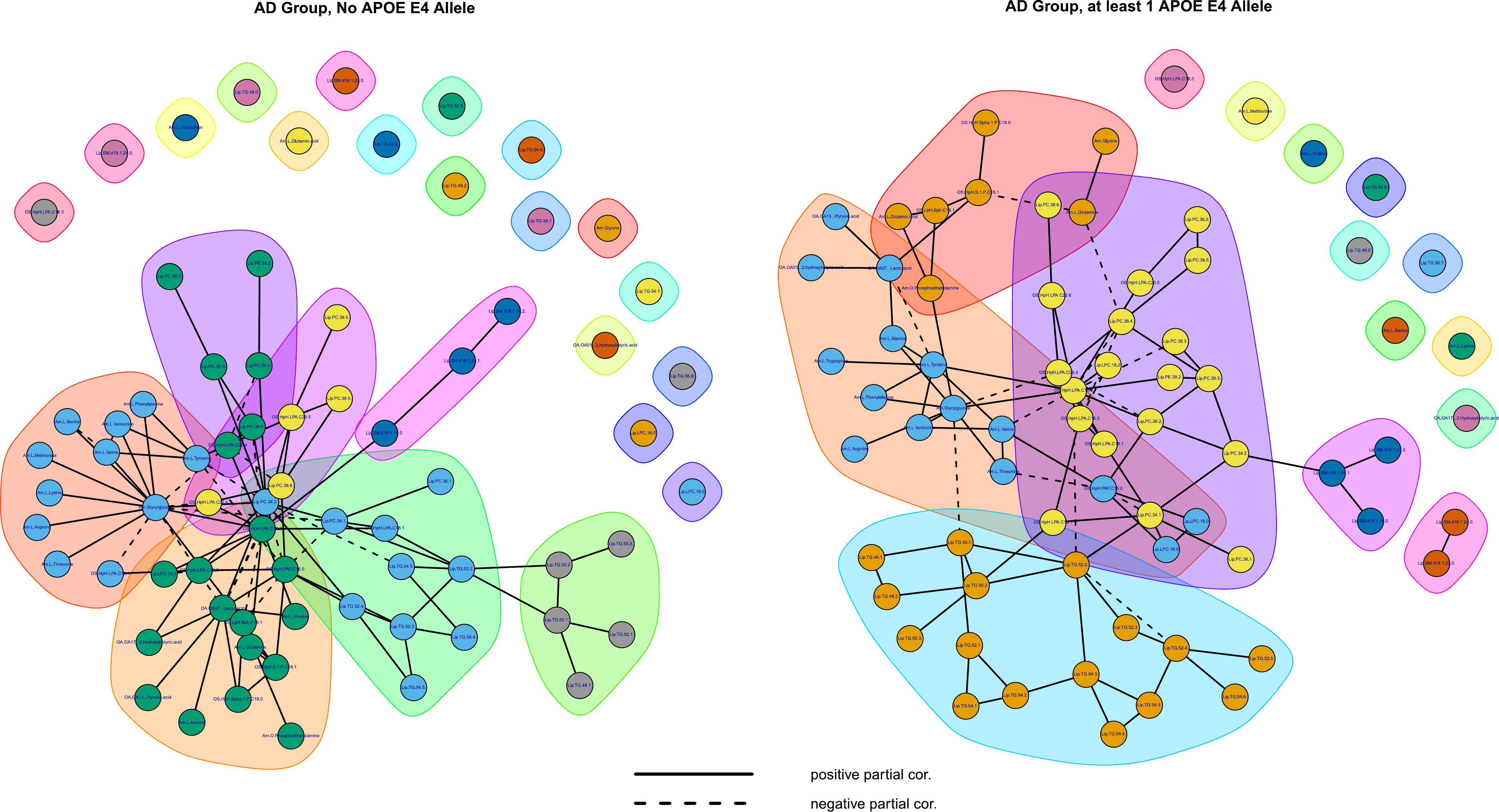}
    \caption{\footnotesize{Class-specific \emph{semi-pruned} networks visualized with their community structure.
    The left-hand panel contains the network for the AD group with no APOE $\epsilon$4 allele.
    The right-hand panel contains the network for the AD group with at least 1 APOE $\epsilon$4 allele.
    Solid edges represent positive partial correlations while dashed edges represent negative partial correlations.}}
  \label{FIG:Module2}
\end{figure}
\end{landscape}

Figure \ref{FIG:Coreness} depicts the $k$-core decomposition of the topologies of interest.
The inner-core of the $\mathrm{SCD}^{\neg\epsilon4}$ topology consists exclusively of lipid and oxidative stress compounds.
The inner-core of the $\mathrm{AD}^{\epsilon4}$ topology includes the amines Glycylglycine, Tyrosine, Isoleucine, Threonine, and Valine.
The inner-cores of the $\mathrm{SCD}^{\epsilon4}$ and $\mathrm{AD}^{\neg\epsilon4}$ topologies include metabolites from all compound-families and are thus less compound-family centered.
Figure \ref{FIG:Module1} contains the $\mathrm{SCD}^{\neg\epsilon4}$ and $\mathrm{SCD}^{\epsilon4}$ topologies visualized with their community structure.
Figure \ref{FIG:Module2} then contains the $\mathrm{AD}^{\neg\epsilon4}$ and $\mathrm{AD}^{\epsilon4}$ topologies visualized with their community structure.
The colored borders demarcate communities within the respective topologies.
The $\mathrm{SCD}^{\neg\epsilon4}$ and $\mathrm{AD}^{\epsilon4}$ topologies are indeed the most modular ones.
The $\mathrm{SCD}^{\neg\epsilon4}$ has clear organic acid, triglyceride, Phosphatidylcholine, and amine communities.
The Lyso-phosphatidic acids (oxidative stress compounds) largely form a community with the Phosphatidylcholines.
It contains 2 loosely connected amine communities.
The $\mathrm{AD}^{\epsilon4}$ topology, on the other hand, seems to have larger and more densely connected amine communities that include organic acid compounds and that have ties to oxidative stress
compounds.

\begin{landscape}
\begin{figure}
\centering
  \includegraphics[scale = .4]{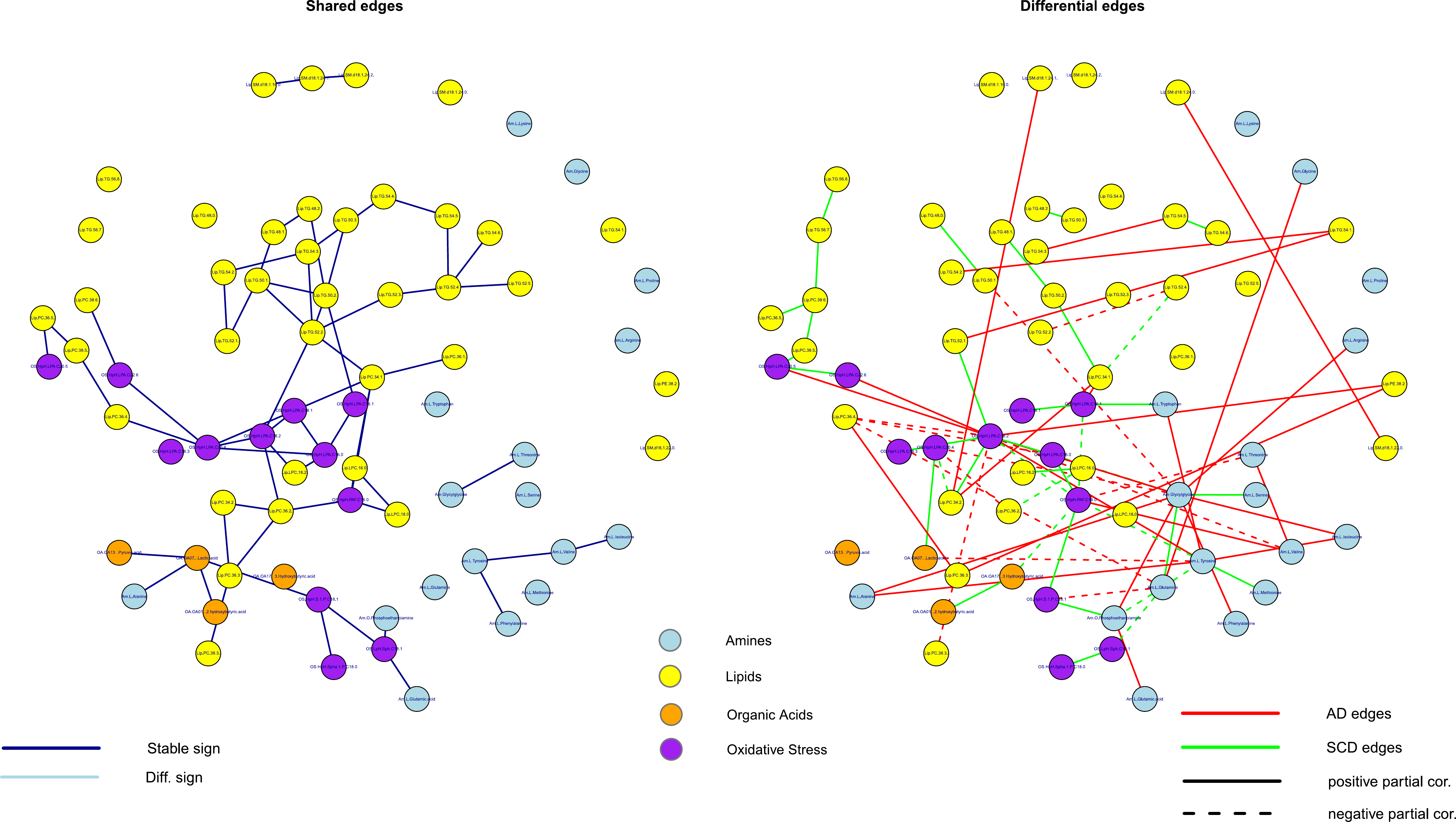}
    \caption{\footnotesize{Common and differential networks for the SCD group with no APOE $\epsilon$4 allele versus the AD group with at least 1 APOE $\epsilon$4 allele.
    The left-hand panel contains the network consisting of the edges (solid and colored blue) that are shared between these groups.
    The right-hand panel contains the network consisting of the edges that are unique for either of the groups.
    Red edges represent connections that are present in the AD group with at least 1 APOE $\epsilon$4 allele only.
    Green edges represent connections that are present in the SCD group with no APOE $\epsilon$4 allele only.
    Solid edges represent positive partial correlations while dashed edges represent negative partial correlations.
    The metabolite compounds are colored according to metabolite family: Blue for amines, yellow for lipids, orange for organic acids, and purple for oxidative stress.
    Note that the nodes in these networks have coordinates concordant with the node-placing of Figure \ref{FIG:Coords1}.
    Appears as Figure 4 in the main text.}}
  \label{FIG:DIFFgraphs1}
\end{figure}
\end{landscape}

The $\mathrm{SCD}^{\neg\epsilon4}$ and $\mathrm{AD}^{\epsilon4}$ topologies represent the most structured graphs.
Moreover, they represent (relatively) homogenous groups (in terms of AD pathology).
In assessing differential graph structures, we thus focus on these two topologies.
Figure \ref{FIG:DIFFgraphs1} then contains in the left-hand panel the network of shared connections and in the right-hand panel the network of unique connections between the $\mathrm{SCD}^{\neg\epsilon4}$ and $\mathrm{AD}^{\epsilon4}$ groups.
Table \ref{TABLE:CentralDiffSCDnaADa} then contains a list of metabolites with the highest degrees in the differential network (right-hand panel of \ref{FIG:DIFFgraphs1}).
These differential degrees indicate the metabolites that change their regulatory function (in terms of differential connections) the most between the $\mathrm{SCD}^{\neg\epsilon4}$ and $\mathrm{AD}^{\epsilon4}$ groups.
Overall, the most differentially wired metabolites across their topologies belong exclusively to the oxidative stress and amine compound-families.
The oxidative stress compound LPA C18:2 is well-connected in both the $\mathrm{SCD}^{\neg\epsilon4}$ and $\mathrm{AD}^{\epsilon4}$ topologies, but is wired very differently between them.
From this perspective LPA C18:2 is implied in the loss of normal and the gain of abnormal connections in the AD state driven by APOE $\epsilon$4.
The oxidative stress compound PAF C16:0 seems to be well-connected in the $\mathrm{SCD}^{\neg\epsilon4}$ group mostly.
Hence, this compound is implied in the loss of normal regulatory connections in the AD state driven by APOE $\epsilon$4.
The amine Glycylglycine seems to be well-connected in the $\mathrm{AD}^{\epsilon4}$ group mostly.
This compound is thus implied in the gain of abnormal regulatory connections in the AD state driven by APOE $\epsilon$4.
These connections are amongst amines predominantly.
This latter observation also, to a lesser degree, holds for the amines Tyrosine and Glutamine.

\begin{table}[h]
\centering
\caption{Most differentially wired metabolic features in the differential network for the SCD group with no APOE $\epsilon$4 allele versus the AD group with at least 1 APOE $\epsilon$4 allele.}
\label{TABLE:CentralDiffSCDnaADa}
\begin{tabular}{llr}
\hline
\hline
Feature             	& Compound class  	& Degree \\ \hline
LPA C18:2 		& Oxidative stress: Lyso-phosphatidic acid&          13\\
Glycylglycine       	& Amines		&          11\\
Tyrosine          	& Amines		&           8\\
PAF C16:0 		& Oxidative stress: Platelet activating factor&           8\\
Glutamine         	& Amines		&           7\\ \hline
\end{tabular}
\end{table}

\begin{table}[h]
\begin{scriptsize}
\centering
\caption{Centrality measures for the SCD group with no APOE $\epsilon$4 allele.}
\label{TABLE:CentralSCnoAPOE}
\begin{tabular}{lrllrllrllr}
\hline
\hline
\multicolumn{2}{c}{Degree} &  & \multicolumn{2}{c}{Betweenness} &  & \multicolumn{2}{c}{Closeness}   &  & \multicolumn{2}{c}{Eigenvalue}  \\
\cline{1-2} \cline{4-5} \cline{7-8} \cline{10-11}
LPA C18:2	     & 10  &  & PAF C16:0	 & 585.19 &  & PAF C16:0	  & 0.416 &  & LPA C18:2	& .405 \\
PAF C16:0	     & 10  &  & LPA C20:4	 & 507.93 &  & LPA C18:2	  & 0.406 &  & PAF C16:0	& .376 \\
TG(50:2)	     &  7  &  & LPA C18:2	 & 409.96 &  & LPA C16		  & 0.379 &  & LPA C16		& .330 \\
PC(34:1)	     &  7  &  & PC(34:1)	 & 363.92 &  & PC(34:1)		  & 0.375 &  & LPA C18:1	& .270 \\
LPA C16		     &  7  &  & Tyrosine 	 & 248.58 &  & LPA C20:4	  & 0.371 &  & LPA C20:4	& .254 \\
LPA C20:4	     &  7  &  & LPA C22:6	 & 243.67 &  & TG(52:2)		  & 0.354 &  & PC(34:1)		& .245 \\
TG(52:2)	     &  6  &  & LPA C16		 & 232.80 &  & LPA C18:1	  & 0.348 &  & PC(36:2)		& .233 \\
\hline
\end{tabular}
\end{scriptsize}
\end{table}

\begin{table}[h]
\begin{scriptsize}
\centering
\caption{Centrality measures for the SCD group with at least 1 APOE $\epsilon$4 allele.}
\label{TABLE:CentralSCAPOE}
\begin{tabular}{lrllrllrllr}
\hline
\multicolumn{2}{c}{Degree} &  & \multicolumn{2}{c}{Betweenness}     &  & \multicolumn{2}{c}{Closeness}     &  & \multicolumn{2}{c}{Eigenvalue}      \\
\cline{1-2} \cline{4-5} \cline{7-8} \cline{10-11}
Glycylglycine 		& 13 &  & Glycylglycine 	& 365.75 &  & Glycylglycine 		& .443 &  & Glycylglycine 	& .360 \\
LPA C18:2	 	& 12 &  & LPA C18:2		& 305.64 &  & LPA C18:2		 	& .429 &  & Tyrosine 		& .346 \\
PAF C16:0	 	& 12 &  & TG(52:2)		& 294.73 &  & Tyrosine 			& .427 &  & LPA C18:2 		& .290 \\
Tyrosine 		& 11 &  & Phenylalanine 	& 292.56 &  & Phenylalanine 		& .419 &  & PAF C16:0		& .281 \\
Phenylalanine 		&  9 &  & Tyrosine 		& 267.47 &  & TG(52:2)		 	& .411 &  & Phenylalanine 	& .279 \\
L-Lactic acid	 	&  9 &  & PAF C16:0		& 241.17 &  & PAF C16:0		 	& .410 &  & Glutamine 		& .272 \\
TG(52:2)	 	&  9 &  & L-Lactic acid		& 166.26 &  & Glutamine 		& .394 &  & TG(52:2)		& .255 \\
Glutamine 		&  7 &  & SPH C18:1		& 140.86 &  & L-Lactic acid	 	& .381 &  & L-Lactic acid	& .209 \\
PC(36:2)	 	&  6 &  & TG(50:2)		& 120.13 &  & PC(36:2)		 	& .368 &  & PC(36:2)		& .204 \\
SPH C18:1	 	&  6 &  & Glutamine 		& 111.86 &  & SPH C18:1		 	& .354 &  & SPH C18:1		& .184 \\
\hline
\end{tabular}
\end{scriptsize}
\end{table}

\begin{table}[h]
\begin{scriptsize}
\centering
\caption{Centrality measures for the AD group with no APOE $\epsilon$4 allele.}
\label{TABLE:CentralADnoAPOE}
\begin{tabular}{lrllrllrllr}
\hline
\multicolumn{2}{c}{Degree} &  & \multicolumn{2}{c}{Betweenness}     &  & \multicolumn{2}{c}{Closeness}     &  & \multicolumn{2}{c}{Eigenvalue}      \\
\cline{1-2} \cline{4-5} \cline{7-8} \cline{10-11}
Glycylglycine 		& 16 &  & Glycylglycine 	& 418.71 &  & Glycylglycine 		& .455 &  & LPA C18:2		 	& .369 \\
LPA C18:2 		& 14 &  & LPA C18:2 		& 376.70 &  & LPA C18:2 		& .446 &  & Glycylglycine 		& .356 \\
L-Lactic acid	 	& 12 &  & PC(34:2)		& 319.04 &  & PC(34:2)		 	& .436 &  & PC(34:2)		 	& .353 \\
PC(34:2)	 	& 12 &  & TG(52:2)		& 250.23 &  & L-Lactic acid	 	& .419 &  & L-Lactic acid 		& .331 \\
PAF C16:0	 	& 10 &  & L-Lactic acid		& 216.33 &  & PAF C16:0		 	& .405 &  & PAF C16:0		 	& .288 \\
Tyrosine 		&  6 &  & PC(34:1)		& 184.65 &  & PC(34:1)		 	& .367 &  & SPH C18:1		 	& .196 \\
TG(52:2)	 	&  6 &  & PAF C16:0		& 165.12 &  & SPH C18:1		 	& .360 &  & LPA C16		 	& .189 \\
PC(34:1)	 	&  6 &  & TG(50:1)		& 103.00 &  & LPA C16		 	& .348 &  & PC(34:1)		 	& .178 \\
PC(38:6) 		&  6 &  & SM(d18:1/16:0)	& 102.00 &  & LPA C20:4		 	& .348 &  & LPA C20:4		 	& .167 \\
SPH C18:1	 	&  6 &  & LPA C18:1		&  78.31 &  & PC(38:6)		 	& .342 &  & PC(38:6)		 	& .161 \\
LPA C16		 	&  6 &  & SPH C18:1		&  77.82 &  & Tyrosine		 	& .339 &  & S1P C18:1			& .158 \\
\hline
\end{tabular}
\end{scriptsize}
\end{table}

\begin{table}[h]
\begin{scriptsize}
\centering
\caption{Centrality measures for the AD group with at least 1 APOE $\epsilon$4 allele.}
\label{TABLE:CentralADAPOE}
\begin{tabular}{lrllrllrllr}
\hline
\multicolumn{2}{c}{Degree} &  & \multicolumn{2}{c}{Betweenness}    &  & \multicolumn{2}{c}{Closeness}      &  & \multicolumn{2}{c}{Eigenvalue}    \\
\cline{1-2} \cline{4-5} \cline{7-8} \cline{10-11}
LPA C18:2	 	& 13 &  & LPA C18:2		& 682.98 &  & LPA C18:2		      	& .452 &  & LPA C18:2   	& .473 \\
Glycylglycine 		&  9 &  & TG(52:2)		& 537.91 &  & Glycylglycine      	& .399 &  & LPA C20:4   	& .336 \\
Tyrosine 		&  7 &  & Glycylglycine 	& 388.77 &  & TG(52:2)      		& .398 &  & Glycylglycine   	& .279 \\
TG(52:2)	 	&  7 &  & TG(50:1)		& 232.91 &  & LPA C20:4		      	& .385 &  & PC(36:4)	   	& .236 \\
PC(34:1)	 	&  7 &  & PC(34:1)		& 231.83 &  & Tyrosine      		& .363 &  & PC(36:2)	   	& .230 \\
LPA C20:4	 	&  7 &  & Tyrosine 		& 231.23 &  & PC(34:1)		      	& .354 &  & LPA C18:1   	& .218 \\
PC(36:4)	 	&  6 &  & PC(36:4)		& 217.96 &  & PC(36:4)		      	& .354 &  & Tyrosine   		& .211 \\
PAF C16:0	 	&  6 &  & PC(34:2)		& 179.58 &  & Valine      		& .349 &  & Valine   		& .206\\
\hline
\end{tabular}
\end{scriptsize}
\end{table}

\end{bibunit}

\cleardoublepage

\renewcommand{\theequation}{S3.\arabic{equation}}
\renewcommand{\thefigure}{S3.\arabic{figure}}
\renewcommand{\thetable}{S3.\arabic{table}}
\renewcommand{\bibnumfmt}[1]{[S3.#1]}
\renewcommand{\citenumfont}[1]{S3.#1}
\renewcommand{\thesection}{\arabic{section}}

\setcounter{section}{0}
\setcounter{subsection}{0}
\setcounter{equation}{0}
\setcounter{figure}{0}
\setcounter{table}{0}
\setcounter{page}{1}

\phantomsection
\addcontentsline{toc}{section}{Supplementary Material}
\begin{center}
{\huge SUPPLEMENTARY TEXT 3\\~\\
Basic Results on all Cases}
\end{center}

\begin{bibunit}
\vspace{2cm}
This supplementary text presents, for purposes of comparison, basic information on the obtained expression and classification signatures when considering data from all patients.
That is, patients whose clinical diagnosis was discordant from their CSF-biomarker status were not excluded from the analyzes described below.
Processing of the data was analogous to the steps described in Section 2.4 of the main text and \emph{Section 1.2} of \emph{Supplementary Text 2}.
Again, metabolites with more than 10\% missing observations were removed, leading to the removal of the same 5 metabolites mentioned in \emph{Section 1.2} of \emph{Supplementary Text 2}.
Also, again three data samples were removed as their (plasma) quality was deemed unsure and an additional twelve data samples were removed due to instrumental errors in one or more MS platforms.
The final metabolic data set for the analyzes below thus contained $n = 285$ data samples (141 AD and 144 SCD) and $p = 230$ metabolic features.
Section \ref{SEC:SMT3:DES} contains information on the differential expression signature.
Section \ref{SEC:SMT3:CS} then contains information on the classification signature.
Section \ref{SEC:SMT3:reflect} concludes with some reflections on the findings.

\section{Differential Expression Signature}\label{SEC:SMT3:DES}
The approach for the evaluation of differential metabolic expression between AD and SCD subjects was described in \emph{Section 2.1.1} of \emph{Supplementary Text 2}.
The list of metabolic features that survive multiple testing correction when only sex and age are used as possible confounders can be found in Table \ref{Table:DiffExpressACsaONLY}.
Table \ref{Table:DiffExpressAC} then contains the list of metabolic features that survive multiple testing correction when correcting for all clinical variables of interest (see Table 1 of the main text).
All compounds in the latter table appear to be underexpressed in the AD group relative to the control group, expect for the Sphingomyelin SM(d18:1/20:1).

\begin{table}[t!]							
\begin{tiny}
\caption{Differentially expressed metabolites when correcting for sex and age only.}
\centering
\label{Table:DiffExpressACsaONLY}
\begin{tabular}{llrr}
 \hline\hline
 Metabolite & Compound class & $p$-value & Adjusted $p$-value\\
 \hline
2-Aminoadipic acid	          & Amines			 &6.085903e-08& 1.399758e-05  \\
Methyldopa                     	  & Amines			 &2.583401e-07& 2.970911e-05  \\
Valine                       	  & Amines			 &6.498148e-07& 4.981913e-05  \\
Tyrosine                     	  & Amines			 &1.697832e-06& 9.762531e-05  \\
Lysine                       	  & Amines			 &1.101781e-05& 5.068193e-04  \\
S-3-Hydroxyisobutyric acid	  & Organic acids		 &2.749051e-05& 1.053803e-03  \\
TG(54:6)	                  & Lipids: Triglycerides	 &5.209223e-05& 1.539997e-03  \\
TG(48:0)	                  & Lipids: Triglycerides	 &5.356511e-05& 1.539997e-03  \\
8-iso-PGF2a (15-F2t-IsoP)         & Oxidative stress: Isoprostane&7.698703e-05& 1.967446e-03  \\
TG(50:4)	                  & Lipids: Triglycerides	 &8.835841e-05& 2.032243e-03  \\
Methylmalonic acid	          & Organic acids		 &1.000548e-04& 2.092055e-03  \\
TG(52:4)                     	  & Lipids: Triglycerides	 &1.316578e-04& 2.523440e-03  \\
TG(48:2)                      	  & Lipids: Triglycerides	 &1.728144e-04& 2.798978e-03  \\
TG(50:3)                      	  & Lipids: Triglycerides	 &1.761190e-04& 2.798978e-03  \\
Leucine                      	  & Amines			 &1.825420e-04& 2.798978e-03  \\
TG(56:8)	                  & Lipids: Triglycerides	 &2.126884e-04& 2.886597e-03  \\
TG(51:3)	                  & Lipids: Triglycerides	 &2.317948e-04& 2.886597e-03  \\
TG(50:1)                      	  & Lipids: Triglycerides	 &2.324632e-04& 2.886597e-03  \\
TG(48:3)                      	  & Lipids: Triglycerides	 &2.384581e-04& 2.886597e-03  \\
TG(52:5)                      	  & Lipids: Triglycerides	 &2.827883e-04& 3.182738e-03  \\
TG(50:2)                          & Lipids: Triglycerides	 &2.905978e-04& 3.182738e-03  \\
TG(46:2)                      	  & Lipids: Triglycerides	 &3.557556e-04& 3.610453e-03  \\
TG(48:1)                     	  & Lipids: Triglycerides	 &3.610453e-04& 3.610453e-03  \\
TG(50:0)                     	  & Lipids: Triglycerides	 &4.318757e-04& 4.138809e-03  \\
TG(52:3)                      	  & Lipids: Triglycerides	 &5.484129e-04& 5.045398e-03  \\
TG(56:7)                      	  & Lipids: Triglycerides	 &8.909861e-04& 7.881800e-03  \\
Isoleucine                   	  & Amines			 &9.404737e-04& 8.011442e-03  \\
PGD2                       	  & Oxidative stress: Prostaglandins&9.789025e-04& 8.040984e-03  \\
LPC(18:1)                         & lipids: Lysophosphatidylcholine&1.210008e-03& 9.506953e-03  \\
TG(52:1)                      	  & Lipids: Triglycerides	 &1.240037e-03& 9.506953e-03  \\
SM(d18:1/20:1)                	  & Lipids: Sphingomyelins	 &1.592411e-03& 1.176462e-02  \\
1-Methylhistidine	          & Amines			 &1.636817e-03& 1.176462e-02  \\
5-iPF2a VI 			  & Oxidative stress: Isoprostane&2.014948e-03& 1.404357e-02  \\
2-hydroxybutyric acid		  & Organic acids		 &2.242011e-03& 1.506750e-02  \\
TG(54:5)                      	  & Lipids: Triglycerides	 &2.292880e-03& 1.506750e-02  \\
SM(d18:1/23:0)                	  & Lipids: Sphingomyelins	 &2.583397e-03& 1.607458e-02  \\
TG(51:1)                      	  & Lipids: Triglycerides	 &2.639813e-03& 1.607458e-02  \\
TG(58:10)                     	  & Lipids: Triglycerides	 &2.655801e-03& 1.607458e-02  \\
TG(51:2)                      	  & Lipids: Triglycerides	 &2.892979e-03& 1.706116e-02  \\
TG(46:1)                      	  & Lipids: Triglycerides	 &3.683727e-03& 2.118143e-02  \\
LPA C14:0                  	  & Oxidative stress: Lyso-phosphatidic acid&4.021010e-03& 2.255689e-02  \\
3-Hydroxyisovaleric acid	  & Organic acids		 &4.778686e-03& 2.616899e-02  \\
TG(58:9)                      	  & Lipids: Triglycerides	 &5.031803e-03& 2.691430e-02  \\
Histidine                    	  & Amines			 &6.498831e-03& 3.397116e-02  \\
DG(36:3)	                  & Lipids: Diacylglycerol	 &7.216737e-03& 3.688554e-02  \\
Phenylalanine	                  & Amines			 &7.730540e-03& 3.812873e-02  \\
TG(52:2)                     	  & Lipids: Triglycerides	 &7.791523e-03& 3.812873e-02  \\
SM(d18:1/24:2)                	  & Lipids: Sphingomyelins	 &8.365093e-03& 4.008274e-02  \\
PC(O-44:5)                        & Lipids: Plasmalogen Phosphatidylcholine&8.717054e-03& 4.091679e-02  \\
8,12-iPF2a IV             	  & Oxidative stress: Isoprostane&9.768359e-03& 4.493445e-02  \\
TG(46:0)                      	  & Lipids: Triglycerides	 &1.042619e-02& 4.702009e-02  \\
Methionine                   	  & Amines			 &1.133966e-02& 4.952164e-02  \\
LPA C20:1	                  & Oxidative stress: Lyso-phosphatidic acid&1.141151e-02& 4.952164e-02  \\	
 \hline							
\end{tabular}						
\end{tiny}					
\end{table}						

\begin{table}[t!]
\begin{tiny}
\caption{Differentially expressed metabolites when correcting for all clinical variables.}
\centering
\label{Table:DiffExpressAC}
\begin{tabular}{llrr}
 \hline\hline
 Metabolite & Compound class & $p$-value & Adjusted $p$-value\\
 \hline
Tyrosine                     	  & Amines			& 1.293884e-05 & 0.001941711   \\
2-Aminoadipic acid	          & Amines			& 2.118905e-05 & 0.001941711   \\
8-iso-PGF2a (15-F2t-IsoP)         & Oxidative stress: Isoprostane& 2.647672e-05 & 0.001941711   \\
Methyldopa                     	  & Amines			& 3.376889e-05 & 0.001941711   \\
TG(54:6)	                  & Lipids: Triglycerides	& 7.936780e-05 & 0.003650919   \\
TG(56:8)	                  & Lipids: Triglycerides	& 1.064855e-04 & 0.003864998   \\
Valine                       	  & Amines			& 1.176304e-04 & 0.003864998   \\
TG(50:4)	                  & Lipids: Triglycerides	& 1.415809e-04 & 0.004070451   \\
S-3-Hydroxyisobutyric acid	  & Organic acids		& 2.261572e-04 & 0.005413816   \\
TG(52:4)                     	  & Lipids: Triglycerides	& 2.353833e-04 & 0.005413816   \\
TG(51:3)	                  & Lipids: Triglycerides	& 2.779465e-04 & 0.005697531   \\
PGD2                       	  & Oxidative stress: Prostaglandins& 2.972625e-04 & 0.005697531   \\
Lysine                       	  & Amines			& 3.235788e-04 & 0.005724855   \\
TG(52:5)                      	  & Lipids: Triglycerides	& 3.734782e-04 & 0.005825832   \\
TG(56:7)                      	  & Lipids: Triglycerides	& 3.799456e-04 & 0.005825832   \\
TG(48:3)                      	  & Lipids: Triglycerides	& 4.059878e-04 & 0.005836075   \\
TG(50:3)                      	  & Lipids: Triglycerides	& 6.422663e-04 & 0.008689485   \\
TG(46:2)                      	  & Lipids: Triglycerides	& 8.831594e-04 & 0.011284814   \\
TG(52:3)                      	  & Lipids: Triglycerides	& 9.803097e-04 & 0.011866907   \\
TG(58:10)                     	  & Lipids: Triglycerides	& 1.166642e-03 & 0.013416381   \\
SM(d18:1/23:0)                	  & Lipids: Sphingomyelins	& 1.259544e-03 & 0.013795004   \\
TG(48:2)                      	  & Lipids: Triglycerides	& 1.442328e-03 & 0.014514992   \\
TG(58:9)                    	  & Lipids: Triglycerides	& 1.451499e-03 & 0.014514992   \\
TG(48:0)	                  & Lipids: Triglycerides	& 1.516374e-03 & 0.014531917   \\
5-iPF2a VI 			  & Oxidative stress: Isoprostane& 1.584192e-03 & 0.014574562   \\
SM(d18:1/20:1)                	  & Lipids: Sphingomyelins	& 1.899763e-03 & 0.016805592   \\
Methylmalonic acid	          & Organic acids		& 2.206887e-03 & 0.018156831   \\
3-Hydroxyisovaleric acid	  & Organic acids		& 2.210397e-03 & 0.018156831   \\
Ornithine                      	  & Amines			& 3.068127e-03 & 0.024137545   \\
TG(50:2)                          & Lipids: Triglycerides	& 3.148375e-03 & 0.024137545   \\
Leucine                      	  & Amines			& 3.330067e-03 & 0.024706952   \\
TG(54:5)                      	  & Lipids: Triglycerides	& 3.448227e-03 & 0.024784134   \\
DG(36:3)	                  & Lipids: Diacylglycerol	& 3.614235e-03 & 0.025190125   \\
TG(48:1)                     	  & Lipids: Triglycerides	& 3.740068e-03 & 0.025300461   \\
TG(50:0)                     	  & Lipids: Triglycerides	& 3.923739e-03 & 0.025784569   \\
Phenylalanine	                  & Amines			& 4.059826e-03 & 0.025937777   \\
TG(50:1)                      	  & Lipids: Triglycerides	& 4.237719e-03 & 0.026342577   \\
O-Acetylserine                    & Amines			& 4.959665e-03 & 0.030019024   \\
8,12-iPF2a IV             	  & Oxidative stress: Isoprostane& 5.360289e-03 & 0.031324259   \\
TG(51:2)                      	  & Lipids: Triglycerides	& 5.447697e-03 & 0.031324259   \\
SM(d18:1/25:0)	                  & Lipids: Sphingomyelins	& 7.146770e-03 & 0.039619151   \\
NO2-aLA (C18:3)                   & Oxidative stress: Nitro-Fatty acid& 7.234801e-03 & 0.039619151   \\
LPA C18                  	  & Oxidative stress: Lyso-phosphatidic acid& 7.416156e-03 & 0.039667810   \\
LPA C14:0                  	  & Oxidative stress: Lyso-phosphatidic acid& 8.866032e-03 & 0.046345167   \\
 \hline							
\end{tabular}						
\end{tiny}					
\end{table}

\section{Classification Signature}\label{SEC:SMT3:CS}
\begin{figure}[]
\centering
  \includegraphics[width=\textwidth]{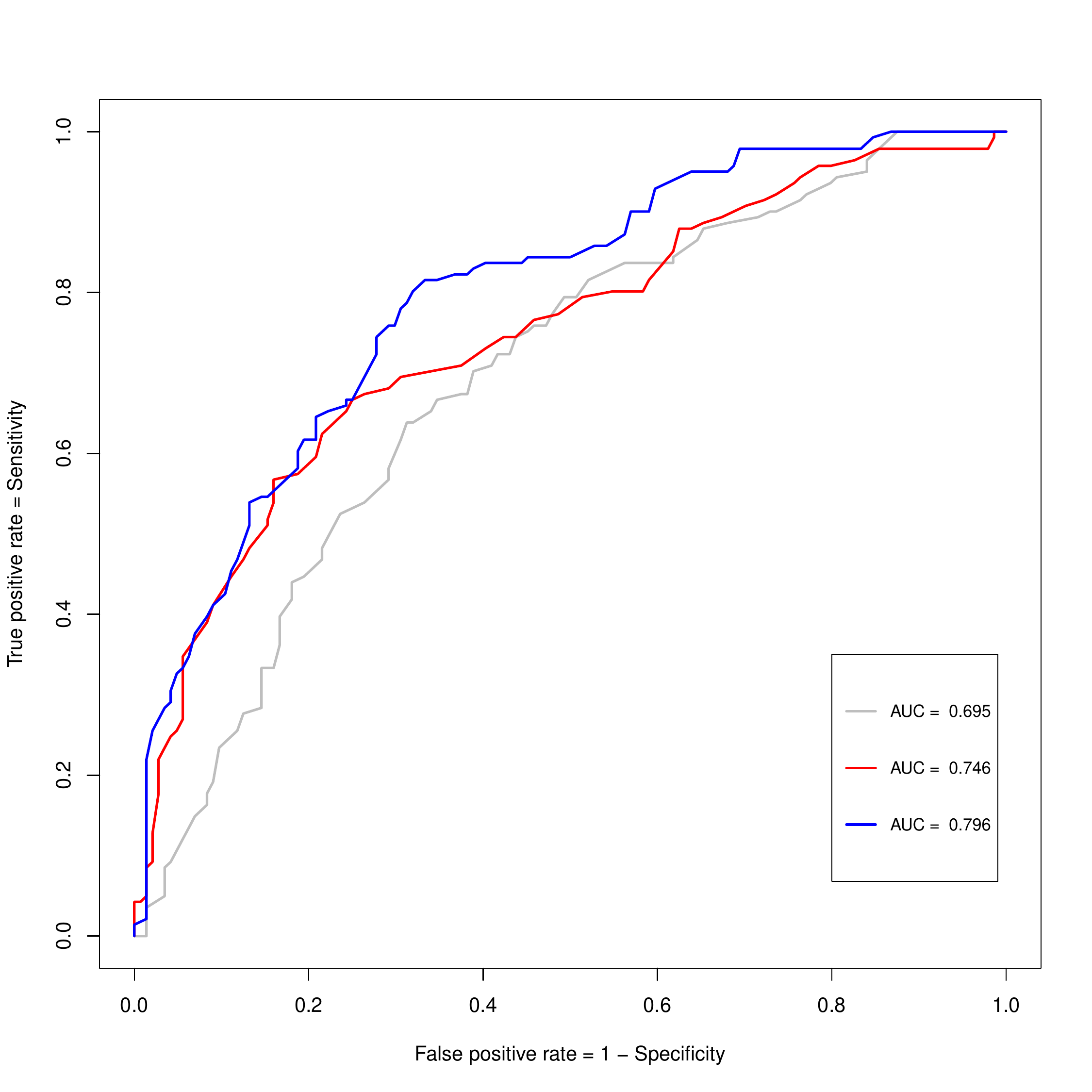}
    \caption{\footnotesize{ROC curves for the classification models.
    The grey line represents the ROC curve for the unpenalized logistic regression model that entertains the clinical characteristics only.
    The red line represents the ROC curve for the logistic model in which the Lasso performed variable selection amongst the metabolites (and that does not consider the clinical characteristics).
    The blue line represents the ROC curve of the logistic model in which the clinical characteristics are present while the Lasso may select amongst the metabolites.}}
  \label{FIG:ROCsACw}
\end{figure}

The approach for the construction of classification signatures was described in in \emph{Section 2.2.1} of \emph{Supplementary Text 2}.
Model performances can be found in Figure \ref{FIG:ROCsACw}.
The prediction model carrying the clinical variables only resulted in an AUC of .695.
The model that used the Lasso for selection amongst the metabolites sorts a somewhat better classification performance, yielding an AUC of .746.
The model that adds a (Lasso-based) selection of metabolites to the clinical variables then improves predictive performance along the full false positive rate range, sorting a AUC of .796.
Table \ref{Table:ClassSITU1AC} contains the metabolites selected in the selection-amongst-metabolites-only situation.
Table \ref{Table:ClassSITU2AC} then contains the metabolites selected in the selection-amongst-metabolites-whilst-clinical-variables-present situation.
We see, for the compounds that also occur in the differential expression signature, that the signs of their effects concur with the pattern of AD-associated under- and overexpression.

\begin{table}[t!]
\begin{tiny}
\caption{Selected metabolites and parameter estimates when considering only metabolites as potential predictors.}
\centering
\label{Table:ClassSITU1AC}
\begin{tabular}{llr}
 \hline\hline
 Metabolite & Compound class & $\hat{\beta}$\\
 \hline
LPC(18:1)                      	  & lipids: Lysophosphatidylcholine	&$ 0.3775846437$	\\
Methyldopa                     	  & Amines				&$-0.3280999265$	\\
PGD2                       	  & Oxidative stress: Prostaglandins	&$-0.3136491690$	\\
NO2-aLA (C18:3)                   & Oxidative stress: Nitro-Fatty acid	&$-0.2042163853$	\\
O-Acetylserine                    & Amines				&$-0.1811852503$	\\
SM(d18:1/23:0)                	  & Lipids: Sphingomyelins		&$-0.1711821392$	\\
Tyrosine                     	  & Amines				&$-0.1588698635$	\\
SM(d18:1/20:1)                	  & Lipids: Sphingomyelins		&$ 0.1375561855$	\\
5-iPF2a VI 			  & Oxidative stress: Isoprostane	&$-0.1296136239$	\\
Glyceric acid        		  & Organic acids			&$-0.1271578713$	\\
PC(O-34:3)	                  & Lipids: Plasmalogen Phosphatidylcholine&$-0.1154877989$ \\
TG(54:6)	                  & Lipids: Triglycerides		&$-0.1099736486$	\\
8,12-iPF2a IV             	  & Oxidative stress: Isoprostane	&$-0.1091228644$	\\
Methylmalonic acid	          & Organic acids			&$-0.1021936840$	\\
TG(48:2)                      	  & Lipids: Triglycerides		&$-0.0880997011$	\\
8-iso-PGF2a (15-F2t-IsoP)         & Oxidative stress: Isoprostane	&$-0.0859696149$	\\
LPA C16	                  	  & Oxidative stress: Lyso-phosphatidic acid&$-0.0712895571$ \\
2,3-dinor-8-iso-PGF2a		  & Oxidative stress: Isoprostane	&$-0.0660850151$	\\
TG(O-50:0)	                  & Lipids: Triglycerides		&$ 0.0631492242$	\\
TG(48:0)	                  & Lipids: Triglycerides		&$-0.0594940757$	\\
PC(O-38:6)	                  & Lipids: Plasmalogen Phosphatidylcholine&$-0.0573684699$ \\
LPA C14:0                  	  & Oxidative stress: Lyso-phosphatidic acid&$-0.0462058983$ \\
Serine                    	  & Amines				& $0.0427658066$	\\
Putrescine                  	  & Amines				&$-0.0402777537$	\\
LPA C22:4              		  & Oxidative stress: Lyso-phosphatidic acid& $0.0336067651$ \\
Lysine                       	  & Amines				&$-0.0308966941$	\\
3-Methoxytyramine	          & Amines				&$-0.0171013575$	\\
cLPA C18:1 			  & Oxidative stress: Cyclic-lyso-phosphatidic acid&$-0.0127374834$ \\
PE(O-38:5)	                  & Lipids: Plasmalogen Phosphatidylethanolamine&$-0.0079640017$ \\
2-Aminoadipic acid	          & Amines				&$-0.0069680882$	\\
Carnosine                 	  & Amines				&$ 0.0022778247$	\\
LPC(20:4)	                  & Lipids: Lysophosphatidylcholine	&$ 0.0006342159$	\\
 \hline			     							
\end{tabular}		     							
\end{tiny}	     							
\end{table}		     							
			     							   							
\begin{table}[t!]				
\begin{tiny}
\caption{Selected metabolites and parameter estimates when considering metabolites as potential predictors on top of the clinical variables.}
\centering
\label{Table:ClassSITU2AC}
\begin{tabular}{llr}
 \hline\hline				
 Metabolite & Compound class & $\hat{\beta}$\\				
 \hline				
PGD2                       	  & Oxidative stress: Prostaglandins			&-0.48204926   \\
SM(d18:1/20:1)                	  & Lipids: Sphingomyelins				& 0.33329421   \\
O-Acetylserine                    & Amines						&-0.28384663   \\
Methyldopa                     	  & Amines						&-0.27782490   \\
8-iso-PGF2a (15-F2t-IsoP)         & Oxidative stress: Isoprostane			&-0.25085235   \\
NO2-aLA (C18:3)                   & Oxidative stress: Nitro-Fatty acid			&-0.24269712   \\
Methylmalonic acid	          & Organic acids					&-0.22969766   \\
Tyrosine                     	  & Amines						&-0.17320260   \\
TG(52:4)	                  & Lipids: Triglycerides				&-0.15033254   \\
Gamma-glutamylalanine      	  & Amines						& 0.13963417   \\
Putrescine                        & Amines						&-0.13769196   \\
8,12-iPF2a IV             	  & Oxidative stress: Isoprostane			&-0.13656275   \\
TG(51:3)	                  & Lipids: Triglycerides				&-0.13258986   \\
SM(d18:1/23:0)                	  & Lipids: Sphingomyelins				&-0.11398902   \\
Citrulline                        & Amines						&-0.10005829   \\
Glyceric acid        		  & Organic acids					&-0.09690281   \\
LPC(18:1)                         & lipids: Lysophosphatidylcholine			& 0.09252004   \\
PC(O-34:3)	                  & Lipids: Plasmalogen Phosphatidylcholine		&-0.09149649   \\
3-Hydroxyisovaleric acid	  & Organic acids					&-0.08574230   \\
5-iPF2a VI 			  & Oxidative stress: Isoprostane			&-0.08026972   \\
LPA C14:0                  	  & Oxidative stress: Lyso-phosphatidic acid		&-0.07726029   \\
2,3-dinor-8-iso-PGF2a		  & Oxidative stress: Isoprostane			&-0.07467916   \\
TG(58:10)	                  & Lipids: Triglycerides				&-0.06409276   \\
Cysteine                          & Amines						& 0.04297397   \\
Carnosine                     	  & Amines						& 0.03408938   \\
PC(O-34:2)	                  & Lipids: Plasmalogen Phosphatidylcholine		&-0.03066565   \\
SM(d18:1/25:0)	                  & Lipids: Sphingomyelins				&-0.01019553   \\			
 \hline					
\end{tabular}				
\end{tiny}				
\end{table}

\section{Some Reflections}\label{SEC:SMT3:reflect}
Of the 285 subjects included in the analyzes above a total of 37 had a CSF-biomarker status discordant with their clinical diagnosis.
That is, these subjects were either clinically diagnosed with AD while their CSF-markers were normal ($\mbox{t-tau}/\mathrm{A}\beta_{42} \leq 0.52 $) or clinically diagnosed as normal while their CSF-markers indicated AD ($\mbox{t-tau}/\mathrm{A}\beta_{42} > 0.52 $).
The disease status of these subjects is thus unsure as it is unclear which diagnosis (clinical or biomarker-based) should take precedence.
Hence, the main analyzes revolved around those cases in which the clinical and biomarker-based diagnoses were concordant.
Below we reflect on the findings above in relation to the main analyzes.

The larger sample size implies that we have more power in finding differentially expressed metabolites.
Hence, Tables \ref{Table:DiffExpressACsaONLY} and \ref{Table:DiffExpressAC} list more metabolites than their corresponding tables in \emph{Supplementary Text 2} (\emph{S2.3} and \emph{S2.2}) and the Main text (Table 3).
As the increase in power comes from subjects whose disease status is somewhat unsure, the results in Tables \ref{Table:DiffExpressACsaONLY} and \ref{Table:DiffExpressAC} are somewhat tentative.
These results are, however, concordant with the results from the main analyzes, in the sense that they overlap, i.e.: The metabolites listed in Table \emph{S2.3} of \emph{Supplementary Text 2} (see also column 3 of Table 3 of the Main text) are a proper subset (except for Proline, PC(O-34:1), LPC(20:4), SM(18:1/16:0), and Ornithine) of the metabolites listed in Table \ref{Table:DiffExpressACsaONLY}; and the metabolites listed in Table \emph{S2.2} of \emph{Supplementary Text 2} (see also column 4 of Table 3 of the Main text) are a proper subset of the metabolites listed in Table \ref{Table:DiffExpressAC}.

The main model of interest -- selecting metabolites on top of clinical predictors -- has a concordant performance on the full and CSF-confirmed data: in both instances the model sorts an AUC of approximately .79.
Moreover, most of the (top) compounds in Table \emph{S2.5} of \emph{Supplementary Text 2} appear as selected (top) compounds in Table \ref{Table:ClassSITU2AC}.
The insecurity regarding the status of subjects with discordant diagnoses is, however, reflected somewhat in the classification signatures.
For example, if we would take, for the main model of interest, the logistic cut-off at the optimal cut-off in terms of accuracy (.42, as determined by 10-fold CV), then we see that those clinically diagnosed with AD while having a normal CSF-status (14) tend to be predominantly classified as AD cases while those clinically diagnosed as normal while having an AD CSF-status (23) seem to be randomly classified as either AD or control cases.

Insecurity regarding the true status of those with a discordant CSF-biomarker status implies that the class-specific (AD or SCD) samples are heterogeneous.
Heterogeneity can lead to the dilution of partial correlations and, hence, may hamper network extraction \cite{Wessel15}.
Results on the regulatory signature are, for reasons of brevity, not included.


\end{bibunit}

\addresseshere

\end{document}